%% file: Main.tex
\documentclass[12pt]{article}
\pdfoutput=1

\include{Header}

\begin{document}

\input{Title}
	
\input{Introduction}

\input{Preliminaries}

\input{QuantumMoney}

\input{QuantumAlgorithms}
	
\input{Conclusion}

\newpage

\def\bibfont{\footnotesize}
{\linespread{1.0}
\bibliographystyle{econ}
\bibliography{References}
}
\appendix
\input{Appendix}

\end{document}

%% file: Header.tex
\usepackage[table]{xcolor}
\definecolor{lightgray}{gray}{0.9}
\usepackage[onehalfspacing]{setspace}
\usepackage[linesnumbered,ruled,vlined]{algorithm2e}
\usepackage{amsmath,amssymb,amsfonts,amsthm}
\usepackage[final]{graphicx}
\usepackage{pgfpages}
\usepackage{rotating}
\usepackage{pdflscape}
\usepackage{tablefootnote}
\usepackage{pdfpages}
\usepackage{float}
\usepackage{bm}
\usepackage{bbm}
\usepackage{multirow}
\usepackage{ulem}
\usepackage{multicol}
\usepackage{booktabs}

\usepackage{lscape}
\usepackage{enumitem}
\usepackage[font=footnotesize]{caption}
\usepackage{physics}
\usepackage{siunitx}
\usepackage{natbib}
\usepackage{geometry}
\usepackage{authblk}
\usepackage{tikzsymbols}
\usepackage{fontawesome}
\usepackage{pifont}
\usepackage{scrextend}
\usepackage[hidelinks,draft=false,pageanchor,hyperindex,breaklinks]{hyperref}

\usepackage{qcircuit}
\usepackage{adjustbox}

\usepackage{textcomp} 
\usepackage{ifdraft,color}
\ifdraft{\newcommand{\authnote}[3]{{\color{#3} {\bf  #1:} #2}}}{\newcommand{\authnote}[3]{}}

\newtheorem{theorem}{Theorem}

  \theoremstyle{definition}
  
  \theoremstyle{remark}

  \theoremstyle{plain}

\newcommand*\circled[1]{\tikz[baseline=(char.base)]{
            \node[shape=circle,draw,inner sep=2pt] (char) {#1};}}
\newcommand{\rot}[1]{\rotatebox[origin=c]{-90}{#1}}
\usepackage[weather]{ifsym}

\newcommand{\Bill}{\framebox[4.0\width]{\$}}
\newcommand{\Coin}{\circled{$\, $¢$\, $ }}
\newcommand{\Bitcoin}{\circled{\faBtc}}

\newcommand{\cmark}{\ding{51}}%
\newcommand{\xmark}{\ding{55}}%

\usepackage{titlesec}

\titleformat{\paragraph}
{\normalfont\normalsize\bfseries}{\theparagraph}{1em}{}
\titlespacing*{\paragraph}
{0pt}{3.25ex plus 1ex minus .2ex}{1.5ex plus .2ex}

%% file: Title.tex
\begin{titlepage}

\title{\vspace{-20mm} Quantum Technology for Economists\footnote{The opinions expressed in this article are the sole responsibility of the authors and should not be interpreted as reflecting the views of Sveriges Riksbank.}}

\author[1]{Isaiah Hull\footnote{Correspondence Address: Research Division, Sveriges Riksbank, SE-103 37, Stockholm, Sweden. Email: isaiah.hull@riksbank.se. Tel: +46 076 589 0661. Fax: +46 8 0821 05 31.}}
\author[2]{Or Sattath}
\author[3]{Eleni Diamanti}
\author[4]{G{\"o}ran Wendin}

\affil[1]{Research Division, Sveriges Riksbank, Stockholm, Sweden}
\affil[2]{Department of Computer Science, Ben-Gurion University, Beersheba, Israel}
\affil[3]{LIP6, CNRS, Sorbonne Universit{\'e}, 75005 Paris, France}
\affil[4]{Department of Microtechnology and Nanoscience, Chalmers University of Technology, Gothenburg, Sweden}

\date{}
\clearpage

\date{December 2020}

\maketitle

\vspace{-8mm}

\singlespacing
\begin{center}
\textbf{Abstract}
\end{center}
\footnotesize
\noindent Research on quantum technology spans multiple disciplines: physics, computer science, engineering, and mathematics. The objective of this manuscript is to provide an accessible introduction to this emerging field for economists that is centered around quantum computing and quantum money. We proceed in three steps. First, we discuss basic concepts in quantum computing and quantum communication, assuming knowledge of linear algebra and statistics, but not of computer science or physics. This covers fundamental topics, such as qubits, superposition, entanglement, quantum circuits, oracles, and the no-cloning theorem. Second, we provide an overview of quantum money, an early invention of the quantum communication literature that has recently been partially implemented in an experimental setting. One form of quantum money offers the privacy and anonymity of physical cash, the option to transact without the involvement of a third party, and the efficiency and convenience of a debit card payment. Such features cannot be achieved in combination with any other form of money. Finally, we review all existing quantum speedups that have been identified for algorithms used to solve and estimate economic models. This includes function approximation, linear systems analysis, Monte Carlo simulation, matrix inversion, principal component analysis, linear regression, interpolation, numerical differentiation, and true random number generation. We also discuss the difficulty of achieving quantum speedups and comment on common misconceptions about what is achievable with quantum computing.

\vspace{5mm}


\end{titlepage}

%% file: Introduction.tex
\section{Introduction}
\label{Introduction}

The field of quantum technology is divided into four broad areas: quantum computing, quantum simulation, quantum communication, and quantum sensing. Quantum computing centers around the exploitation of quantum physical phenomena, such as superposition and entanglement, to perform computation. Quantum simulation involves the development and use of specialized devices to simulate a specific quantum physical processes. Quantum communication studies how quantum phenomena can be used to securely transmit information between parties. And quantum sensing and metrology make use of quantum phenomena to produce more accurate sensors and measurement devices than could be created using existing classical technologies.

Research on quantum technology has largely been confined to a discussion between computer scientists, physicists, engineers, and mathematicians. Our objective in this manuscript is to widen the conversation to include economists, focusing on two areas in which quantum technology is likely to have relevance for the discipline: (1) the use of quantum computing to solve and estimate economic models; and (2) the use of quantum communication to construct forms of currency called ``quantum money,'' which have novel properties that cannot be achieved without exploiting quantum phenomena.\footnote{Those who have a broader interest in quantum technology may want to see \citet{JCJ14} for an overview of quantum simulation; \citet{GT07} for an overview of quantum communication; and \citet{GLM11} and \citet{DRP17} for an overview of quantum sensing and metrology.}

Our examination of quantum computing will focus primarily on quantum algorithms, but will also provide a brief overview of the experimental efforts to develop quantum computing devices. Similarly, our discussion of quantum money will center on theoretical constructions, but will also review experimental progress in its implementation. Throughout the manuscript, we will assume that the reader has no knowledge of physics, but is familiar with probability theory and linear algebra. Furthermore, we will provide a sufficient amount of low-level detail to enable economists to identify points of entry into the existing literature and to contribute with novel research. An econometrician, for instance, will be able to identify what problems remain in the construction of quantum versions of familiar classical algorithms, such as ordinary least squares (OLS) and principal component analysis (PCA).

We will start our exploration of quantum computing and quantum money with an overview of preliminary material, limiting ourselves to a narrow selection of topics that will provide a foundation for understanding basic algorithms and money schemes. This material covers a mathematical and notational description of quantum computers and their functions, including descriptions of the creation and manipulation of quantum states. It also covers theory relevant to the construction of quantum money schemes. Part of the purpose of this section is to communicate how quantum physical phenomena, such as superposition, entanglement, and the no-cloning theorem provide new computational and cryptographic resources.

We will also see how quantum states can be manipulated using quantum operations to perform computation. This includes a description of what types of operations are permissible and how this translates into mathematics. Once the operations have been executed and the computation is complete, we must offload the results to a classical computer. Since the results take the form of a quantum state, we will need to perform measurement, a process that yields a classical state and is analogous to sampling. Understanding measurement will clarify why quantum computers are not simply classical computers with a special capacity for parallel computation: we can only output as much classical information as we input.

In addition to providing a mathematical and notational description of quantum computation, we will also discuss the practicalities of implementing computations in the form of quantum circuits. Such circuits can be simulated classically or run on a quantum computer. We will also introduce the notion of oracles from computer science, which we will use in some cases to determine the size of a quantum speedup; and the no-cloning theorem \citep{WZ82}, which will be essential for the construction of quantum money schemes.

After covering the preliminary material, we will discuss two topics of interest for economists: quantum money and quantum algorithms. The original motivation for quantum money, as given in~\cite{Wie83}, was to construct a form of currency that was ``physically impossible to counterfeit.'' This differs categorically from existing forms of money, which do not exploit quantum phenomena and are therefore vulnerable to attack from counterfeiters. In addition to reverse-engineering threads and inks, and breaking encryption schemes, an attacker could, in principle, copy any ``classical'' form of money bit-by-bit or even atom-by-atom, as no physical law prohibits it.\footnote{Here, we use the term ``classical'' to indicate that the money or payment instrument does not make use of quantum physical phenomena.}

Indeed, such attacks are not merely of theoretical interest. Counterfeiting necessitates costly periodic note and coin re-designs, and forces the general public to do currency checking~\citep{QS15}. State actors have also used counterfeiting to circumvent international sanctions and conduct economic warfare.\footnote{Large-scale counterfeiting has been attempted as a means of undermining public confidence in the monetary system. During World War II, for instance, a Nazi plot called ``Operation Bernhard'' exploited Jewish prisoners in an attempt to counterfeit large quantities of British pounds, with the intention to circulate them via an airdrop~\citep{Pir62}. The Bank of England responded by withdrawing notes above \pounds 5 from circulation. See: \url{https://www.bankofengland.co.uk/museum/online-collections/banknotes/counterfeit-and-imitation-notes}. There are also historic records of mass counterfeiting attempts by England during the French revolution~\citep[p. 33]{Dil1877}.} Prior to the development of fiat currencies, gold and other forms of commodity money relied on intrinsic worth, natural scarcity, and widespread cognizability to safeguard their value against attacks. Even with these natural advantages, high-quality counterfeits were still produced and passed to uninformed merchants.\footnote{Certain forms of commodity money can also be synthesized from other materials. Gold, for instance, can be synthesized, but not yet cost-effectively~\citep{AML+81}. Even if a form of commodity money's value is safeguarded against large-scale synthesization, the discovery of new deposits or improvements in extraction technology constitute supply shocks that could lead to substantial devaluations.} In contrast to existing forms of currency, quantum money is protected by the no-cloning theorem, which makes it impossible to counterfeit by the laws of physics. Along with post-quantum cryptography and quantum key distribution, it also provides a means of protecting the payments system against future quantum attacks.\footnote{\citet{Sho94} introduced a near-exponential quantum speedup to prime factorization, which compromises the commonly-used RSA encryption algorithm. Cryptocurrencies, such as Bitcoin, are also subject to attack from quantum computers \citep{ABL+18}.}

Our overview of quantum money starts with a full description of the original scheme, which was introduced circa 1969, but only published later in \citet{Wie83}. We will see that it achieves what is called ``information-theoretic security,'' which means that an attacker with unbounded classical and quantum resources will not be able to counterfeit a unit of the money.\footnote{Technically, such an adversary might successfully counterfeit the money, but this happens with exponentially small probability in the size of the quantum system. Therefore, by constructing a large enough quantum system, the success probability could easily be made $2^{-100}$, which is effectively $0$ for all practical concerns.} Since this original scheme was proposed, the term ``quantum money'' has come to refer to a broad variety of different payment instruments, including credit cards, bills, and coins, all of which use of quantum physical phenomena to achieve security. 

The real promise of quantum money is that it offers the possibility of combining the beneficial features of both physical cash and digital payments, which is not possible without the use of the higher standard of security quantum money offers. In particular, a form of currency called ``public-key'' quantum money would allow individuals to verify the authenticity of bills and coins publicly and without the need to communicate with a trusted third party. This is not possible with any classical form of digital of money, including cryptocurrencies, which at least require communication with a distributed ledger. Thus, quantum money could restore the privacy and anonymity associated with physical money transactions, while maintaining the convenience of digital payment instruments.

While quantum money offers features that are unachievable in any classical form of currency, implementing a full quantum money scheme requires additional advances in quantum technology. However, recent progress in the development of quantum money has moved us closer to a full implementation. Partial schemes have already been experimentally implemented for variants of private-key quantum money \citep{BOV+17, GAA+18,BCC+17,BBP17}. In all cases, quantum memory remains the primary bottleneck to a full implementation, as existing technologies are not able to retain a quantum state for longer than a fraction of a second. Furthermore, the challenges to implementation are even more substantial for public-key money schemes, which have not yet been partially implemented in an experimental setting. Public-key quantum money also faces formidable theoretical challenges, which may be of particular interest to those working on mechanism design.

In addition to quantum money, we also examine quantum algorithms, which offer the possibility of achieving speedups over their classical counterparts. Since quantum computing makes use of different computational resources than classical computing, we must create entirely new algorithms to achieve such speedups; and cannot simply rely on the parallelization of classical algorithms, as we have, for instance, with GPU (graphics processing unit) computing. This suggests that it will be necessary to develop literatures on quantum econometrics and quantum computational economics. Fortunately, outside of economics, the literature on quantum algorithms has already produced quantum versions of several econometric and computational-economic routines. These routines, however, typically have limitations that do not apply to their classical counterparts. We will both discuss those limitations and also indicate where economists may be able to contribute to the literature.

Our objective in the quantum algorithms section will be to provide a complete review of relevant algorithms for economists, including function approximation, linear systems analysis, Monte Carlo simulation, matrix inversion, principal component analysis, linear regression, interpolation, numerical differentiation, and true random number generation. In some cases, quantum algorithms will achieve an exponential speedup over their classical counterparts, rendering otherwise intractable problems into something that may eventually be feasible to perform on a quantum computer. In other cases, quantum algorithms will alleviate memory constraints that may render certain problems intractable on classical computers by allowing them to be performed with exponentially fewer input resources. In each case, we will describe the original or most commonly-used version of the algorithm in low-level detail, along with its limitations, and then provide an up-to-date description of related work in the literature.

We also go beyond a description of individual quantum algorithms to discuss the underlying mechanism for generating speedups in many quantum algorithms, which is non-trivial. This involves a discussion of phase kickback, phase estimation, and the quantum Fourier transform, which are three common ingredients in many quantum algorithms that achieve a speedup over their classical counterparts. Economists who wish to develop future literatures on quantum econometrics or quantum computational economics will need to understand these concepts.

In addition to reviewing quantum algorithms that have relevance for economists, we also provide an overview of experimental progress in the development of quantum computers. Benchmarking quantum advantage typically involves computational problems that require large amounts of memory and logical operations on classical high-performance computers (HPC). As such, a quantum algorithm may need to run anywhere from minutes to days to demonstrate a speedup over its classical equivalent. At present, quantum algorithms must be executed on noisy intermediate scale quantum (NISQ) devices \citep{Pre18} with up to 54 qubits \citep{AAB+19} and a few hundred gates. While such devices are on the threshold of exceeding the memory capacity of present and future HPCs, demonstrating a quantum advantage will also typically require the execution of a large sequence of operations. This, in turn, will require considerably longer coherence times in quantum circuits or efficient quantum error correction (QEC). Consequently, in the near-term, applications of quantum computing will be limited to proof-of-principle demonstrations and to the development of quantum awareness and education. Moreover, the challenge of achieving quantum speedups is likely to contribute to the development of more efficient classical algorithms.

The remainder of the paper is organized as follows. Section \ref{Preliminaries} provides an overview of preliminary material that is needed to fully understand quantum money and quantum algorithms. This includes mathematical and notational descriptions of quantum physical phenomena, as it is used to perform computation. Section \ref{QuantumMoney} introduces the concept of quantum money, including the complete technical details for the first quantum money scheme and an overview of all of the major theoretical and experimental contributions to the literature. Section \ref{QuantumAlgorithms} provides an exhaustive literature review of quantum algorithms that can be employed to solve and estimate economic models, along with descriptions of how such algorithms can be implemented and whether they face limitations relative to their classical counterparts. It also describes the current status of quantum hardware and software. Finally, Section \ref{Conclusion} concludes, and the Appendix provides additional technical detail on quantum money and quantum algorithms.

%% file: Preliminaries.tex
\section{Preliminaries}
\label{Preliminaries}

This section will provide an overview of concepts from quantum computing and quantum money that assume knowledge of linear algebra and statistics, but not of physics or computer science. The set of topics covered is intended to be as narrow as is possible while still providing readers with a basis for understanding simple quantum algorithms and quantum money schemes. We will discuss how information is encoded in physical systems, how the states of such systems are represented mathematically, and what operations can be performed on them. For a more complete overview of quantum computing, see the section on ``Fundamental Concepts'' in  \cite{NC00}, the section on ``Quantum Building Blocks'' in \cite{RP11}, or the lecture notes for John Watrous's introductory course on Quantum Computing \citep{Wat06}.

\subsection{Quantum States}

In this subsection, we will discuss quantum states, which are the media in which information in quantum computers is stored. States may be acted on by operations to perform computation.

\subsubsection{Quantum Bits}
\label{Quantum Bits}

A binary digit or ``bit'' is the fundamental unit of classical computing. Bits can be in either a 0 or 1 position and may be encoded physically using classical systems with two states. In modern computers, it is common to encode the 0 position with a low voltage level and the 1 position with a high voltage level. The choice to use bits allows for the direct application of Boolean logical operations. Table \ref{ClassicalGates} shows a selection of such operations. As proven by \citet{She13}, universal computation can be performed using only the NOT-AND (NAND) operation and pairs of bits.\footnote{A universal computer is capable of simulating any other computer.} Thus, a classical computer that uses complex logical operations with many inputs will not be capable of performing different operations than a computer that exclusively performs NANDs on pairs of bits.

\begin{table}[!ht]
\vspace{4mm}
\setlength{\tabcolsep}{15pt}
\begin{center}
\scalebox{1.0}{
\begin{tabular}{|c|c|c|c|c|}
\hline
\textbf{State} & \textbf{AND} & \textbf{OR} & \textbf{XOR} & \textbf{NAND} \\
\hline 
00 & 0 & 0 & 0 & 1 \\
01 & 0 & 1 & 1 & 1 \\
10 & 0 & 1 & 1 & 1 \\
11 & 1 & 1 & 0 & 0 \\
\hline
\end{tabular}}
\caption{The table above shows a selection of 2-bit logical operations. In circuits, such operations will be implemented using objects called gates. The AND operation is equal to 1 if both input bits are equal to 1, but is 0 otherwise. The (inclusive) OR operation is 1 if at least one input bit is equal to 1. The exclusive OR or XOR operation is equal to 1 if exactly 1 input bit is equal to 1, but 0 otherwise. The NOT-AND or NAND operation is the negation of the AND operation.}
\label{ClassicalGates}
\end{center}
\end{table}

A quantum bit or ``qubit'' is the fundamental unit of quantum computing. As with classical bits, quantum bits are encoded in two-level systems; however, unlike classical bits, qubits are encoded in quantum systems, such as photon polarizations, electron spins, and energy levels. The use of quantum systems allows for the exploitation of quantum physical properties. For instance, rather than being restricted to either the 0 or 1 position (like classical bits), quantum bits may be in a superposition of both 0 and 1 simultaneously. We will discuss how such properties -- superposition, entanglement, and interference -- may be used to provide an advantage over what is achievable using classical bits in Sections \ref{Superposition} and \ref{Entanglement}.

Importantly, quantum computing can be performed without a deep understanding of the physical processes that underlie it, just as it is possible to perform classical computing without an understanding of the underlying physical systems that encode information. While an understanding of quantum physics may improve intuition, it will be sufficient to understand how states are represented and what operations we may perform on them. The purpose of this section will be to provide such information through a primarily mathematical description.

\subsubsection{Vector Representation}
\label{Vector Representation}

Individual classical bits are limited to two configurations: 0 and 1. If, for instance, we have five bits -- 0, 1, 1, 0, and 0 -- we can represent the underlying state of the system using a five-digit bit string: 01100. More generally, if we have $n$ bits -- 0, 1, ..., 1 -- we can represent the underlying state with the n-digit bit string: 01...1. This is not true for qubits. In addition to being in the two ``classical'' states, 0 or 1, a two-level quantum system may also be in an uncountably infinite number of superposition states. For this reason, we represent an individual qubit using a column vector, as shown in Equation (\ref{eqn:vector_representation}).

\begin{equation}\label{eqn:vector_representation}
\begin{pmatrix}
\alpha \\
\beta \\
\end{pmatrix}
\end{equation} 

\noindent Note that $\alpha$ and $\beta$ are referred to as ``amplitudes'' and lie in $\mathbb{C}$. Furthermore, as shown in Equation (\ref{eqn:modulus}), the modulus squared of each of the elements sums to one.\footnote{If $z \in \mathbb{C}$, then we may decompose $z$ into real and imaginary parts, $z=x+iy$, where $i=\sqrt{-1}$. The modulus of a complex number is $|z| = \sqrt{x^{2}+y^{2}}$.}

\begin{equation}\label{eqn:modulus}
1 = |\alpha|^{2} + |\beta|^{2}
\end{equation}

\noindent If we have a second two-level system, the joint state of the first and second systems is given by their tensor product, as shown in Equation (\ref{eqn:tensor_product}).

\begin{equation}\label{eqn:tensor_product}
\begin{pmatrix}
\alpha \gamma \\
\alpha \delta \\
\beta \gamma \\
\beta \delta \\
\end{pmatrix}
=
\begin{pmatrix}
\alpha  

\begin{pmatrix}
\gamma \\
\delta \\
\end{pmatrix} \\

\beta 

\begin{pmatrix}
\gamma \\
\delta \\
\end{pmatrix} \\

\end{pmatrix}
=
\begin{pmatrix}
\alpha \\
\beta \\
\end{pmatrix}
\otimes
\begin{pmatrix}
\gamma \\
\delta \\
\end{pmatrix}
\end{equation}

\noindent As we have seen, a single qubit state is in $\mathbb{C}^{2}$ and a two-qubit state is in $(\mathbb{C}^{2})^{\otimes2} = \mathbb{C}^{4}$. More generally, an n-qubit state will lie in $(\mathbb{C}^{2})^{\otimes n} = \mathbb{C}^{2^{n}}$. This means that an $n$-qubit quantum system is capable of representing $2^{n}$ complex numbers; whereas, an $n$-bit classical system is only capable of representing $n$ binary digits. This exponential scaling in computational resources that arises from a linear scaling in the number of qubits provides the basis for quantum speedups. Importantly, however, as we discuss in Section \ref{Sub:Limitations}, this will not lead to general exponential gains with respect to resource requirements or run times.

\subsubsection{Dirac Notation}
\label{Dirac Notation}

While it is possible to represent quantum states using column vectors, it will often be more convenient to use bra-ket notation, introduced by \citet{Dir39}. This is because the size of the column vector needed to represent the state of a system scales exponentially with the number of qubits. A 20-qubit system, for instance, will require a 1,048,576-element column vector.

As you may have noticed in the previous subsection, our column vector representation may be reformulated in terms of basis vectors, as shown in Equation (\ref{eqn:basis_vectors}). Dirac notation simplifies the more cumbersome column vector representation by using ``kets'' to represent quantum states.

\begin{equation}\label{eqn:basis_vectors}
\begin{pmatrix}
\alpha \\
\beta \\
\end{pmatrix}
=
\alpha
\begin{pmatrix}
1 \\
0 \\
\end{pmatrix}
+
\beta
\begin{pmatrix}
0 \\
1 \\
\end{pmatrix}
\end{equation} 
In Equation (\ref{eqn:basis_vectors}), we have used what is referred to as the ``computational basis,'' which is given in Equation (\ref{eqn:computational_basis}).

\begin{equation}\label{eqn:computational_basis}
\left\{ \begin{pmatrix}
1 \\
0 \\
\end{pmatrix},
\begin{pmatrix}
0 \\
1 \\
\end{pmatrix}
\right\}
\end{equation}
In Dirac notation, we will use the ket, $\ket{\phi}$, to represent the underlying state of the system. We can also decompose the state using basis vector kets: $\{\ket{0},\ket{1}\}$. This reformulation is given in Equation (\ref{eqn:basis_vectors_reformulation}).
\begin{equation}\label{eqn:basis_vectors_reformulation}
\ket{\phi}
=
\alpha
\begin{pmatrix}
1 \\
0 \\
\end{pmatrix}
+
\beta
\begin{pmatrix}
0 \\
1 \\
\end{pmatrix}
=
\alpha \ket{0} + \beta \ket{1}
\end{equation}
Furthermore, if we have multiple qubits, $\ket{\phi}$ and $\ket{\psi}$, their state will be the tensor product of the two individual qubit states, which may be written in any of the ways shown in Equation (\ref{eqn:multi_qubit_states_1}).

\begin{equation}\label{eqn:multi_qubit_states_1}
\ket{\phi} \otimes \ket{\psi} = \ket{\phi} \ket{\psi} = \ket{\phi \psi} = (\phi_0 \ket{0} + \phi_1 \ket{1}) \otimes (\psi_0 \ket{0} + \psi_1 \ket{1}) 
\end{equation}
Since tensor products satisfy the distributive property, we may re-express Equation (\ref{eqn:multi_qubit_states_1}) as Equation (\ref{eqn:multi_qubit_states_2}).\footnote{Tensor products satisfy the distributive and associative properties, but not the commutative property.}

\begin{equation}\label{eqn:multi_qubit_states_2}
\phi_0 \psi_0 \ket{00} +  \phi_0 \psi_1 \ket{01} + \phi_1 \psi_0 \ket{10} + \phi_1 \psi_1 \ket{11}
\end{equation}
In addition to the computational basis, it will often be convenient to use other bases, including the Hadamard basis: $\{\ket{+},\ket{-}\}$. Note that $\ket{+}$ and $\ket{-}$ are defined in Equations (\ref{eqn:plus_state}) and (\ref{eqn:minus_state}).

\begin{equation}\label{eqn:plus_state}
\ket{+} = \frac{1}{\sqrt{2}} \ket{0} + \frac{1}{\sqrt{2}} \ket{1}
\end{equation}

\begin{equation}\label{eqn:minus_state}
\ket{-} = \frac{1}{\sqrt{2}} \ket{0} - \frac{1}{\sqrt{2}} \ket{1}
\end{equation}
In addition to the ket, Dirac notation also introduces the bra, which is the conjugate transpose of the ket. Equation (\ref{eqn:bra}) defines the bra that corresponds to $\ket{\phi}$.

\begin{equation}\label{eqn:bra}
\bra{\phi} 
=
\begin{pmatrix}
\alpha^{*} & \beta^{*} \\
\end{pmatrix}
\end{equation}
Note that $\alpha^{*}$ and $\beta^{*}$ are the complex conjugates of $\alpha$ and $\beta$.\footnote{If $\alpha = x + iy$, then $\alpha^{*} = x - iy$.} The bra will be useful notationally when we wish to express an inner or outer product, which will be used frequently. The inner product of $\ket{0}$ and $\ket{1}$ is expressed in Equation (\ref{eqn:inner_product}).

\begin{equation}\label{eqn:inner_product}
\langle 0 | 1 \rangle = 
\begin{pmatrix}
1 & 0 \\
\end{pmatrix}
\begin{pmatrix}
0 \\
1 \\
\end{pmatrix}
= 0
\end{equation}
The equivalent outer product is defined in Equation (\ref{eqn:outer_product}).

\begin{equation}\label{eqn:outer_product}
\ket{0} \bra{1} = 
\begin{pmatrix}
1 \\
0 \\
\end{pmatrix}
\begin{pmatrix}
0 & 1 \\
\end{pmatrix}
=
\begin{pmatrix}
0 & 1 \\
0 & 0 \\
\end{pmatrix}
\end{equation}

\subsubsection{Superposition}
\label{Superposition}

In the computational basis, a superposition is a linear combination of the two basis states, $\ket{0}$ and $\ket{1}$: $\alpha \ket{0} + \beta \ket{1}$. Pure states in the computational basis, $\ket{0}$ and $\ket{1}$, are also superpositions in different bases. Note that we may write $\ket{0}$, which is a pure state in the computational basis, as a superposition in the Hadamard basis, as shown in Equation (\ref{eqn:pure_and_superposition}).

\begin{equation}\label{eqn:pure_and_superposition}
\ket{0} = \frac{1}{2}(\ket{0}+\ket{1})+\frac{1}{2}(\ket{0}-\ket{1}) = \frac{\sqrt{2}}{2} \ket{+} + \frac{\sqrt{2}}{2} \ket{-}
\end{equation}
The ability to create quantum superpositions will provide us with a computational resource that is not available in classical computing. While a classical bit must be in either the 0 or 1 position, a qubit may be in an uncountably infinite number of linear combinations of the $\ket{0}$ and $\ket{1}$ states. 

As we will discuss in Section \ref{Sec:Quantum Measurement}, we cannot observe the amplitudes associated with a superposition. Rather, we are restricted to performing measurement on a state in a particular basis, which will cause the superposition to collapse into a basis state. For instance, upon measurement in the computational basis, $\alpha \ket{0} + \beta \ket{1}$ would yield $\ket{0}$ with probability $|\alpha|^{2}$ and $\ket{1}$ with probability $|\beta|^{2}$.

\subsubsection{Entanglement}
\label{Entanglement}

Most multi-qubit states can be written as tensor products, such as $\ket{00}$ or $(\alpha \ket{0}  + \beta \ket{1}) \ket{1}$. However, some states, which are said to be ``entangled,'' cannot be expressed in such a way. Rather, these states exhibit ``correlation'' in the sense that measurement of one qubit yields information about the states of the remaining unmeasured qubit(s).

There are four maximally-entangled two-qubit states, which are referred to as the Bell states and were introduced in \cite{Bel64} as a resolution to the paradox in \cite{EPR35}. One such Bell state is $\ket{\phi^{+}}$ is defined in Equation (\ref{eqn:phi_plus}).

\begin{equation}\label{eqn:phi_plus}
\ket{\phi^{+}}=\frac{1}{\sqrt{2}}(\ket{00}+\ket{11})
\end{equation}
Notice that no choice of amplitudes -- $\alpha$, $\beta$, $\gamma$, and $\delta$ -- will satisfy Equation (\ref{eqn:phi_plus_impossible}).
\begin{equation}\label{eqn:phi_plus_impossible}
\ket{\phi^{+}} = (\alpha \ket{0} + \beta \ket{1}) \otimes (\gamma \ket{0} + \delta \ket{1}) = \alpha \gamma \ket{00} + \alpha \delta \ket{01} + \beta \gamma \ket{10} + \beta \delta \ket{11}
\end{equation}
Importantly, entanglement does not reduce to the concept of correlation in classical probability. Rather, if we measure the first qubit in $\ket{\phi^{+}}$ using the computational basis, we will get either $\ket{0}$ or $\ket{1}$ with probability $\frac{1}{2}$. If our measurement returns $\ket{0}$ for the first qubit, we will also get $\ket{0}$ for the second with certainty. Alternatively, if we get $\ket{1}$ for the first qubit, we will get $\ket{1}$ for the second qubit.

This property of entangled quantum states remains true even if we separate the qubits in space and perform measurement simultaneously. It is sometimes referred to as non-locality, since the speed of interactions does not appear to depend on physical proximity. Entanglement plays an important role in several quantum algorithms. It also used in the quantum teleportation protocol and features prominently in proposals for secure communication via a quantum internet.

\subsection{Quantum Dynamics}
\label{Sec:Quantum Dynamics}

The evolution of quantum states over time can be described using unitary operations. The necessary and sufficient condition for the unitarity of a matrix, $U$, is that $U^{\dagger}U = I$. Note that $\dagger$ is the adjoint operator, defined in Equation (\ref{eqn:adjoint_operator}), which transposes $U$ and takes the complex conjugate of each of its elements.
\begin{equation}\label{eqn:adjoint_operator}
U^{\dagger} 
=
\begin{pmatrix}
u_{0,0}^{*} & u_{1,0}^{*} \\
u_{0,1}^{*} & u_{1,1}^{*} \\
\end{pmatrix}
\end{equation}
Unitary operations preserve the Euclidean norm, which ensures that quantum states maintain a Euclidean length of 1 post-transformation. Furthermore, unitary operations are trivially invertible through the use of the adjoint operator, allowing for reversibility, which is a requirement of quantum computation.\footnote{This is a consequence of the Schr{\"o}dinger equation, which describes the time evolution of a quantum system and is reversible. Note that there is no such requirement for classical computing. If, for instance, we apply an AND gate to 0 and 1 input bits in a classical circuit, we get the output 0. Without additional information, we cannot reverse that 0 to recover 0 and 1, since inputs of 0 and 0 would also yield a 0.}

The simplest operations are single-qubit unitaries. Among these, commonly-applied unitaries include the identity operator, $I$, and the Pauli operators -- $X$, $Z$, and $Y$ -- which are defined in Equation (\ref{eqn:pauli operators}).

\begin{equation}\label{eqn:pauli operators}
I 
=
\begin{pmatrix}
1 & 0 \\
0 & 1 \\
\end{pmatrix},
X
=
\begin{pmatrix}
0 & 1 \\
1 & 0 \\
\end{pmatrix},
Z 
=
\begin{pmatrix}
1 & 0 \\
0 & -1 \\
\end{pmatrix},
Y = 
\begin{pmatrix}
0 & -i \\
i & 0 \\
\end{pmatrix}
\end{equation}
The identity operator leaves the quantum state unchanged. The Pauli X unitary applies a ``bit flip'' or NOT operation, as shown in Equation (\ref{eqn:pauli x}). That is, $\ket{0}$ becomes $\ket{1}$, $\ket{1}$ becomes $\ket{0}$, and, more generally, $\alpha \ket{0} + \beta \ket{1}$ becomes $\beta \ket{0} + \alpha \ket{1}$. 

\begin{equation}\label{eqn:pauli x}
X \ket{\psi} = 
\begin{pmatrix}
0 & 1 \\
1 & 0 \\
\end{pmatrix}
\begin{pmatrix}
\alpha \\
\beta \\
\end{pmatrix}
=
\begin{pmatrix}
\beta \\
\alpha \\
\end{pmatrix}
\end{equation}
The Pauli Z unitary applies a relative phase flip. That is, it changes the sign of the the second amplitude. In the computational basis, this will shift a $\ket{+}$ state to a $\ket{-}$ state and a $\ket{-}$ state to a $\ket{+}$ state, as shown in Equation (\ref{eqn:pauli z}).

\begin{equation}\label{eqn:pauli z}
Z \ket{+} = 
\begin{pmatrix}
1 & 0 \\
0 & -1 \\
\end{pmatrix}
\begin{pmatrix}
\frac{1}{\sqrt{2}} \\
\frac{1}{\sqrt{2}} \\
\end{pmatrix}
= 
\begin{pmatrix}
\frac{1}{\sqrt{2}} \\
-\frac{1}{\sqrt{2}} \\
\end{pmatrix}
=
\ket{-}
\end{equation}
As shown in Equations (\ref{eqn:pauliY_computational}) and (\ref{eqn:pauliY_hadamard}), the Pauli Y unitary acts as a NOT operation in both the computational and Hadamard bases.

\begin{equation}\label{eqn:pauliY_computational}
Y \ket{0} = 
\begin{pmatrix}
0 & -i \\
i & 0 \\
\end{pmatrix}
\begin{pmatrix}
1 \\
0 \\
\end{pmatrix}
= 
\begin{pmatrix}
0 \\
i \\
\end{pmatrix}
=
i\ket{1}
\end{equation}

\begin{equation}\label{eqn:pauliY_hadamard}
Y \ket{+} = 
\begin{pmatrix}
0 & -i \\
i & 0 \\
\end{pmatrix}
\begin{pmatrix}
\frac{1}{\sqrt{2}} \\
\frac{1}{\sqrt{2}} \\
\end{pmatrix}
= 
\begin{pmatrix}
-\frac{i}{\sqrt{2}} \\
+\frac{i}{\sqrt{2}} \\
\end{pmatrix}
=
-i\ket{-}
\end{equation}
Note that the $i$ in Equation (\ref{eqn:pauliY_computational}) and the $-i$ in Equation (\ref{eqn:pauliY_hadamard}) are referred to as global phases. Since quantum states are only unique up to a global phase, we may treat $i\ket{1}$ as $\ket{1}$ and $-i\ket{-}$ as $\ket{-}$.

In addition to the Pauli operators, the Hadamard operator, $H$, is also frequently used in quantum algorithms, including the quantum Fourier transform (QFT).\footnote{See \ref{Quantum Fourier Transform} in the Appendix for an overview of the quantum Fourier transform.} When applied to computational basis states, the Hadamard operator creates an equal superposition of the $\ket{0}$ and $\ket{1}$ states, as shown in Equations (\ref{eqn:hadamard_plus}) and (\ref{eqn:hadamard_minus}). The Hadamard operator also transforms states in the Hadamard basis to the computational basis: $H\ket{+} = \ket{0}$, and $H\ket{-} = \ket{1}$. 

\begin{equation}\label{eqn:hadamard_plus}
H \ket{0} = 
\frac{1}{\sqrt{2}}  \begin{pmatrix}
1 & 1 \\
1 & -1 \\
\end{pmatrix}
\begin{pmatrix}
1 \\
0 \\
\end{pmatrix}
=
\ket{+}
\end{equation}

\begin{equation}\label{eqn:hadamard_minus}
H \ket{1} = 
\frac{1}{\sqrt{2}}  \begin{pmatrix}
1 & 1 \\
1 & -1 \\
\end{pmatrix}
\begin{pmatrix}
0 \\
1 \\
\end{pmatrix}
=
\ket{-}
\end{equation}
Beyond the I, X, Z, Y, and H single-qubit unitaries, many quantum algorithms will require the use of rotation matrices, including the phase operation, $S$, and the $\pi/\ 8$ operation, $T$, which are defined in Equation (\ref{eqn:phase_operators}).

\begin{equation}\label{eqn:phase_operators}
S = 
\begin{pmatrix}
1 & 0 \\
0 & e^{\frac{i \pi}{2}} \\
\end{pmatrix},
T = 
\begin{pmatrix}
1 & 0 \\
0 & e^{\frac{i \pi}{4}} \\
\end{pmatrix}
\end{equation}

To create entanglement between qubits, we will use the controlled-NOT or CNOT operation. This two-qubit operation has a control qubit and a target qubit. If the control qubit is in the $\ket{1}$ position, then the $X$ operation will be applied to the target qubit. In Equation (\ref{eqn:cnot}), we apply a CNOT, $cX$, to the quantum state, $\ket{11}$.

\begin{equation}\label{eqn:cnot}
cX \ket{11} = 
\begin{pmatrix}
1 & 0 & 0 & 0 \\
0 & 1 & 0 & 0 \\
0 & 0 & 0 & 1 \\
0 & 0 & 1 & 0 \\
\end{pmatrix}
\begin{pmatrix}
0 \\
0 \\
0 \\ 
1 \\
\end{pmatrix}
=
\begin{pmatrix}
0 \\
0 \\
1 \\ 
0 \\
\end{pmatrix}
= \ket{10}
\end{equation}
In addition to the $cX$ operations, we will use $cU$ to denote arbitrary controlled-unitary operations, such as $cZ$, $cY$, or $cH$. However, in order to perform any quantum computation, $I$, $X$, $Y$, $Z$, $H$, $S$, $T$, and $cX$ will be sufficient, since they form a universal set of operations.

More generally, we may perform $n$-qubit operations using the tensor products of each of the single-qubit operations. For instance, applying an $I$ operation to qubit 1 and an $X$ operation to qubit 2 in the state $\ket{01}$ is equivalent to applying $I \otimes X$ to $\ket{01}$, as is shown in Equation~\ref{eqn:two_qubit_ops}.

\begin{equation}\label{eqn:two_qubit_ops}
I X \ket{01} = 
\begin{pmatrix}
1 & 0 \\
0 & 1 \\
\end{pmatrix}
\otimes
\begin{pmatrix}
0 & 1 \\
1 & 0 \\
\end{pmatrix}
\begin{pmatrix}
0 \\
1 \\
0 \\ 
0 \\
\end{pmatrix}
=
\begin{pmatrix}
0 & 1 & 0 & 0 \\
1 & 0 & 0 & 0 \\
0 & 0 & 0 & 1 \\
0 & 0 & 1 & 0 \\
\end{pmatrix}
\begin{pmatrix}
0 \\
1 \\
0 \\ 
0 \\
\end{pmatrix}
=
\begin{pmatrix}
1 \\
0 \\
0 \\ 
0 \\
\end{pmatrix}
= \ket{00}
\end{equation}

Finally, while we have discussed everything in terms of unitary operations thus far, we will also use the term ``gates'' to refer to the implementation of such operations in quantum circuits. We will introduce quantum circuits and gates in Section \ref{Sec:Quantum Circuits}.

\subsection{Quantum Measurement}
\label{Sec:Quantum Measurement}

In quantum algorithms, measurement is performed before reading out the results to a classical computer.  As shown in \citet{Bor26}, this triggers a ``collapse of the wavefunction.'' For an arbitrary state, such as $\ket{\psi} = \alpha \ket{0} + \beta \ket{1}$, this means that the superposition will collapse into a pure state in the basis in which it is measured. If the computational basis is selected, the probability of outcome $\ket{0}$ will be $|\alpha|^{2}$ and the probability of outcome $\ket{1}$ will be $|\beta|^{2}$. These probabilities are computed for $\ket{0}$ using $\langle \psi | 0 \rangle \langle 0 | \psi \rangle$ and $\ket{1}$ using $\langle \phi | 1 \rangle \langle 1 | \phi \rangle$. Note that these are not observable, but we can infer them by repeatedly preparing the same state and then performing measurement.

Measurements on multi-qubit states also work the same. Consider, for instance, an equal superposition of two qubits, as shown in Equation (\ref{eqn:equal_superposition}). The probability of measuring any of the four possible states is given by $|\frac{1}{2}|^{2}=\frac{1}{4}$. Furthermore, upon measurement, the superposition will collapse into $\ket{00}$, $\ket{01}$, $\ket{10}$, or $\ket{11}$.

\begin{equation}\label{eqn:equal_superposition}
\ket{\phi} = \frac{1}{2} \ket{00} + \frac{1}{2} \ket{01} + \frac{1}{2} \ket{10} + \frac{1}{2} \ket{11}
\end{equation}

It is also possible to perform a partial measurement on just one qubit. If, for instance, we measure just the first qubit, the probability of getting $\ket{0}$ will be $\frac{1}{2}$. This is because there are two states for qubit two associated with a $\ket{0}$ state for qubit one. The probabilities associated with these states sum to $\frac{1}{2}$. Furthermore, if we measure qubit one and find it to be in state $\ket{0}$, then the state of the second, unmeasured qubit will be $\frac{1}{\sqrt{2}}(\ket{0}+\ket{1})$. We can see this by rewriting $\ket{\phi}$ as $\frac{1}{2}\ket{0}(\ket{0}+\ket{1})+\frac{1}{2}\ket{1}(\ket{0}+\ket{1})$. Note that the amplitude $\frac{1}{2}$ must be renormalized to $\frac{1}{\sqrt{2}}$ after measurement, since the probabilities will otherwise not sum to one.

\subsection{Quantum Circuits}
\label{Sec:Quantum Circuits}

A quantum circuit is a model of quantum computation that consists of initial states, gates, wires, and measurement. Circuits are typically initialized with $\ket{0}$ as the state for each qubit. Gates, which are unitary operations acting non-trivially only on a constant number of qubits, are then applied in sequence. Finally, a measurement is performed at the end of the circuit, collapsing the superpositions of the measured qubits. Note that circuit diagrams should be read from left to right.

\begin{figure}
\begin{center}
\scalebox{1.5}{\mbox{
\Qcircuit
{
\ket{0}&  &\gate{H} & \meter
}
}
}
\caption{The diagram above shows a circuit with one qubit. The qubit is initialized in the state $\ket{0}$. A Hadamard gate is then applied to the qubit, putting it the $\ket{+}=\frac{1}{\sqrt{2}}(\ket{0}+\ket{1})$ state. Finally, a measurement is performed. Unless otherwise stated, a measurement is assumed to be in the standard basis (i.e. $\{0,1 \}$). In this case, the outcome is 0 or 1 with equal probability.}
\label{fig:rng_circuit}

\end{center}
\end{figure}
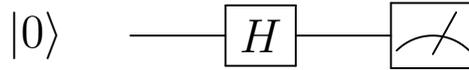

Figure \ref{fig:rng_circuit} provides a diagram of a circuit that performs true random number generation from a Bernoulli distribution with parameters $q=p=1/2$. On the left side of the circuit, one qubit is initialized in state $\ket{0}$. The Hadamard gate, indicated by the H inscribed within a rectangle, is then applied to the qubit, putting it in an equal superposition: $\frac{1}{\sqrt{2}}\ket{0} + \frac{1}{\sqrt{2}}\ket{1}$. Finally, measurement is applied, which collapses the superposition into either $\ket{0}$ or $\ket{1}$ with equal probability. The result of the measurement will be read out to a classical bit for storage and use on classical computers.

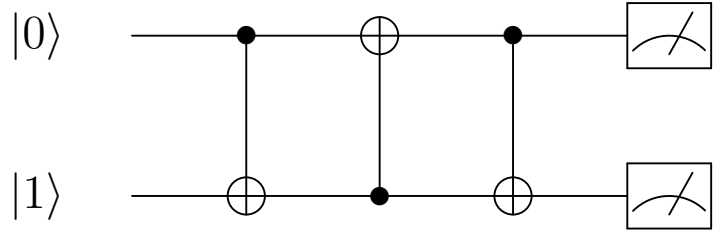
\begin{figure}
\begin{center}
\scalebox{1.5}{\mbox{
\Qcircuit
{
\ket{0}&  &\ctrl{1} & \targ & \ctrl{1} & \meter \\
\ket{1}&  &\targ & \ctrl{-1} & \targ & \meter
}
}
}
\caption{The diagram above shows a circuit with two qubits, which are initialized in state $\ket{01}$. The circuit then applies three CNOT gates in sequence: CNOT(0,1), CNOT(1,0), and CNOT(0,1). This swaps the qubits in the 0 and 1 position, yielding state $\ket{10}$ when measurement is performed.}
\label{fig:swap_circuit}
\end{center}
\end{figure}

Figure \ref{fig:swap_circuit} illustrates a simple two-qubit circuit: the SWAP circuit. Executing a SWAP will change qubit 0's state to qubit 1's and vice versa.\footnote{In practice, we will often need to employ SWAP gates to deal with the architectural constraints of quantum computers. If two qubits are not located sufficiently close in physical space, we may not be able to apply two-qubit gates to them. For this reason, we may execute a SWAP to move the relevant qubits closer together.} Note that we have initialized the system to be in the $\ket{01}$ state and should expect $\ket{10}$ as the circuit's output. The first CNOT gate uses qubit 0 as the control and qubit 1 as the target. We will indicate this by CNOT(0,1). The state of the system remains unchanged at $\ket{01}$, since qubit 0 is in the $\ket{0}$ position. CNOT(1,0), which is applied next, changes the state to $\ket{11}$. The final CNOT, which uses qubit 0 as a target and qubit 1 as a control, changes the state of the system to $\ket{10}$. Finally, we perform measurement, yielding the state $\ket{10}$. Note that we could have performed this particular SWAP operation using just the last two CNOT gates, since we knew the underlying state of the system; however, if the initial state had instead been $\ket{10}$, we would have needed the first two CNOT gates instead. Using the full set of three CNOTs in the sequence will allow us to perform a SWAP on two qubits in an arbitrary state, such as $\ket{\psi \phi}$.

\begin{figure}
\begin{center}
\scalebox{0.75}{\mbox{
\Qcircuit
{
\ket{\psi} & & \qw & \qw & \qw & \ctrl{2} & \qw & \qw & \qw & \ctrl{2} & \qw & \ctrl{1} & \gate{T} & \ctrl{1} & \qw \\
\ket{\phi} & & \qw & \ctrl{1} & \qw & \qw & \qw & \ctrl{1} & \qw & \qw & \gate{T} & \targ & \gate{T^\dagger} & \targ & \qw \\
\ket{1} & & \gate{H} & \targ & \gate{T^{\dagger}} & \targ & \gate{T} & \targ & \gate{T^\dagger} & \targ & \gate{T} &  \gate{H} & \qw & \qw & \qw \\
}
}
}
\caption{All quantum computations can be performed using a universal set of two-qubit gates, such as the set we introduced earlier: I, X, Y, Z, H, S, T, and CNOT. This means that the NAND operation we performed using $X$ gates and a Toffoli gate can also be performed using two-qubit gates. This circuit shows the implementation given in \citet{NC00}.}
\label{fig:2_qubit_gate_toffoli}
\end{center}
\end{figure}
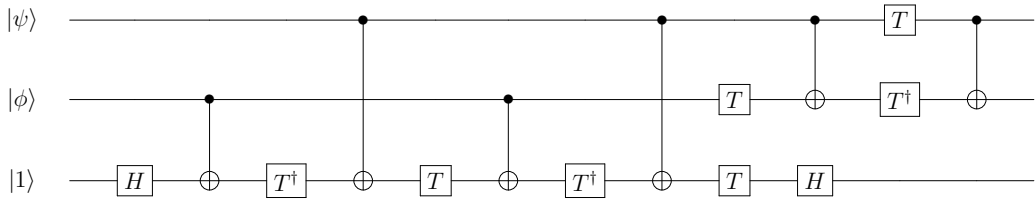

Quantum circuits often require the use of extra qubits called ``ancillas.'' This is because quantum computation must be reversible, which often requires us to retain information after a gate has been applied. In classical circuits, for instance, we may implement a NAND gate by taking two input bits, applying a NAND, and then outputting a single bit. However, in a quantum circuit, we must use a Toffoli gate (controlled-controlled-NOT), coupled with an ancilla bit initialized in the $\ket{1}$ position, as shown in Figure \ref{fig:3_qubit_gate_toffoli}. Notice that we initialize qubit 2, the ancilla qubit, in the $\ket{1}$ state, and qubits 0 and 1 in the $\ket{\psi}$ and $\ket{\phi}$ states. In the special case where $\ket{\psi} = \ket{\phi} = \ket{1}$, an $X$ gate is applied to the target qubit, yielding the state $\ket{110}$. If $\ket{\psi}$ and $\ket{\phi}$ are instead in arbitrary superpositions, then applying the gate maps the state $\psi_0 \phi_0 \ket{001} + \psi_0 \phi_1 \ket{011} + \psi_1 \phi_0 \ket{101} + \psi_1 \phi_1 \ket{111}$ to the state $\psi_0 \phi_0 \ket{001} + \psi_0 \phi_1 \ket{011} + \psi_1 \phi_0 \ket{101} + \psi_1 \phi_1 \ket{110}$. While we have chosen to use a three-qubit gate to implement this circuit, it is always possible to do it with a longer sequence of two-qubit gates. Figure \ref{fig:2_qubit_gate_toffoli} shows how the same NAND operation can be performed using a Toffoli operation that has been broken down into two-qubit gates.

\begin{figure}
\begin{center}
\scalebox{1.5}{\mbox{
\Qcircuit
{
\ket{\psi}&  &\ctrl{1}&\qw  \\
\ket{\phi}& &\ctrl{1} &\qw\\
\ket{1}& &\targ   &\qw
}
}
}
\caption{The quantum circuit diagram above shows the application of the NAND operation to the $\ket{\psi \phi}$ state using a Toffoli gate. The Toffoli gate performs a controlled-controlled-NOT operation: $\ket{\psi,\phi,\gamma}$ is mapped to $\ket{\psi,\phi,\gamma\oplus AND(\psi,\phi)}$, where $\oplus$ denotes an exclusive OR operation. Since $\gamma$ is initialized as $\ket{1}$, the third qubit is mapped to $NAND(\psi,\phi)$ Furthermore, since the input state, $\ket{\psi \phi}$ is also retained, the circuit is reversible.}
\label{fig:3_qubit_gate_toffoli}
\end{center}
\end{figure}
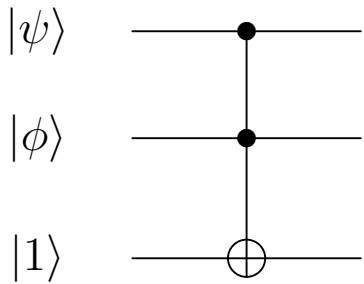

When we discuss improvements that quantum algorithms provide over other quantum algorithms or over their classical counterparts, one metric we will often use is called ``gate complexity.'' Given a circuit (or more broadly, a quantum algorithm), the gate complexity of the circuit is the number of elementary gates that are used in that circuit. This measures the minimum number of elementary steps needed to perform a given computation. Comparing Figure \ref{fig:3_qubit_gate_toffoli} and Figure \ref{fig:2_qubit_gate_toffoli}, we can see that the two-qubit gate Toffoli requires many more elementary operations than the three-qubit gate Toffoli.\footnote{The two-qubit version of the Toffoli gate will typically be implemented on quantum computers due to architecture restrictions that require qubits to be close together physically.} 

\subsection{Oracles}
\label{Sec:Oracles}

 
\citet{Tur39} introduced the concept of an ``oracle,'' describing it as an ``...unspecified means of solving number theoretic problems.'' For our purposes, the term oracle will typically refer to a black-box (classical) function that can be applied in a quantum circuit. Crucially, it is impossible to ``look inside'' this black-box; the only operation it permits is to apply the function to a quantum state. In many cases, we will not know whether such an oracle can be implemented; however, assuming the existence of an oracle will allow us to construct a quantum circuit and evaluate its properties. As we will discuss later, several quantum money schemes rely on an oracle in the form of an unknown black-box function that performs a particular operation. Furthermore, when comparing quantum algorithms to each other and their classical counterparts, we will often describe performance improvements in terms of measures of complexity.\footnote{See \ref{Computational Complexity} in the Appendix for a brief overview of terms and notation related to computational complexity.} One such measure is ``query complexity,'' which computes the number of times an algorithm queries an oracle.

\subsection{No-Cloning Theorem}
\label{Sec:No-Cloning Theorem}

The original concept of quantum money, as introduced in \cite{Wie83}, achieved information-theoretic security by making use of the ``the no-cloning theorem.'' This theorem, which was proven by \cite{WZ82}, demonstrates that it is not possible to clone an unknown quantum state. With respect to quantum money, this means that a counterfeiter with access to unlimited resources will still not be able to copy a quantum bill. This, of course, is not true for physical forms of money and classical digital currencies.

\cite{NC00} provide a simple, alternative formulation of the no-cloning theorem proof, which we reproduce here. It starts by assuming the existence of a unitary operation, $U$, that can copy a qubit in an unknown state. We then apply $U$ to two qubits, $\ket{\psi}$ and $\ket{\phi}$. Note that $\ket{0}$ is an ancilla qubit into which the copy is computed.

\begin{equation}
U(\ket{\psi} \otimes \ket{0}) = \ket{\psi} \otimes \ket{\psi}
\label{no_cloning_1}
\end{equation}
\begin{equation}
U(\ket{\phi} \otimes \ket{0}) = \ket{\phi} \otimes \ket{\phi}
\label{no_cloning_2}
\end{equation}
If we take the inner product of equations (\ref{no_cloning_1}) and (\ref{no_cloning_2}), we get the following:

\begin{equation}
\langle \psi | \phi \rangle = (\langle \psi | \phi \rangle)^{2}
\end{equation}
Since $0 \leq \langle \psi | \phi \rangle \leq 1$, this can only be true if the states are identical or orthogonal. If the former is true, then $U$ can only be used to clone a single quantum state. If the latter is true, then $U$ can only be used to clone orthogonal states. In either case, $U$ is not capable of cloning an arbitrary, unknown quantum state.

An alternative formulation of the proof exploits the linearity of quantum operations. If $U$ is a quantum operation that clones arbitrary quantum states, then the following should be true:

\begin{equation}
U(\ket{0} \otimes \ket{0}) = \ket{0} \otimes \ket{0}
\label{no_cloning_3}
\end{equation}
\begin{equation}
U(\ket{1} \otimes \ket{0}) = \ket{1} \otimes \ket{1}
\label{no_cloning_4}
\end{equation}
\begin{equation}
U(\ket{\phi} \otimes \ket{0}) = \ket{\phi} \otimes \ket{\phi}
\label{no_cloning_5}
\end{equation}
Now, let $\ket{\phi} = \alpha \ket{0} + \beta \ket{1}$. We may rewrite Equation (\ref{no_cloning_5}) as follows:

\begin{equation}
U(\ket{\phi} \otimes \ket{0}) = U (\alpha \ket{0} \ket{0} + \beta \ket{1} \ket{0})
\label{no_cloning_6}
\end{equation}

\begin{equation}
 = \alpha U (\ket{0} \ket{0}) + \beta U(\ket{1} \ket{0})
\label{no_cloning_7}
\end{equation}

\begin{equation}
 = \alpha \ket{0} \ket{0} + \beta \ket{1} \ket{1}
\label{no_cloning_8}
\end{equation}
Our objective was to clone $\ket{\psi}$ using $U$, which should have produced the following quantum state:

\begin{equation}
\ket{\phi} \ket{\phi} = (\alpha \ket{0} + \beta \ket{1}) \otimes (\alpha \ket{0} + \beta \ket{1})
\label{no_cloning_9}
\end{equation}

\begin{equation}
= \alpha^{2} \ket{0} \ket{0} + \alpha \beta \ket{0} \ket{1} + \beta \alpha \ket{1} \ket{0} + \beta^{2} \ket{1} \ket{1}
\label{no_cloning_10}
\end{equation}
As with the previous proof, this will only be true when either $\alpha = 1$ or $\beta = 1$. Otherwise, Equations (\ref{no_cloning_8}) and (\ref{no_cloning_10}) are not equivalent. Thus, it will not be possible to clone an arbitrary unknown state, since $\ket{\phi}$ may be in a superposition of $\ket{0}$ and $\ket{1}$.

%% file: QuantumMoney.tex
\section{Quantum Money}
\label{QuantumMoney}

\newlength\nesw
\settowidth{\nesw}{$\nearrow$}
\def\neswarrow{\nearrow\hspace{-\nesw}\swarrow}

\newlength\nwse
\settowidth{\nwse}{$\nwarrow$}
\def\nwsearrow{\nwarrow\hspace{-\nwse}\searrow}

In this section, we will provide a complete overview of quantum money that is intended for economists, following similar efforts for Bitcoin \citep{BCEM15} and Distributed Ledger Technologies \citep{Tow18}.\footnote{More generally, we contribute to the growing descriptive literature on new payment technologies that attempts to identify useful entry points through which economists can make meaningful research contributions, including \cite{BCEM15}, \cite{Dwy15}, \cite{Dyh16}, \cite{CK17}, \cite{HLM17}, \cite{BL17}, \cite{Tow18}, and \cite{CG19}.} This will include descriptions of existing quantum money schemes, a summary of the progress in the experimental implementation of quantum money, and a discussion of potential future relevance for economists and central banks. Our intention is to cover important concepts at a high level, but also provide enough low level detail to (1) enable sufficiently motivated economists to find points of entry into the literature; and (2) assist central banks that are exploring digital currency issuance and are open to quantum money as a (distant) future development path.

Our examination of quantum money will begin with a description of \cite{Wie83}, which proposed a form of currency that is protected by the laws of physics, rather than through security features or computational assumptions. Wiesner's scheme is a simple form of ``private-key'' quantum money that has the advantage of being explicable entirely in terms of the concepts introduced in Section \ref{Preliminaries}. We will, however, need to go beyond \citet{Wie83} to fully realize the benefits of quantum money. In particular, we will discuss new varieties of quantum money introduced within the last decade called ``public-key'' quantum money and ``quantum lightning.'' Such varieties have novel and desirable properties that cannot be achieved with any classical form of money or payment instrument. We will document these properties (and others) for different varieties of quantum money in Table \ref{tab:properties}.

\subsection{The First Quantum Money Scheme}
\label{FirstQuantumMoneyScheme}

The first quantum money scheme was introduced in \citet{Wie83}. It makes use of the no-cloning theorem, proven in \citet{WZ82},\footnote{See Section \ref{Sec:No-Cloning Theorem} for two proofs of the no-cloning theorem.} which states that it is not possible to clone an unknown quantum state. To construct a unit of Wiesner's money, the central bank must generate a classical serial number and a random classical bill state. The classical serial number is unique and is publicly-known. The classical bill state is known exclusively by the central bank, which encodes it in a quantum state that is hidden from the bill holder.

The procedure for generating and encoding the quantum bill state starts with the random drawing of $n$ pairs of binary numbers. The first element of each pair corresponds to the bill's classical state. The second element corresponds to a basis used for encoding or measurement. For instance, 0 might correspond to the computational basis and 1 might correspond to the Hadamard basis. The central bank encodes each element of the classical bill state in a two-level quantum system, using the corresponding basis. A draw of 00, for instance, would be encoded as a $\ket{0}$. The scheme is given below for the $n = 5$ case.

\begin{enumerate}
\item \textbf{Classical serial number.} E57804SG.
\item \textbf{Randomly-generated binary pairs.} 01 11 00 10 11.
\item \textbf{Classical bill state.} 01011.
\item \textbf{Bases.} 11001.
\item \textbf{Quantum state.} $\ket{+} \ket{-} \ket{0} \ket{1} \ket{-}$
\end{enumerate}

The central bank records the classical serial number, classical bill state, and the measurement bases for each bill. If a merchant wishes to verify the authenticity of a bill, she may send it to the central bank, which will identify the bill using the classical serial number and perform measurement on the quantum state using the specified bases. If the measurement results match the recorded classical bill states, then the central bank will verify the bill's quantum state as valid. Otherwise, it will reject it as invalid.

Recall that the no-cloning theorem prohibits the copying of unknown quantum states. A counterfeiter who wishes to recover the state of a bill will need to perform measurement on each qubit, just as the central bank does during the verification process. Unlike the central bank, however, the counterfeiter does not know the bases in which the information is encoded. In our simple example, for instance, the counterfeiter would have to correctly guess that the first qubit was encoded in the Hadamard basis. Otherwise, she would incorrectly apply measurement in the computational basis.

Now, recall that the Hadamard basis states, $\ket{+}$ and $\ket{-}$, are in equal superpositions of the computational basis states, $\ket{0}$ and $\ket{1}$, and vice-versa. This means that measuring the first qubit in the computational basis would induce a change in the quantum state to a $\ket{0}$ or $\ket{1}$ with equal probability. This would also be reflected in the classical readout of the measurement. Rather than yielding a 0 with probability 1, the measurement result would be either a 0 or a 1 with equal probability. Consequently, guessing bears the risk of destroying the quantum state.

\cite{Aar09}, \cite{Lut10}, \cite{MVW12}, and \cite{NSBU16} show that \cite{Wie83} and its early extensions were subject to adaptive attacks. Such attacks modify one qubit at a time and then attempt authentication to try to uncover the underlying quantum state.\footnote{\cite{MVW12} analyzed the optimal forging strategy for Wiesner's scheme in the non-adaptive setting and proved that the probability of successfully counterfeiting a note decreases exponentially fast in the number of qubits.} \cite{Aar09} and \cite{Lut10} suggest that adaptive attacks can be prevented by not returning bills that fail the verification process. This, however, is still not sufficient, according to \cite{NSBU16}, which instead recommends replacing the old quantum money state with a new quantum money state after every valid verification. See Section \ref{Attacks on Wiesner} in the Appendix for a complete description of an adaptive attack.

\cite{Wie83} was the first scheme to achieve information-theoretic security. This means that an attacker with unbounded classical and quantum resources would still be unable to counterfeit a unit of Wiesner's money. Since \cite{Wie83}, at least eight additional schemes have been introduced that achieve information-theoretic security.\footnote{See \cite{TOI03}, \cite{MS10}, \cite{Gav12}, \cite{MVW12}, \cite{PYJ+12}, \cite{AC12} (Section 5), \cite{BS16} (Section 6), and \cite{AGKZ20}.} Relative to any digital money or payment scheme, information-theoretic security is a categorical improvement. There are, however, at least three drawbacks to Wiesner's money. First, it requires online verification, which makes it unattractive relative to cash. Second, it uses a private-key scheme, which requires the issuer to conceal information that is used for verification purposes. And third, it is currently technologically infeasible without substantial improvements in the development of quantum memory.

\subsection{Properties of Modern Schemes}

In the previous subsection, we discussed the construction of the first quantum money scheme, along with its properties. While Wiesner's money achieved information-theoretic security -- a standard not possible for any form of payment that does not exploit quantum phenomena -- it failed to provide additional improvements over existing payment systems. 

In this section, we will discuss the properties of modern quantum money schemes, building on the criteria originally outlined in \cite{MS10} and \cite{Aar09}. This will include a broad categorization of forms of money into bill, coin, and lightning schemes and into private and public schemes. It will also include a discussion of security, anonymity, reliance on an unspecified algorithm (oracle), production and verification efficiency, classical verifiability and mintability, and noise tolerance. Table \ref{tab:properties} provides a compact summary of the properties of all quantum money schemes.

\subsubsection{Bills, Coins, and Anonymity}
\label{Sec:bills vs coins}

Quantum money schemes differ in the degree to which they allow anonymity. \cite{MS10} define anonymity in terms of the difficulty of tracing how a unit of money is received and spent. We will refer to this version of anonymity as ``untraceability.'' With Wiesner's money, for instance, the use of a classical serial number eliminates the possibility of retaining anonymity, since the same unit of money is identifiable across the  transactions in which it was used. We refer to forms of quantum money with serial numbers as quantum bills, using the analogy to physical bills, which also have serial numbers, and are also not untraceable for essentially the same reason. 

Classical coins -- or the ideal version of them -- are indistinguishable and, therefore, provide anonymity for users. \cite{MS10} introduced the notion of a quantum coin, which is a form of quantum money in which all quantum money states are exact copies of each other and are, thus, untraceable. The scheme introduced by \citet{TOI03} also achieves untraceability, but through a different underlying mechanism.

Notice that constructing a coin scheme is conceptually harder than a bill scheme: the no-cloning theorem (see Section~\ref{Sec:No-Cloning Theorem}) states that it is impossible to clone a quantum state, given a \textit{single} copy of it. To prove unforgeability for quantum coins, we need a strengthened version of this theorem in which polynomially many copies of the state are available to the counterfeiter. 

\subsubsection{Public Quantum Money}
\label{Sec:public verifiability}
In Wiesner's scheme, quantum bills are transmitted to the central bank for verification. This is similar to a credit card transaction, where the payment terminal sends information to a trusted third party for verification. In an analogy to private-key cryptography, \cite{Aar09} called such schemes ``private-key'' quantum money, since verification requires the bank's private key.\footnote{Note that the private key must be kept secret, as it allows minting of new money.}


\newgeometry{margin=2.25cm}

\begin{table}[!ht]
\captionsetup{font=scriptsize}
\centering
\rowcolors{1}{}{lightgray}
\normalsize
\resizebox{0.85\linewidth}{!}{%
\begin{tabular}{lclllllllc}

  & \rot{Bill/Coin/Lightning (Sec.~\ref{Sec:bills vs coins}-\ref{Sec:lightning})} & \rot{Security (Sec.~\ref{Sec:Security})\tablefootnote{\scriptsize IT: Information-theoretic security; C: Computational security from standard assumption; N: No security proof or computational security based on a non-standard assumption.}} & \rot{Public (Sec.~\ref{Sec:public verifiability})\tablefootnote{\scriptsize $\diagup$: Does not provide full public verifiability.}} & \rot{Oracle not required (Sec.~\ref{Sec:money oracles})} & \rot{Efficient (Sec.~\ref{Sec:Efficiency})} & \rot{Classically verifiable (Sec.~\ref{Sec:quantum money with classical communication})} & \rot{Classically mintable (Sec.~\ref{Sec:quantum money with classical communication})} & \rot{Noise tolerant (Sec.~\ref{Sec:noise tolerance})} & \rot{Unbroken (Sec.~\ref{Sec:Security})\tablefootnote{\scriptsize \cmark: Unbroken; \xmark: Broken; $\diagup$: Broken in some cases.}} \\
\cite{Wie83} & \Bill & IT & \xmark & \cmark & \cmark & \xmark & \xmark & \cmark & $\diagup$ \citet{NSBU16} \\
\citet{BBBW82} & \Bill & C & $\diagup$ & \cmark & \cmark & \xmark & \xmark & \cmark & \xmark\ \citet{Sho94} \\
\citet{TOI03} & \Coin\tablefootnote{\scriptsize Untraceable for users, but not for the bank.} & IT & \xmark & \cmark & \cmark & \xmark & \xmark & \xmark & \cmark \\
\citet{Aar09} & \Bill & N & \cmark & \cmark & \cmark & \xmark & \xmark & \xmark & \xmark\ \citet{LAF+10} \\
\citet[Sec. 4]{MS10} & \Coin & IT & \xmark & \cmark & \xmark & \xmark & \xmark & \xmark & \cmark \\
\citet[Sec. 5]{MS10} & \Coin & IT & \cmark & \xmark & \xmark & \xmark & \xmark & \xmark & \cmark \\
\citet{Gav12} & \Bill & IT & \xmark & \cmark & \cmark & \cmark & \xmark & \xmark & \cmark \\
\citet[Sec. 5]{AC12} & \Bill & IT & \cmark & \cmark & \cmark & \xmark & \xmark & \xmark & \cmark \\
\citet[Sec. 6]{AC12} & \Bill & N & \cmark & \xmark & \cmark & \xmark & \xmark & \xmark & \xmark\ \citet{PDF+18} \\
\citet{FGH+12} & \Lightning & N & \cmark & \cmark & \cmark & \xmark & \cmark & \xmark & \cmark \\
\citet[Sec. 4]{MVW12}\tablefootnote{\scriptsize Combines \citet{Wie83} with classical verifiability.\label{MVW12PJH+12}} & \Bill & IT & \xmark & \cmark & \cmark & \cmark & \xmark & \cmark & \cmark \\
\citet{PYJ+12}\footref{MVW12PJH+12} (CV-qticket, p.2) & \Bill & IT & \xmark & \cmark & \cmark & \cmark & \xmark & \cmark & \cmark \\
\citet[Sec. 4]{GK15} & \Bill & IT & \xmark & \cmark & \cmark & \cmark & \xmark & \cmark & \cmark \\
\citet[Sec. 6]{BS16}\tablefootnote{\scriptsize Combines \citet{AC12} with classifical verification.\label{BS16Note}} & \Bill & IT & \cmark & \xmark & \cmark & \cmark\footref{BS16RS19} & \xmark & \xmark & \cmark \\
\citet[Sec. 7]{BS16}\footref{BS16Note} & \Bill & N & \cmark & \cmark & \cmark & \cmark\tablefootnote{\scriptsize Provides classical verification with the bank, but not with other users.\label{BS16RS19}} & \xmark & \xmark & \cmark \\
\citet{AA17} & \Bill & IT & \xmark & \cmark & \cmark & \cmark & \xmark & \cmark & \cmark \\
\citet{JLS18} & \Coin & C & \xmark & \cmark & \cmark & \xmark & \xmark & \xmark & \cmark \\
\citet[Sec. 5]{Zha19}\tablefootnote{\scriptsize \citet{Zha19} fixes the attack on \citet{AC12}.} & \Bill & N\tablefootnote{\scriptsize The security proof is based on the existence of post-quantum indistinguishability obfuscation, for which there are no constructions based on standard assumptions.} & \cmark & \cmark & \cmark & \xmark & \xmark & \xmark & \cmark \\
\citet[Sec. 4]{Zha19} & \Lightning & N\tablefootnote{\scriptsize The construction is based on collision resistant non-collapsing hash function. There are no candidate constructions for such a function, and therefore it cannot be instantiated.} & \cmark & \cmark & \cmark & \xmark & \cmark & \xmark & $\diagup$ \\
\citet[Sec. 6]{Zha19} & \Lightning & N & \cmark & \cmark & \cmark & \xmark & \cmark & \xmark & \xmark\ \citet{Rob19} \\
\citet{RS19proc} & \Bill & C & \xmark & \cmark & \cmark & \cmark & \cmark & \xmark & \cmark \\
\citet[Sec. 2]{RS19arxiv} & \Lightning & N & \cmark & \cmark & \cmark & \cmark\footref{BS16RS19} & \cmark & \xmark & $\diagup$\tablefootnote{\scriptsize The construction could be based on \citet[Section 4]{Zha19} or \citet[Section 6]{Zha19}.} \\
\citet{AGKZ20} & \Lightning & IT & \cmark & \xmark & \cmark & \cmark & \cmark & \xmark & \cmark \\
\citet{CS20} & \Lightning + \Bitcoin & N & \cmark & \cmark & \cmark & \cmark & \cmark & \xmark & $\diagup$\tablefootnote{\scriptsize The construction could be based on \citet{FGH+12}, \citet[Section 4]{Zha19} or \citet[Section 6]{Zha19}.} \\
\citet{BS20} & \Coin & C\tablefootnote{\scriptsize Security proof only in a weak adversarial model.\label{BS20RZ20}} & $\diagup$ & \cmark & \cmark & \xmark & \xmark & \xmark & \cmark \\
\citet{RZ20} & \Lightning & C\footref{BS20RZ20} & $\diagup$ & \cmark & \cmark & \xmark & \cmark & \xmark & \cmark
\vspace{2mm}
\end{tabular}
}
\caption{The table above classifies quantum money schemes according to nine properties. In cases where a paper introduces multiple schemes, we include separate entries and provide section references. Additional information about the classification system for each property is provided in the footnotes.}
\label{tab:properties}
\end{table}

\restoregeometry

In a ``public-key'' quantum money scheme, the bank generates both a private key and a public key. The private key is used to mint money. The public key is sent to all users. The public key allows users to efficiently verify the authenticity of a unit of quantum money. This eliminates the need for a user to communicate with the central bank to perform verification, as is done in Wiesner's scheme. Rather, verification can be performed ``locally.'' 

It is important to emphasize that no public-key scheme can achieve information-theoretic security. That is, unlike \citet{Wie83} and other private-key schemes, public-key schemes cannot use the no-cloning theorem alone to rule out the possibility of counterfeiting, and must instead base their security on computational hardness assumptions~\citep{Aar09}.

Several public quantum money schemes have been proposed since~\cite{Aar09} originally introduced the concept. As shown in Table \ref{tab:properties}, none of these schemes has a security proof based on standard hardness assumptions, which we discuss in Section~\ref{Sec:Security}. Constructing such a scheme is considered to be an important open question.

\subsubsection{Quantum Lightning}
\label{Sec:lightning}

In a public quantum money scheme, the central bank can prepare many instances of the quantum state associated with a given serial number. A quantum lightning scheme has all the properties of public quantum money, but also guarantees that \textit{even the central bank itself} cannot generate multiple bills with the same serial number. The notion of ``quantum lightning'' was first defined in \cite{Zha19}; however, \cite{FGH+12}, which predates \cite{Zha19}, also offers a construction that satisfies the definition. We highlight the construction by \cite{FGH+12} in Appendix \ref{Knot-Based Quantum Money}, which requires the use of concepts from knot theory. A detailed overview of \cite{Zha19} is beyond the scope of the paper.

From a transparency perspective, the impossibility of constructing multiple bills with the same serial number could be used to provide a demonstrable guarantee on the amount of money in circulation. If a bill's serial number is required for verification and the list of all serial numbers is made publicly available, then it would be possible for anyone to verify an upper-bound on the amount of money in circulation. This is not, of course, true for physical cash, since a rogue central bank could produce multiple bills with the same serial number. This property eliminates the need for one dimension of trust in the central bank, which could be valuable in countries with recent histories of high inflation.

\subsubsection{Security}
\label{Sec:Security}

As we will see in this subsection, minor changes in the definition of ``unforgeability'' can have important implications for the security of a quantum money scheme. We will demonstrate this by examining the concept of unforgeability through a sequence of examples where an adversary is attempting to perform verification in a way that was not intended by the central bank. We will then provide a full definition of security for public quantum money.

We start by defining forgery as an act through which an adversary successfully receives money from a bank without passing the bank's verification scheme. This simple definition might appear to be sufficiently broad, but it actually fails to capture certain forms of forgery. Consider, for instance, an adversary who received one quantum money state from the bank and passed two verifications. Clearly, that is forgery as well. Or perhaps an adversary needs $n$ money states to produce $n+1$ states that pass verification, which we would also define as undesirable and a form of forgery. We would also say that an adversary performs forgery if she starts with $n$ money states and generates $m$ states for which strictly more than $n$ pass verification. Note that these types of forgeries are listed in a decreasing order of hardness. We, of course, want of all of these to be impossible for the adversary, so we will typically try to rule out the easiest form.

We may also want to consider the case where the adversary succeeds with the forgery attempt, but only with some small probability, such as $\frac{1}{4}$. We would like to prevent this as well. Unfortunately, it is impossible to guarantee a success probability of $0$, since brute force attacks have non-zero success probability. The standard way to formalize this in cryptography is to use the notion of a ``negligible'' function. A function is negligible if it decays faster than an inverse polynomial. Formally, a function $f:\mathbb N \to \mathbb R^+$ is said to be negligible if for every $c\in \mathbb N$ there exists $N_c$ such that for all $x\geq N_c$, $f(x)\leq \frac{1}{x^c}$. Therefore, we say that the scheme is secure if a forger's success probability is negligible.

Another issue which needs to be specified is whether the adversary can request that money be returned after a verification attempt. If the scheme is secure even under this condition, we say that it is secure against adaptive attacks (see Section \ref{Attacks on Wiesner} of the Appendix). Some schemes are not secure by this definition, which means that new money must be minted and delivered after a successful verification. Wiesner's scheme, for instance, is secure against non-adaptive attacks~\citep{PYJ+12,MVW12}, but requires modification to secure it against adaptive attacks~\citep{Aar09,Lut10,NSBU16}. \cite{Gav12} proposed alternative private-key schemes that also achieved unconditional security, even against adaptive attacks.

Now that we have examined the different varieties of attack an adversary may conduct, we will construct a full definition of security for public quantum money. Like many cryptographic schemes, unforgeability is defined by a security game between a challenger and an adversary. The challenger generates a public-key and a private-key, and sends the public-key to the adversary. The adversary asks for $n$ quantum money states. The bank then applies the minting algorithm to produce $\ket{\$_1},\ldots,\ket{\$_n}$ and sends those money states to the adversary. The adversary prepares $m$ (possibly entangled) quantum states and sends them to the challenger. The challenger verifies these $m$ states using the verification algorithm. We say that the adversary wins if the number of successful verifications is strictly larger than $n$. Furthermore, the scheme is said to be secure for all adversaries that run in polynomial time if the probability of winning this game is negligible.\footnote{Here, we mean negligible in the ``security parameter.'' In most cases, it means the number of qubits of the quantum money state, which is a parameter which can be chosen by the central bank: as the security parameter increases, it becomes increasingly difficult to forge.}

Note that all public-key schemes, including those that predate the modern literature \citep{BBBW82}, rely on complexity-theoretic notions of security \citep{AC12}, which must make explicit assumptions about the resources available to an adversary. In the security definition above, for instance, we assume that an adversary operates in polynomial time. This differs from certain private-key schemes, such as \cite{Wie83}, which achieve information-theoretic security and are unconditionally secure against adversaries.

One notable attempt to construct public-key quantum money using a complexity-theoretic notion of security was proposed in \cite{FGH+12}, which used exponentially large superpositions and knot theory to generate quantum bill states. The security of this scheme rested on the computational intractability of generating a valid quantum bill state, as well as the impossibility of copying unknown quantum states. Unfortunately, it is not possible to fully analyze the scheme's security properties without first achieving advances in knot theory. See Section \ref{Knot-Based Quantum Money} in the Appendix for a full description of the scheme.

\subsubsection{Oracles}
\label{Sec:money oracles}

Certain public money schemes, such as \citet[Sec. 4]{MS10}, rely on the use of an \textit{oracle}. As discussed in Section \ref{Sec:Oracles}, an oracle is a black-box function, which we will assume is universally available to users for the purpose of this section. The main advantage of an oracle is that users cannot ``look-inside'' of it. Rather, the only way to access it is through the input-output behavior of the function. If this were not the case, a potential forger could gain information by analyzing the circuit that implements the oracle, rather than its input-output behavior. Therefore, constructing a public money scheme with an oracle is substantially easier than constructing a scheme without one.

There are two ways to interpret quantum money constructions that rely on oracles. The first is that the oracle construction could be an intermediate step towards a full public quantum money scheme. \citet{AC12}, for instance, start with a scheme based on an oracle (Section 5) and later show how it can be removed (Section 6). Alternatively, an oracle could be interpreted as a technology, such as an application programming interface (API), that the central bank provides to external users. Of course, using the oracle would require quantum communication with the central bank, which would void the main advantage of using public quantum money.

\paragraph{Complexity-Theoretic No-Cloning Theorem}

We will now briefly outline how oracles enable the construction of \textit{public} quantum money scheme. We will approach this by discussing two useful theorems proven in the paper that provide a basis for constructing certain public-key quantum money schemes. The first is a statement of existence for the oracle upon which the scheme relies.

\begin{theorem}[\cite{Aar09}]
There exists a quantum oracle, U, relative to which publicly-verifiable quantum money exists.
\end{theorem}
\vspace{3mm}

\noindent The second theorem, which \cite{Aar09} refers to as the ``complexity-theoretic no-cloning theorem," explains the properties of the oracle, $U$, and provides strict guarantees regarding unforgeability. In this theorem, the counterfeiter is given the quantum oracle used for verification, $U_{\psi}$, and $k$ quantum bills, $\ket{\psi}^{\otimes k}$, each of which consists of an $n$-qubit pure state, $\ket{\psi}$. The counterfeiter then attempts to use the $k$ quantum bill states to generate $k+\delta$ valid quantum bill states. \cite{Aar09} proves that this will require $\Omega \Big(\frac{\delta^{2}\sqrt{2^{n}}}{l^{2}klogk}-l \Big)$ queries to $U_{\psi}$. Even for $\delta = 1$, this quickly becomes intractable as the number of qubits, $n$, increases.
\\
\begin{theorem}[Complexity-Theoretic No-Cloning \citep{Aar09}, proof in ~\cite{AC12} Theorem B.1]
Let $\ket{\psi}$ be an n-qubit pure state sampled uniformly at random (according to the Haar measure). Suppose we are given the initial state, $\ket{\psi^{\otimes k}}$, for some $k \geq 1$, as well as an oracle, $U_{\psi}$, such that $U_{\psi} \ket{\psi} = -\ket{\psi}$ and $U_{\psi} \ket{\phi} = \ket{\phi}$ for all $\ket{\phi}$ orthogonal to $\ket{\psi}$. Then for all $l>k$, to prepare l registers $\rho_{1},...,\rho_{l}$ such that:

\begin{equation}
\sum_{i=1}^{l} \bra{\psi} \rho_{i} \ket{\psi} \geq k + \delta
\end{equation}

\noindent \textit{We need:}

\begin{equation}
\Omega \Big(\frac{\delta^{2}\sqrt{2^{n}}}{l^{2}klogk}-l \Big)
\end{equation} 

queries to $U_{\psi}$.
\end{theorem}

\noindent The complexity-theoretic no-cloning theorem combines two elements: (1) the original no-cloning theorem; and (2) the quadratic upper-bound on the quantum speedup achievable in unstructured search problems \citep{Gro96, BBBV97}. It shows that a counterfeiter who has access to $k$ random, valid bills needs to perform $\propto \sqrt{2^{n}}$ queries to successfully counterfeit a bill. This does not improve substantially over using Grover's algorithm to identify valid bill states. Consequently, if bills have a high number of qubits, $n$, then the probability of a successful counterfeit will be negligible.

The complexity-theoretic no-cloning theorem was also used in later public-key quantum money schemes, including \cite{AC12} and the quantum coin construction \cite{MS10} (see Section~\ref{Sec:bills vs coins}). These schemes make use of the oracle from \cite{Aar09} in their verification algorithm.

\subsubsection{Efficiency}
\label{Sec:Efficiency}

Efficiency requires that all of the protocols used to mint and verify units of money can be executed in polynomial time on a quantum computer. For instance, a scheme that has 1000 qubits and a verification process run time with exponential scaling could take millions of years to verify a single state and, therefore, would not be considered efficient. Inefficient schemes can, however, prove useful as milestones for efficient, practical schemes. \cite{MS10}, for instance, is an inefficient scheme, but served as a building block for \cite{JLS18}, which is efficient, but has the downside of reducing the unforgeability level to computational security.

\subsubsection{Classical Verification and Mintability}
\label{Sec:quantum money with classical communication}
Assume we have a unit of quantum money and want to verify its validity. This could be, for instance, a quantum subway token, as envisioned by \cite{BBBW82}, which would permit entrance into the subway after verification was performed at the turnstile. Alternatively, it could be a form of quantum money that is used to make payment in e-commerce transactions. In the former case, we could physically deposit the token; however, in the latter case, we are transacting at a distance and would need access to a communication channel to make the payment. 

\cite{Wie83} relies on a \textit{quantum} communication channel to verify quantum money states; however, as \cite{Gav12} demonstrated in a paper that introduced the concept of ``classical verifiability,'' this is not strictly necessary. Classical verifiability means there is no need for a quantum communication channel to verify a form of quantum money. Instead, verification is performed using an interactive protocol between the bank and the payer. There are at least two advantages to using a classical channel for verification. First, such schemes do not require the creation of quantum communication channels between merchants and the central bank to perform verification. Instead, existing classical communication channels, such as the classical internet, can be used. And second, an attacker will not be able to modify the bill's underlying quantum state by intercepting communications and applying transformations to the qubits, as they could with Wiesner's money.

Certain quantum money schemes, such as \citet{FGH+12} and \cite{RS19proc}, include a procedure to mint quantum money through the use of purely classical interactions between the bank and the receiver. Such schemes necessarily rely on computational assumptions~\citep{RS20}. The benefit of adopting a scheme that permits classical mintability is that quantum communication is not needed to mint and distribute money. Furthermore, when classical minting is combined with classical verification, no quantum communication infrastructure is needed.

\subsubsection{Noise Tolerance}
\label{Sec:noise tolerance}

One of the greatest remaining challenges to implementing quantum technologies, including quantum money, is the noise that arises from a quantum system as it interacts with its surrounding environment and decoheres.\footnote{For more details on the technical challenge of noise in quantum technologies, see Section~\ref{Sec:Experimental Implementation}.} The most straightforward way to deal with noise is to use quantum error-correction; however, some quantum technologies, including certain varieties of quantum money, are designed to build noise tolerance into the system, rather than relying on quantum error-correction. Since quantum error-correction is prohibitively hard to implement at large scale with current technologies, experimental work on quantum money, which is discussed in detail in Section~\ref{Sec:Quantum Money Implementation}, has relied on the noise-tolerant schemes that were introduced in \cite{PYJ+12} and \citet{AA17}.

\subsection{Experimental Implementation}
\label{Sec:Quantum Money Implementation}

When the concept of quantum money was originally introduced in \cite{Wie83}, it was clear that it would not be technologically feasible to implement in the foreseeable future. As proposed, it relied on the physical encoding of classical states into properties of single photons, the elementary particles of light, such as their polarization. Quantum optical systems are indeed the privileged experimental platform for the implementation of quantum cryptographic schemes like quantum money because of the maturity of the techniques for the manipulation of photons and their capacity to be transferred along physical communication channels like optical fibers or free space. However, even if a central bank has the means to construct such quantum bills and perform the encoding at a low cost, the states would have to be stored in a memory for some time without substantial decoherence before their retrieval, use, and verification. Quantum memories constitute a challenging technology and despite important progress in their development in the recent years, their characteristics -- namely the storage time, the retrieval efficiency of the quantum state after storage, and the fidelity of the retrieved state with respect to the initial one -- cannot in general be optimized within a single system and are currently not suitable for practical use \citep{HEH+16}.

There has, however, been substantial progress related to the resources required by theoretical quantum money schemes, which has enabled researchers to partially implement some forms of quantum money. Since public-key schemes present additional technical challenges, both from a theoretical and practical point of view, research thus far has focused on the construction of private-key forms of money. While this does not offer the possibility of public verification, it could provide a payment instrument similar to CBDCs, but with information-theoretic security, a standard which is unachievable with purely digital forms of money. We briefly review the experimental implementation of two such schemes.

\subsubsection{Quantum Optical Money}

The first scheme we consider is the realization of elementary quantum optical bills, which were first experimentally demonstrated in \cite{BCC+17}. We note that the authors refer in their paper to banknotes but, in fact, use a verification process that requires interaction with the bank, which means that it is not a public-key scheme. The scheme is based on \cite{Wie83}, but with the following modification: rather than encoding a randomly-drawn classical bit string in qubits using randomly-drawn bases, each bill encodes a grayscale image using a matrix of polarized photons. The polarization states are chosen to correspond to the colors in a grayscale image. Using three sets of encoding bases, for instance, would allow for the use of six photon polarizations, each corresponding to a different color between white and black.

Their proof-of-principle experimental implementation is performed using encoding in photon polarizations in a laboratory setting. They were able to successfully demonstrate the creation of quantum money states under the altered version of \cite{Wie83} they proposed. They also showed how optimal cloning attacks could be used to perform counterfeiting, focusing on attacks directed against individual photons, which they argue are the most plausible in the near-term.

\subsubsection{Quantum Credit Cards}

\cite{Gav12}, \cite{GK15}, and \cite{AA17} propose schemes that can be categorized as ``quantum credit cards,'' which rely on quantum retrieval games \citep{BJK04, AKL16}. \cite{BOV+17} were the first to experimentally implement such a scheme. Their approach made use of polarized weak coherent states of light and allowed for classical verification, which eliminates the need to establish a quantum communication channel with the verifier. They were also able to rigorously demonstrate unforgeability, yielding improved security over existing forms of digital payment, such as credit cards. They subsequently also examined the security of this scheme in more detail in a practical setting and provided numerical bounds for realistic loss and noise parameter regions \citep{BDG19}.

As with the other experimental implementations of quantum money, \cite{BOV+17} is constrained by progress in the development of quantum memory. They do, however, construct their quantum money scheme to be compatible with recent developments in the implementation of quantum memory, enabling such proof-of-principle demonstrations in the near-term.

In closely related work, \cite{GAA+18} experimentally implement a quantum money scheme that is based on \cite{AA17}. They demonstrate how each part of the scheme can be executed, including bill state preparation and verification, but also encounter the same quantum memory bottleneck. Both \cite{BOV+17} and \cite{GAA+18} take experimental imperfections into account when evaluating the security of their schemes.

\subsubsection{Remaining Challenges}

The main impediment to the full-scale implementation of private-key quantum money is the difficulty of storing quantum states. For the scheme implemented in \cite{BOV+17}, if we assume that only the storage mechanism is responsible for the loss in the system, it can be shown that the memory must achieve an 85\% storage/retrieval efficiency, giving rise to less than a 2\% error rate upon verification, given an average photon number in the coherent state equal to 1 \citep{Boz19}. This is within experimental reach, as it has already been demonstrated that using quantum memories based on cold atomic clouds can provide up to 90\% storage/retrieval efficiency \citep{HTH+18}. Additionally, the error rate due to state preparation can also be reduced and the quantum memory state fidelity can approach 99\% for an average photon number greater than 1 \citep{VHC+18}. In the long term, using superconducting nanowire single-photon detectors (which can achieve detection efficiency around 90\%) and further optimization of the storage/retrieval efficiency could allow a full demonstration of the scheme under security conditions where the client terminal would be trusted. Note, however, that the storage times for such quantum memories are on the order of microseconds currently (although other techniques can reach milliseconds or even seconds) and that multiplexing techniques would have to be used to store multiple qubits simultaneously in the same memory, using, for instance, spatial modes as in \cite{VHC+18}. This discussion illustrates the various trade-offs that need to be considered with respect to the use of quantum money as means of performing financial transactions in practice.

%% file: QuantumAlgorithms.tex
\section{Quantum Algorithms}
\label{QuantumAlgorithms}

This section provides an overview of quantum algorithms that may have near-term future relevance for econometricians and computational economists. It is divided into two subsections. The part first covers theoretical developments in the construction of quantum algorithms and the second part describes experimental progress in their implementation on quantum computing devices. In the theory subsection, we will focus primarily on providing a high-level examination of algorithms, but will also discuss low-level detail where useful. For each algorithm, we will try to identify prospective applications within economics, determine computational speedups achievable with existing algorithms, and identify whether an algorithm has additional restrictions that do not apply to its classical counterpart. In the experimental section, we will provide a history of the development of quantum computers, including a review of the most recent progress in their development. We will also discuss their limitations.

Many of the algorithms we present involve the use of phase kickback, phase estimation, and the quantum Fourier transform (QFT). Those who wish to understand the details of these subroutines should see Sections \ref{Phase Kickback}, \ref{Phase Estimation}, and \ref{Quantum Fourier Transform} in the Appendix. Interested readers may also wish to see \citet{Mon16} for a high-level survey of quantum algorithms, \citet{Chi17} for detailed lecture notes on the same subject, or the Quantum Zoo for a regularly-updated database of quantum algorithms.\footnote{See \href{http://quantumalgorithmzoo.org}{http://quantumalgorithmzoo.org}, which is a regularly-updated list of quantum algorithms maintained by Stephen Jordan.}

\subsection{Theoretical Progress}

As was the case for classical algorithms and classical computers, theoretical progress in quantum computing tends to lead experimental implementation. In this subsection, we will provide an overview of quantum algorithms that have relevance for economists and the state of progress in their refinement.

\subsubsection{Numerical Differentiation}

Numerical methods commonly used to solve economic models and perform econometric estimation often rely on the computation of first and second derivatives. The steepest ascent algorithm, for instance, requires the repeated computation of the gradient. Furthermore, the Newton-Raphson method, the hill climbing method, and the family of quasi-Newton methods, including Davidon-Fletcher-Powell (DFP), Broyden-Fletcher-Goldfarb-Shanno (BFGS), and Berndt-Hall-Hall-Hausman \citep{BHH+74}, and \citet{Mar63} require the computation of both the gradient and the Hessian matrix of second derivatives. Such methods are commonly used in financial econometrics,\footnote{The large literature on ARCH and GARCH models makes use of numerical gradient and Hessian computation (\citealp{Bol86}; \citealp{ELR87}; \citealp{Bol87a}; \citealp{Bol87b}; \citealp{Dan94}; \citealp{Zak94}; \citealp{ER98}; \citealp{Gra96}; \citealp{ER98}; \citealp{SG98}; and \citealp{Eng00}). Numerical differentiation is also employed to solve a variety of different models of financial markets (\citealp{Hsi91}; \citealp{HJ94}; \citealp{LS92}; \citealp{DG96}; \citealp{DG97}; \cite{DE00}; \citealp{BKS03}).} structural microeconometrics,\footnote{\citet{AM02} and \citet{SJ12} provide algorithms for structural microeconometric models that make use of numerical derivatives. \citet{AM10} offers a survey of the literature on dynamic discrete choice models, which makes extensive use of gradient-based methods. \cite{BH78}, \cite{Lan79}, and \cite{HM80} employed gradient-based methods to solve structural microeconomic models.} maximum likelihood estimation,\footnote{Maximum likelihood estimation (MLE) is used for a variety of different economic and financial problems, including the estimation of structural models (see, e.g., \citealp{Bel80}; \citealp{Gre82}; \citealp{Whi82}; \citealp{Bun88}; \citealp{RSP05}; \citealp{FR07}; and \citealp{SJ12}). It is often necessary to compute the gradient and Hessian of the likelihood function, which can create a bottleneck in the estimation algorithm for high-dimensional problems.} dynamic stochastic general equilibrium (DSGE) modelling,\footnote{Packages used to solve and estimate DSGE models, such as Dynare, commonly make use of numerical gradients and Hessian matrices.} and large-scale macroeconomic modelling conducted by central banks and government agencies.\footnote{See \cite{CCW10} and \citet{CTW11} for examples of large-scale central bank models that require the computation of a gradient or Hessian.}

For high dimensional models and estimation problems, using analytical gradients may be impossible if there is no closed-form solution or error-prone if the model is sufficiently complicated. For this reason, such computational and econometric routines often employ numerical differentiation. Finite difference methods, for instance, can be used to compute gradients numerically by performing functional evaluations within an approximately-linear neighborhood of a point. The forward difference method, which is the simplest method, approximates the partial derivative of a function, $\frac{\partial f(x)}{\partial x_{j}}$, where $x=(x_{1},x_{2},...,x_{d})$ and uses a Taylor expansion to bound the error size as follows:

\begin{equation}\label{eqn:finite_difference_1}
f(x_{1},...,x_{j}+l,...x_{d}) - f(x) = \frac{\partial f(x)}{\partial x_{j}} l + \frac{\partial^{2} f(x)}{\partial x_{j}^{2}} \frac{l^{2}}{2!} + \frac{\partial^{3} f(x)}{\partial x_{j}^{3}} \frac{l^{3}}{3!} + ...
\end{equation}
We can then rearrange Equation (\ref{eqn:finite_difference_1}) as follows:

\begin{equation}\label{eqn:finite_difference_2}
\frac{\partial f(x)}{\partial x_{j}} = \frac{f(x_{1},...,x_{j}+l,...x_{d}) - f(x)}{l} - \frac{\partial^{2} f(x)}{\partial x_{j}^{2}} \frac{l}{2!} - \frac{\partial^{3} f(x)}{\partial x_{j}^{3}} \frac{l^{2}}{3!} - ...
\end{equation}
For a small $l$, the error associated with using the forward difference will be of order $O(l)$. We can reduce this to $O(l^{2})$ by using the centered difference, $f(x+l)-f(x-l)$. Note that we must perform $d+1$ function evaluations to compute the gradient, $\gradient f(x) = (\frac{\partial f(x)}{\partial x_{1}},...,\frac{\partial f(x)}{\partial x_{d}})$, using the forward-difference method. This is because we must perform one evaluation of $f(x)$ and one evaluation for each of the $d$ components of the gradient. If we use the centered-difference method, we must instead perform $2d$ function evaluations to compute the gradient. Furthermore, numerical computation of the Hessian matrix will require $O(d^{2})$ function evaluations.

\citet{Jor05} introduced a quantum algorithm for numerical gradient computation. To compare the performance of this quantum algorithm with classical gradient algorithms, he employs the concept of query complexity, which we discussed in Section \ref{Sec:Oracles}.\footnote{See Section \ref{Computational Complexity} in the Appendix for an overview of computational complexity and related notation.} Here, query complexity measures the number of functional evaluations needed to compute a gradient with $d$ components to $n$ bits of precision. As we showed earlier, the simplest method of classical numerical gradient computation, forward-differencing, requires $d+1$ queries to compute a gradient with $d$ components. In contrast, Jordan's quantum algorithm requires only one query, regardless of the size of $d$. It is also able to compute $n$th order derivatives using $O(d^{n-1})$ queries, rather than the $O(d^{n})$ queries that would be required by a classical routine.

\begin{algorithm}
\DontPrintSemicolon
Initialize $d$ input registers with $n$ qubits each in the $\ket{0}$ position.\;
Initialize 1 output register with $n_{0}$ qubits in the $\ket{0}$ position.\;
Apply $H$ to all input registers.\;
Apply $X$ to the output register.\;
Apply inverse quantum Fourier transform, yielding:\;
\qquad $\frac{1}{\sqrt(N^{d} N^{0}} \sum_{\delta_{1}=0}^{N-1} \sum_{\delta_{2}=0}^{N-1} ... \sum_{\delta_{d}=0}^{N-1} \ket{\delta_{1}} \ket{\delta_{2}} ... \ket{\delta_{d}} \sum_{\delta_{a}=0}^{N_{0}-1} e^{i2 \pi a / N_{0}} \ket{a} $ \;
Use an oracle to compute $f$.\;
Add the output (modulo $2^{n_{0}}$) to the output register.\;
Apply a quantum Fourier transform to each register, yielding:\;
\qquad $\ket{\frac{N}{m}\frac{\partial f}{\partial x_{1}}} \ket{\frac{N}{m}\frac{\partial f}{\partial x_{2}}} ... \ket{\frac{N}{m}\frac{\partial f}{\partial x_{d}}}$ \;
Measure in computational basis, yielding $\gradient f$. \;
\caption{Quantum Numerical Gradient Computation \citep{Jor05}}\label{alg:Jor05}
\end{algorithm}

The pseudocode for \cite{Jor05} is given in Algorithm \ref{alg:Jor05}. Note that the inputs and outputs to the oracle are integers from a bounded, nonnegative interval, which are represented by binary strings. The inputs and outputs to $f$ are real numbers. The oracle retains the inputs to preserve reversibility. Furthermore, $f$ must be continuous within a vicinity of $x = (x_{1},x_{2},...,x_{d})$, the point at which the gradient is computed. Finally, the number of ancilla qubits, $n_{0}$, can be set as specified in Equation (\ref{eqn:ancilla_qubits}) to ensure that the output is accurate within an $\pm \theta$ interval. Note that the parameter $m$ is the size of the interval that bounds the individual components of the gradient. 

\begin{equation}\label{eqn:ancilla_qubits}
n_{0} = log_{2} \left[ \frac{max(f) - min(f)}{\frac{ml}{2^{n}}\frac{\theta}{2\pi}} \right]
\end{equation}

Interested readers may also wish to see \cite{Bul05}, \cite{Rot09}, and \cite{Mon11}, all of which expand on \cite{Jor05}. Currently, no quantum algorithm provides more than a polynomial speedup over classical algorithms with respect to query complexity. However, for high dimensional models and estimation problems, collapsing the number of functional evaluations from at least $d+1$ to one may substantially reduce program run time. This is especially true in root-finding operations, where the gradient must be repeatedly computed to locate an optimum.

\subsubsection{Interpolation}

Solving dynamic economic models often entails the use of functional equations, such as Bellman equations and Euler equations.\footnote{See \cite{SV98} for an explanation of the attractive convergence properties of value function iteration and \cite{AFR06} for a comparison of solution methods for dynamic equilibrium models.} While global solutions to such problems can be represented by a tensor of values that approximates an unknown function at a discrete set of points, improving the accuracy of such representations will incur a high computational cost due to the curse of dimensionality. In particular, if we have $n$ continuous states, which are each discretized into $s$ nodes, then the tensor product representation of the value function or decision rule will contain $k=s^{n}$ nodes. This means that a doubling of the density of nodes in each state will result in an increase in the size of the state space by a factor of $2^{n}$, which is prohibitive even for relatively small models.\footnote{A wide variety of computational models in economics and finance employ interpolation in the solution method. For a range of applications, see \cite{KW94}, \cite{Ack03}, \cite{Rus97}, and \cite{CS05}. For surveys of problem classes that often employ interpolation, see \cite{HN07}, \cite{AM10}, and \cite{Kea11}.}

Consequently, commonly-used solution methods for high dimensional models do not typically rely on node density to achieve an accurate approximation of the unknown function of interest. \cite{KK04} and \cite{JMMV14}, for instance, make use of the Smolyak method to construct sparse grids, effectively circumventing the curse of dimensionality by avoiding the use of tensor product grids. A more common approach uses tensor product grids, but then interpolates between the nodes.\footnote{See \citet{Jud98} for an overview of interpolation methods.} We will focus on that approach in this subsection.

While interpolation often yields a decrease in run time for a given level of accuracy, it remains one of the most computationally costly routines in many solution methods.\footnote{\citet{HM09} compare run times and Euler equation residuals for an infinite horizon Ramsey model under several different solution methods, including value function iteration with and without interpolation. When the state space contains 5,000 nodes, they find that cubic polynomial interpolation is 32 times faster than value function iteration and also generates small Euler equation residuals.} Consider the case where we wish to interpolate a value function $V$ with a single state variable, which has $k$ nodes, using monomial basis functions: $1$, $x_{i}^{2}$, $x_{i}^{3}$,..., $x_{i}^{d},$ $\forall i \in [k]$. Our objective is to find a $d$-dimensional vector of coefficients, $c$, such that Equation (\ref{eqn:polynomial_interpolation}) is satisfied. Note that $v_{1}, v_{2},..., v_{k}$ are the values of $V$ at each of the $k$ nodes.

\begin{equation}\label{eqn:polynomial_interpolation}
\begin{bmatrix}
v_{1} \\
v_{2} \\
\vdots \\
v_{k} \\
\end{bmatrix}
= \\
\begin{bmatrix}
1 & x_{1} & x_{1}^{2} & \dots & x_{1}^{d} \\
1 & x_{2} & x_{2}^{2} & \dots & x_{2}^{d} \\
\vdots & \vdots & \vdots & \ddots & \vdots \\
1 & x_{k} & x_{k}^{2} & \dots & x_{k}^{d} \\
\end{bmatrix}
\begin{bmatrix}
c_{1} \\
c_{2} \\
\vdots \\
c_{d} \\
\end{bmatrix}
\end{equation}
The collection of monomial terms forms a Vandermonde matrix. If $k=d$, then we may solve the system of linear equations to determine $c_{1},c_{2},...,c_{d}$ exactly. This operation will have a time cost of $O(k^{3})$. Alternatively, if $k>d$, we will instead compute a residual vector and minimize its $L_{2}$ norm.

\begin{equation}\label{eqn:polynomial_interpolation_residual}
\begin{bmatrix}
\epsilon_{1} \\
\epsilon_{2} \\
\vdots \\
\epsilon_{k} \\
\end{bmatrix}
=
\begin{bmatrix}
v_{1} \\
v_{2} \\
\vdots \\
v_{k} \\
\end{bmatrix}
-
\begin{bmatrix}
1 & x_{1} & x_{1}^{2} & \dots & x_{1}^{d} \\
1 & x_{2} & x_{2}^{2} & \dots & x_{2}^{d} \\
\vdots & \vdots & \vdots & \ddots & \vdots \\
1 & x_{k} & x_{k}^{2} & \dots & x_{k}^{d} \\
\end{bmatrix}
\begin{bmatrix}
c_{1} \\
c_{2} \\
\vdots \\
c_{d} \\
\end{bmatrix}
\end{equation}

In general, we will need to use polynomial interpolation for problems with multiple state variables. If we have $n$ state variables and choose to interpolate with a $d$-degree polynomial, inversion of the multivariate Vandermonde matrix will require $\binom{n+d}{d}$ queries in a classical setting. Alternatively, we may think of this as follows: if we have a value function, $V$, with $n$ state variables that takes the form of a $d$-degree polynomial, then we may recover the coefficients of that polynomial from a tensor product grid with $k$ nodes.

Relative to classical algorithms, the current state-of-the-art quantum polynomial interpolation algorithm \citep{CCH17} achieves a $\frac{n+1}{2}$ reduction in queries (nodes) needed to perform interpolation over the real numbers. There are, however, a few special cases for which the number of queries differs, which are given in Equation (\ref{interpolation_cases}).

\begin{equation}\label{interpolation_cases}
k = 
\begin{cases}
2n+2 & d = 2, n \geq 2 \\
[\frac{2}{n+1} \binom{n+d}{d} + 2] & (n,d) = (4,3), (2,4), (3,4), (4,4) \\
[\frac{2}{n+1} \binom{n+d}{d}] & \mbox{otherwise}
\end{cases}
\end{equation}

A rough sketch of the the pseudocode for \citet{CCH17} is given in Algorithm \ref{alg:CCH17}. For the details of the algorithm, along with the relevant mathematical preliminaries, see \citet{CCH17}. For explanations of phase kickback, phase estimation, and the quantum Fourier transform -- important components of the algorithm -- see sections \ref{Phase Kickback}, \ref{Phase Estimation}, and \ref{Quantum Fourier Transform} in the Appendix. It may also be useful to see the preceding work on univariate polynomial interpolation in \citet{CDHS16}. A larger literature explores polynomial interpolation over finite fields, which are useful for cryptographic applications. 

\begin{algorithm}
\DontPrintSemicolon
Given oracle that computes $n$-variable, $d$-degree polynomial: $f(x_{1},...,x_{2})$.\;
Oracle computes $\ket{x,y} \rightarrow \ket{x,y+f(x)}, \forall x, y$.\;
Initialize system in bounded superposition over working region.\;
Exploit phase kickback to recover coefficients of polynomial.\;
Apply $k$ parallel standard queries in the Fourier basis.\;
Encode results in a phase.\;
\caption{Quantum Multivariate Polynomial Interpolation \citep{CCH17}}\label{alg:CCH17}
\end{algorithm}

\subsubsection{Linear Systems}

Linear systems are often used in econometrics and economic models. Within econometrics, regression problems often require the solution of a linear system. Furthermore, both macroeconomic and microeconomic models often consist of nonlinear systems of difference or differential equations that are linearized and studied around a point of interest, such as the model's steady state (see, e.g., \cite{BK80}, \cite{TU90}, \cite{LS07}, \cite{Lud07}, and \cite{Rei09}). While classical solution algorithms for linear systems are already sufficiently fast for most research applications, growth in the use of large microdata sets in econometrics and heterogeneous agent models in macroeconomics is likely to increase the value of run time reductions for linear system solution algorithms in the future.

\paragraph{Quantum Linear Systems Problems}

\cite{HHL09} construct a quantum algorithm for sampling the solution of a linear system of the form $Ax=b$, where $A$ is an NxN Hermitian matrix,\footnote{In the case where $A$ is not Hermitian, the authors point out that we may instead use $C=\begin{bmatrix} 0 & A \\ A^{\dagger} & 0 \\ \end{bmatrix}$ Furthermore, we may replace $x$ with $\begin{bmatrix} 0 \\ x \end{bmatrix}$ and $b$ with $\begin{bmatrix} b \\ 0 \end{bmatrix}$.} $b$ is a unit vector, $x$ is the solution vector, and $x^{\dagger}Mx$ is the sample returned for an arbitrary operator, $M$. Whereas classical algorithms can solve the system and sample it in $poly(N,\kappa)$ time, the quantum algorithm \cite{HHL09} propose is able to do the same in $poly(log(N), \kappa)$ time, where $A$ is a $d$-sparse matrix with condition number $\kappa$. This amounts to an exponential speedup, which is a remarkable, given that writing down $x$ and $A$ would require $N$ and $N^{2}$ steps, respectively.

A rough sketch of \cite{HHL09} is given in Algorithm \ref{alg:HHL09}. For recent work that builds on \cite{HHL09}, see \cite{Amb10} and \cite{CKS17}. \cite{CKS17} provide a further speedup under certain conditions, reducing the solution to time $log(1/\epsilon)$, where $\epsilon$ is the precision of the output state.

\begin{algorithm}
\DontPrintSemicolon
Encode the $b$ vector as a quantum state: $\ket{b} = \sum_{i}^{N} b_{i} \ket{i}$\;
Apply $e^{iAt}$ to $\ket{b}$ for a superposition of $t$ values.\;
Use phase estimation to decompose $\ket{b}$ into the eigenbasis of $A$, $\ket{u_{j}}$, and eigenvalues, $\lambda_{j}$, yielding $\sum_{j=1}^{N} \beta_{j} \ket{u_{j}} \ket{\lambda_{j}}$. \;
Map $\ket{\lambda_{j}}$ to $C\lambda^{-1}_{j} \ket{\lambda_{j}}$, where $C$ is a normalizing constant. \;
Uncompute $\lambda_{j}$, yielding a state proportional to $\sum_{j=1}^{N} \beta_{j} \lambda_{j}^{-1} \ket{u_{j}} = A^{-1} \ket{b} = \ket{x}$.
\caption{Quantum Linear Systems Problem \citep{HHL09}}\label{alg:HHL09}
\end{algorithm}

While the \cite{HHL09} algorithm generates an exponential speedup and constitutes one of the most successful quantum algorithms, it also faces several limitations that do not apply to standard classical algorithms for solving linear systems. In particular, \cite{Aar15} identifies the following four limitations of the algorithm:

\begin{enumerate}
\item It requires the vector $b$ to be loaded into memory quickly. This is currently infeasible for many applications and relies on advancement in the development of quantum random access memory (qRAM).
\item The algorithm must apply $e^{-iAt}$ unitary transformations for many values of $t$. Depending on the computational cost of this operation, the gains from the algorithm could be negated entirely.
\item Strong restrictions must hold on the invertibility of $A$, since the run time for the algorithm grows linearly in the condition number, $\kappa$.
\item The output of the algorithm is $\ket{x}$, rather than $x$. Thus, we can either apply an operator, $M$, and output $x^{\dagger} M x$ or repeatedly solve the system to recover a limited amount of information about $x$. Recovering, $x_j$, for instance, would require $N$ runs.
\end{enumerate}

In many cases, at least one of the four limitations may render the \cite{HHL09} algorithm unsuitable for a particular application. In the following subsections, we will discuss related families of algorithms that work around the limitations of \cite{HHL09} and the peculiarities of quantum computing more generally.

\paragraph{Linear Regression}

\cite{WBL12} proposed an early modification of \cite{HHL09} for the purpose of performing linear regression. They consider the least squares solution to a linear system of the form $x = (A^{\dagger}A)^{-1} A^{\dagger} b$ for the general case where $x \in \mathbb{C}$ and propose using the Moore-Penrose pseudo-inverse, $A^{+}$, which reduces the form of the solution to the following: $x = A^{+} b$. The paper proves that this solution is optimal for the least squares problem and demonstrates how to achieve a quantum speedup. In particular, they show that the query complexity of estimating the model fit -- which may be the most useful component of the algorithm for econometricians -- 
grows only logarithmically in the number of variables, $N$.

The algorithm they propose consists of three steps. First, they obtain the pseudo-inverse of $A$ using a quantum algorithm. Second, they compute a bounded estimate of the quality of the least squares fit. And finally, they estimate $x$. The details of the algorithm's implementation are beyond the scope of this paper and are omitted from our summary. Interested readers should see \cite{WBL12}.

Note that \cite{WBL12} demonstrates the value of an application that works within the limitations of \cite{HHL09}. Namely, it allows us to use the speed of a modified version of \cite{HHL09} to obtain a bounded measure of fit without estimating the parameter values themselves, which is considerably more costly. This would permit an econometrician to compare hundreds of competing models without committing to the costly parameter estimation step until a suitable model is selected. This could be particularly useful for empirical problems that involve the use of large microdata sets with many variables and observations.

Since \cite{WBL12}, several other applications have demonstrated how tasks involving linear regression can be performed within the limitations of \cite{HHL09}. \cite{ZFF15}, for instance, constructs an algorithm for generating conditional mean predictions and variance estimates from Gaussian process regressions. This circumvents the issue with reading out the superposition, $\ket{x}$, by instead sampling the solution. In many cases, \cite{ZFF15} can provide an exponential speedup over equivalent classical algorithms. Similarly, \cite{SSP16} examine how \cite{HHL09} can be used to achieve an exponential speedup for generating predictions, but weaken the dependence of the speedup on the condition number. Finally, \cite{KP17} constructs a quantum gradient descent algorithm for weighted least squares (WLS) that achieves an exponential speedup over classical algorithms.

With respect to implementation, \cite{DSD+18} demonstrate how to construct a 7-qubit circuit that implements a 3-variable ordinary least squares regression. They explain that any linear regression problem is convertible to a Quantum Linear Systems Problem (QLSP) of the form discussed in \cite{HHL09}. For OLS, for instance, we have the following optimality condition:

\begin{equation}\label{quantum_ols_1}
X'X \hat{\beta} = X'y
\end{equation}

Rather than premultiplying both sides by $(X'X)^{-1}$ to get an analytical expression for $\hat{\beta}$, we instead note that Equation (\ref{quantum_ols_1}) satisfies the conditions for a QLSP, where $A = X'X$, $x = \hat{\beta}$, and $b = X'y$. For econometric problems, $X'X$ will contain real-valued elements and will be symmetric. Thus, it will also satisfy the requirement to be Hermitian.

\cite{DSD+18} then describe how to construct a quantum circuit for a three-variable regression problem, where the data is defined as follows:

\begin{equation}\label{quantum_ols_2}
X'X = \frac{1}{4} \begin{bmatrix} 
15 & 9 & 5 & -3 \\
9 & 15 & 3 & -5 \\
5 & 3 & 15 & -9 \\
-3 & -5 & -9 & 15 \\
\end{bmatrix}
\end{equation}

\begin{equation}\label{quantum_ols_3}
X'y = \frac{1}{2} \begin{bmatrix} 
1 \\
1 \\
1 \\
1 \\
\end{bmatrix}
\end{equation}

The purpose of this choice of $X$ and $y$ was to achieve the following: (1) ensure that $X'X$ was Hermitian with four distinct eigenvalues ($\lambda_{1} = 1$, $\lambda_{2} = 2$, $\lambda_{3} = 4$, and $\lambda_{4} = 8$); and (2) ensure that $X'y$ can be be prepared quickly, as is required by \cite{HHL09} for an exponential speedup. In this case, $X'y$ can be prepared as a quantum state by applying two Hadamard gates. The circuit \cite{DSD+18} propose is described in Algorithm \ref{alg:DSD+18}.

\begin{algorithm}
\DontPrintSemicolon
Initialize one ancilla qubit, a two-qubit input register, and a four-qubit clock register.
Encode $X'y$ in the input register as a quantum state: $\ket{X'y} = \frac{1}{2} \ket{00} + \frac{1}{2} \ket{01} + \frac{1}{2} \ket{10} + \frac{1}{2} \ket{11}$. \;
Perform quantum phase estimation on a four-qubit clock register. This entails the application of Hadamard gates, followed by controlled-unitaries, followed again by an inverse quantum Fourier transform.\;
Phase shift the ancilla qubit, $\ket{s}$, based on the clock register state. \;
Perform inverse phase estimation. \;
Conditional on obtaining $\ket{1}$ in the ancilla qubit, the final state, $\ket{x}$, will be as follows after normalization:
$\frac{1}{\sqrt{340}} (-\ket{00} + 7 \ket{01} + 11 \ket{10} + 13 \ket{11})$. \;
This is proportional to the solution: $\frac{1}{32} \begin{bmatrix} -1 & 7 & 11 & 13 \end{bmatrix}$.
\caption{Quantum Ordinary Least Squares \citep{DSD+18}}\label{alg:DSD+18}
\end{algorithm}

\paragraph{Matrix Inversion}

Beyond demonstrating how to sample a linear system's solution exponentially faster, \cite{HHL09} also inspired work on several other closely-related problems. Both \cite{Ta13} and \cite{FL16} build on \cite{HHL09} to construct matrix inversion algorithms that require less space. \cite{Ta13} reduces the amount of space needed from $O(log^{2}(N))$ bits with a classical algorithm to $O(log(N))$ for a quantum algorithm. 
The current state-of-the-art, \cite{FL16}, provides an efficiency improvement over \cite{Ta13} with respect to the space needed to perform inversion and also eliminates the need to perform intermediate measurements.

\paragraph{Finite Element Methods} 

\cite{FR07} show that finite element methods perform well as a solution method for macroeconomic models. In particular, they are stable over a large range of risk aversion parameter values and shock variances.\footnote{See \cite{Hug00} for an introduction to linear finite element analysis.} Quantum algorithms used to implement finite element methods can achieve a polynomial speedup over their classical counterparts \citep{MP16}. Similar to earlier work, \citet{MP16} also relies on the ability to solve large systems of linear equations and is based on \cite{HHL09}.

\paragraph{Computational Finance}

Recent work has proposed algorithms for solving common problems in finance on a quantum computer. Much of this work makes use of specialized devices called quantum annealers, which we cover briefly in Section \ref{sec:Quantum Annealing}. Notably, one recent paper \citep{RL18} introduces an algorithm for solving portfolio optimization problems using gate-and-circuit (universal) quantum computers. It makes use of \cite{HHL09} and achieves a run time of poly(log(N)), where $N$ is the number of assets. Existing classical algorithms require poly(N) time. For an overview of selected methods in computational finance from the teams at IBM Quantum and QC Ware, see \citet{EGM+20} and \citet{BDJ+20}.

\subsubsection{Machine Learning}

The machine learning literature has made an attempt to integrate quantum algorithms into their existing toolkit. Many such algorithms involve the solution of linear systems and build on \cite{HHL09}. \cite{RML14}, for instance, makes use of quantum algorithms to achieve an exponential speedup for the Support Vector Machine (SVM) classifier. \cite{LMR14}, which we will examine further in Section \ref{sec:PCA}, shows how to achieve an exponential speedup in principal component analysis (PCA). \cite{BWP+18} provide a broad overview of the current state of quantum machine learning. Finally, both \citet{EGM+20} and \citet{BDJ+20} review quantum machine learning applications that could be applied to problems in finance.

\subsubsection{Principal Component Analysis}\label{sec:PCA}

PCA is used for a wide variety of applications in economics and finance: (1) performing high dimensional vector autoregressions in macroeconomics \citep{BBE05}; (2) constructing diffusion indices as part of a forecasting exercise \citep{SW02}; (3) measuring connectedness and systemic risk within the financial system \citep{BMAL12}; (4) modeling the determinants of credit spreads \citep{Col01}; (5) reducing model dimensionality (see, e.g., \citealp{Bai03}; and \citealp{BN06}); and (6) pricing financial derivatives \citep{HJM90}.

The \cite{LMR14} algorithm provides an exponential improvement over any classical PCA algorithm. It first requires the preparation of multiple copies of a quantum state in the form of a density matrix, $\rho$, which will represent the underlying classical dataset on which PCA will be performed.\footnote{A density matrix is an alternative way to express a quantum state. It is often used when there is classical uncertainty about the true underlying state. In such cases, we express the density matrix as a mixture of pure states, $\rho = \sum_{i} p_{i} \ket{\phi_{j}} \bra{\phi_{j}}$. Note that the density matrix for each pure state is given by the outer product of its state in ket or vector form.} As the authors show, it is possible to perform density matrix exponentiation on non-sparse matrices in $O(log(d))$ time, where $d$ is the dimension of the Hilbert space. This is already an exponential improvement over the equivalent classical algorithms for non-sparse matrix exponentiation. The paper then builds on the first result to improve quantum state tomography, which is the process by which unknown quantum states are uncovered. Finally, the PCA routine uses the improved tomographic algorithm to extract information about the density matrix--namely, the eigenvectors associated with its largest eigenvalues. These are the principal components and the algorithm is able to recover them in $O(log(d))$ time, which is an exponential speedup over existing classical PCA algorithms.

The \cite{LMR14} algorithm for PCA has already been implemented in small-scale demonstrations. Additionally, recent work by \citet{MCR+19} has shown how it can be applied to financial derivatives pricing and implemented on a 5-qubit IBM quantum computer. There are, however, two caveats. First, it requires quantum state preparation of a density matrix, $\rho$, which may be non-trivial. And second, it works best when several principal components dominate.

\subsubsection{Statistical Distance}

Many problems in economics and finance involve the measurement of statistical distance. In finance, for instance, a large literature attempts to measure the underlying empirical distribution of securities returns.\footnote{See, e.g., \cite{Off72}, \cite{EE76}, \cite{RK76}, \cite{Hag78}, \cite{Cas79}, \cite{Kom84}, \cite{Pen87}, \cite{Sol90}, \cite{ABDE01}, \cite{CDG12}, and \cite{BTL13}.} Similarly, many heterogenous agent macro modelling exercises attempt to determine whether the steady state distribution of outcomes changes with the policy regime.\footnote{Incomplete markets models with many heterogeneous agents were introduced by \cite{Bew77}, \cite{Hug93}, and \cite{Aiy94}. \cite{KS98} provided a tractable solution method for incomplete markets models with aggregate uncertainty. \cite{KMV18} showed how monetary policy could be included in such models. A large and growing literature has made use such models to study the distributional impact of policy. See, e.g., \cite{HL96}, \cite{GP03}, \cite{CDR03}, \cite{KP06}, \cite{CS06}, \cite{CCNR07}, \cite{BPP08}, \cite{HSV10}, \cite{HKV11}, \cite{KV14}, and \cite{GL17}.} The former exercise requires the existence of a sufficient amount of returns data to identify the distance between the empirical and theoretical distributions. And the latter requires the simulation of a sufficient number of agents to perform a comparison of two different distributions.

Given the data requirements of such empirical problems and the computational cost of such theoretical exercises, it would be beneficial to reduce the number of distributional draws needed to perform tests of statistical distance. \cite{BHH11} provide a quantum algorithm that achieves this. In particular, they consider the case where we have two unknown distributions, $p$ and $q$, on a set with $N$ elements. They then consider how many draws are needed to determine the distance between $p$ and $q$ in the $L_{1}$ norm, $\norm{p-q}_{1}$, with constant precision, $\epsilon$. Classical algorithms require $\Omega(N^{1-o(1)})$ draws. They show that a quantum algorithm is can achieve the same precision with just $O(N^{1/2})$ draws.

There are, however, a few details worth discussing. First, the algorithm requires the selection of threshold parameters, $a$ and $b$, where $0 \leq a \leq b \leq 2$. Furthermore, the test takes the form a ``promise problem,'' which decides whether $\norm{p-q}_{1} \leq a$ or $\norm{p-q}_{1} \geq b$. For cases where the promise does not apply, $a < \norm{p-q}_{1} < b$, the algorithm may return any decision or may fail to converge on a decision. Beyond this, \cite{BHH11} also provide tests of uniformity and orthogonality that have query complexities of $O(N^{1/3})$, which is a polynomial speedup over their classical counterparts, which have query complexities of $\Omega(N^{1/2})$.

Finally, \cite{Mon15} also introduces a quantum algorithm that can be used to reduce the query complexity of statistical distance comparisons. The proposed algorithm considers the total variation distance, rather than the $L_{1}$ distance. We refer interested readers to \cite{Mon15} and \cite{BHH11} for the details of the two algorithms.

\subsubsection{Monte Carlo Simulations}

Monte Carlo simulations are often employed when it is not possible to derive a closed-form solution for a statistical object of interest. In economic modeling, Monte Carlo methods are used for a variety of applications, including the simulation of the steady state distribution of wealth in incomplete markets models, the simulation of agent choices over time, and numerical integration.\footnote{See, e.g., \cite{KD78} and \cite{G89}.} In econometrics, Monte Carlo methods are used to perform Markov Chain Monte Carlo (MCMC), which is a computationally-expensive subroutine of estimation algorithms.\footnote{The MCMC algorithm is widely used in estimation problems in economics and finance. See, e.g., \cite{AC93}, \cite{C93}, \cite{Ruu91}, and \cite{CNS02}.} They are also used to evaluate the finite sample properties of estimators and to construct test statistics.\footnote{See \cite{Mac91}, \cite{DM93}, and \cite{McD98} for a discussion of how Monte Carlo methods can be used to compute critical values for unit root and cointegration tests. See \cite{Hen84} for a broad overview of Monte Carlo methods in econometrics.}

\cite{Mon15} provides several quantum algorithms that achieve a speedup in Monte Carlo expected value estimation. We will focus on the simplest algorithm, where we wish to estimate the mean, $\mu$, of some stochastic process, $\nu$, that is bounded between 0 and 1. \cite{Mon15} shows how we can construct $\tilde{\mu}$, where $\abs{\tilde{\mu}-\mu} < \epsilon$ with a 0.99 probability. Algorithm \ref{alg:Mon15} provides a rough sketch of the procedure. 

Note that Algorithm \ref{alg:Mon15} uses the following definitions. First, $W$, is a unitary operator on $k+1$ qubits defined as $\ket{x} \ket{0} \rightarrow \ket{x} (\sqrt{1-\phi(x)} \ket{0} + \sqrt{\phi(x)}\ket{1})$. Second, $\phi(x)$ is a function that maps $\{0,1\}^{k}$ to $\mathbb{R}$. Finally, $P$ is a projector, and $U$ and $V$ are unitary transformations, where $U = 2 \ket{\psi} \bra{\psi} - I$, and $V = I - 2P$.

This algorithm has a complexity of $O(1/\epsilon)$, which is a quadratic improvement over classical algorithms that estimate the mean. Beyond the aforementioned algorithm, \cite{Mon15} also introduces an algorithm for mean estimation when $\nu$ is non-negative, bounded in $L_{2}$, but not necessarily in the $[0,1]$ interval. Finally, the paper also considers the more general case where $v$ is bounded in variance only.

\begin{algorithm}
\DontPrintSemicolon
Define an algorithm, $A$, which consists of variable, $v$, and a parameter estimate constructed from realizations of that variable, $\mu$. \;
Initialize an input state of $\ket{0}^{\otimes n}$. \;
Perform $A \ket{0}^{\otimes n}$, generating the state $\ket{\psi'} = \sum_{x} \alpha_{x} \psi_{x} \ket{x}$. \;
Attach an ancilla qubit and apply the unitary transformation, $W$, yielding state $\ket{\psi}$:
$\ket{\psi} = (I \otimes W) (A \otimes I) \ket{0}^{\otimes n} \ket{0} = \sum_{x} \alpha_{x} \ket{\phi_{x}} \ket{x}  (\sqrt{1-\phi(x)} \ket{0} + \sqrt{\phi(x)}\ket{1})$ \;
Using unitary transformations, $U$ and $V$, perform amplitude estimation, yielding $\tilde{\mu}$. \;
Repeat this process $\delta$ times and retain the median result.
\caption{Quantum Monte Carlo Mean Estimation \citep{Mon15}}\label{alg:Mon15}
\end{algorithm}

It is perhaps also worthwhile to briefly discuss the amplitude estimation algorithm, which \cite{Mon15} relies on and which \cite{BHMT02} introduce. This algorithm takes as inputs a quantum state, $\ket{\phi}$, two unitary transformations, $U = 2 \ket{\psi} \bra{\psi} I$ and $V = I - 2P$, and an integer, $t$. It then returns an estimate of the expectation value of $\mu$, $\bra{\phi} P \ket{\phi}$, where the condition in Equation (\ref{eqn:amplitude_estimation}) holds with a probability of at least $8/\pi^{2}$. Furthermore, the probability can be increased to $1-\delta$ for an arbitrary $\delta$ by repeating the process $O(log(1/\delta))$ times and selecting the median outcome, as demonstrated in \cite{JVV86}.

\begin{equation}\label{eqn:amplitude_estimation}
\abs{\tilde{\mu} - \mu} \leq 2 \pi \frac{\sqrt{a(1-a)}}{t} + \frac{\pi^{2}}{t^{2}}
\end{equation}

\subsubsection{Matrix Powers}

\begin{figure}
\begin{center}
\includegraphics[width=0.50\textwidth, bb=0 0 300 225]{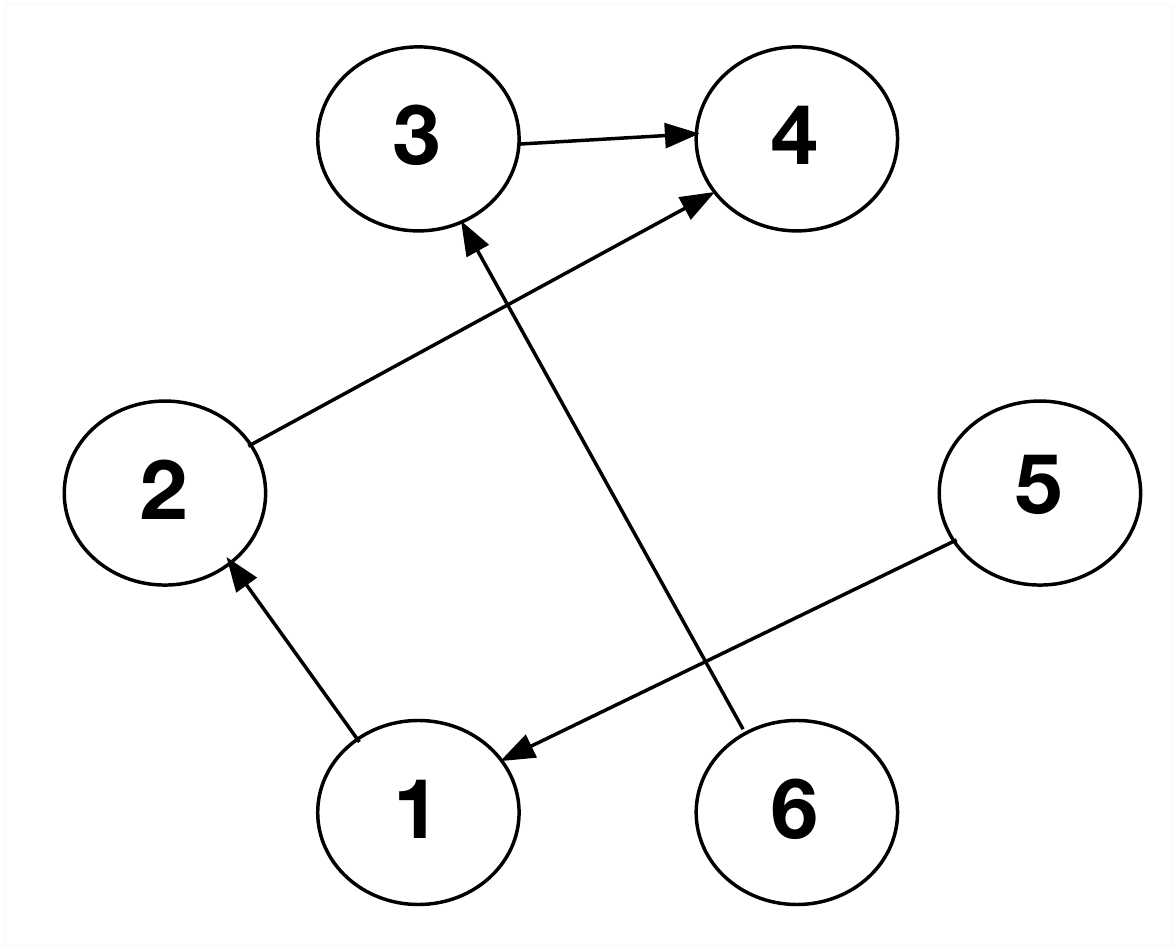}
\caption{The figure above shows a graph that models exposures between financial firms. An arrow pointing from firm $i$ to firm $j$ indicates that $j$ is exposed to firm $i$. Thus, if a financial shock hits firm $i$, it will propagate to firm $j$ after one period.}
\label{fig:adjacency_matrix}
\end{center}
\end{figure}

Many problems that involve networks or interconnectedness can be formulated as adjacency matrix problems.\footnote{For work that makes use of adjacency matrices in finance and economics, see \cite{BM74}, \cite{BCZ06}, \cite{BG11}, \cite{MGS12}, \cite{AGM12}, \cite{DY14}, \cite{LB14}, \cite{GKWZ15}, \cite{AP15}, \cite{GSX+15}, and \cite{GS16}.} Adjacency matrices are typically sparse and may be used to model transitions within a graph. Figure \ref{fig:adjacency_matrix} shows a graph that describes exposures between financial firms in a stylized theoretical model. An arrow pointing from firm $i$ to firm $j$ indicates that $j$ is exposed to $i$. If a shock hits firm $i$, then the shock will propagate to firm $j$ in the following period.

We can express these relationships using the adjacency matrix, $A$, shown in Equation (\ref{adjacency_matrix}). Note that a $1$ in row $j$ of column $i$ indicates that a shock to $i$ will propagate to $j$ in the following period. Furthermore, we may model a shock to firm $i$ by post-multiplying $A$ by $s_{i}$, where $s_{i}$ is a column vector of zeros with a one in the $i$th row. If we wanted to determine the state of the financial system $m$ periods after the shock, we could compute this as $A^{m} s_{i}$.

\begin{equation}\label{adjacency_matrix}
A = \begin{bmatrix} 
0 & 0 & 0 & 0 & 1 & 0 \\
1 & 0 & 0 & 0 & 0 & 0 \\
0 & 0 & 0 & 0 & 0 & 1 \\
0 & 1 & 1 & 1 & 0 & 0 \\
0 & 0 & 0 & 0 & 0 & 0 \\
0 & 0 & 0 & 0 & 0 & 0 \\
\end{bmatrix}
\end{equation}

While the problem we considered was intentionally stylized, computing $A^{m}$ for a large $A$ and $m$ can be computationally costly. \cite{JW06} demonstrate a superpolynomial speedup for problems of this form using a quantum algorithm. There are, however, several limitations of the algorithm:

\begin{enumerate}
\item $A$ must be a symmetric matrix.
\item $A$ must have fewer than polylog(N) non-zero entries per row, where $N$ is the number of columns.
\item We must know the function, $f$, which efficiently maps each row number to the row's non-zero entry values.
\item The number of exponentiations, $m$, must be polylogarithmic in $N$.
\item It is only possible to examine diagonal elements, $(A^{m})_{jj}$.
\end{enumerate}

With respect to the fifth limitation, the algorithm only allows us to recover bounds on diagonal elements by testing whether $(A^{m})_{jj} \geq g +  \epsilon b^{m}$ or $(A^{m})_{jj} \leq g - \epsilon b^{m}$, where $g \in [-b^{m}, b^{m}]$, and $\epsilon = 1 / polylog(N)$. Interested readers should see \cite{JW06} for the details of the algorithm's implementation.

The authors stress that the algorithm is best suited to problems that can be reformulated in terms of a large, sparse matrix. This includes adjacency matrix problems of the style we considered earlier in this subsection, as long as they satisfy the aforementioned criteria. Furthermore, problems that involve large, sparse Markov transition matrices or fixed point algorithms may be able to exploit this routine. The most important limitation is that the research question will need to be answerable by testing bounds on a particular matrix diagonal element.

\subsubsection{Quantum Annealing}
\label{sec:Quantum Annealing}

This paper focuses primarily on ``universal'' quantum computers, which use the gate-circuit model of computation. In this section, however, we will discuss a specialized device called a ``quantum annealer'' that enables scaling with fewer technical difficulties. Whereas constructing a 50-qubit universal quantum computer is challenging with existing technology, commercially-available quantum annealers routinely make use of thousands of qubits. The most recently developed annealer (``Advantage") offered by D-Wave Systems employs 5000 qubits.

In contrast to universal quantum computers, which can perform any computation, quantum annealers are exclusively capable of solving combinatorial optimization problems. This does, however, cover a large number of interesting and computationally-difficult problems in economics and finance. Indeed, recent work has shown that quantum annealers can solve small-scale versions of problems in finance, including the prediction of financial crises through the use of network models \citep{OML19b, DLM+19}. There has also been work in the physics literature that speculates on how quantum annealers could be used more generally to solve problems in finance \citep{OML19a}.

Quantum annealers use a process that is similar to adiabatic quantum computing, which converts combinational optimization problems into quadratic unconstrained binary optimization (QUBO) problems of the form given in Equation (\ref{eqn:QUBO}).

\begin{equation}\label{eqn:QUBO}
H_{0} = \sum_{ij} Q_{ij} x_{i} x_{j} + c_i x_i
\end{equation}

\noindent Note that $Q_{ij}$ and $c_{i}$ are given, and $x_{i}, x_{j} \in \{0,1\}$. $H$ is the Hamiltonian, which expresses the level of energy in the system, the lowest of which is called the ``ground state.''

Adiabatic quantum computing works by embedding the weights, $Q_{ij}$ and $c_{j}$, in a quantum system and then finding the ground state, which corresponds to the global minimum. The system is first initialized in the ground state for an arbitrary and trivial Hamiltonian, $H_{1}$. The parameters of $H_{1}$ are then slowly changed until they become the parameters of $H_{0}$ -- namely, $Q_{ij}$ and $c_{j}$ -- as expressed in Equation (\ref{eqn:time-dependent Hamiltonian}).

\begin{equation}\label{eqn:time-dependent Hamiltonian}
H(t) = A(t)H_{0} + B(t)H_{1}
\end{equation}

According to the quantum adiabatic theorem, the Hamiltonian will remain in the ground state as long as the transition happens sufficiently slowly. Furthermore, theory provides us with a ``speed limit,'' below which the system will remain in the ground state, allowing us to find a global minimum. Importantly, however, the speed limit will not always have an analytical expression and may require exponential time.

While the quantum adiabatic theorem is instructive for understanding a process that is analogous to quantum annealing, commercially-available quantum annealers do not perform adiabatic quantum computing. In particular, they do not always remain in the ground state during the optimization process. It is also not generally believed that they are capable of solving NP-complete problems efficiently.

Although questions remain about the extent to which quantum annealers can provide a quantum advantage, it is clear that they can be applied to a large class of problems in economics and finance. In particular, any optimization problem that can be converted into a QUBO can also be run on a quantum annealer. For an extended overview of the quantum annealing literature, see \citet{HKL+20}.

\subsubsection{Random Number Generation}

Random numbers are frequently used in economics in simulation exercises and estimation routines. For most research applications, it is only important that the numbers generated satisfy statistical test requirements and can be reproduced. For this reason, research applications in economics and finance typically employ random number generators that come with common statistical packages. These are typically pseudo-random number generators (PRNGs) that do not use physical or ``true'' sources of randomness.

\begin{figure}
\begin{center}
\includegraphics[width=0.80\textwidth,bb=0 0 400 200]{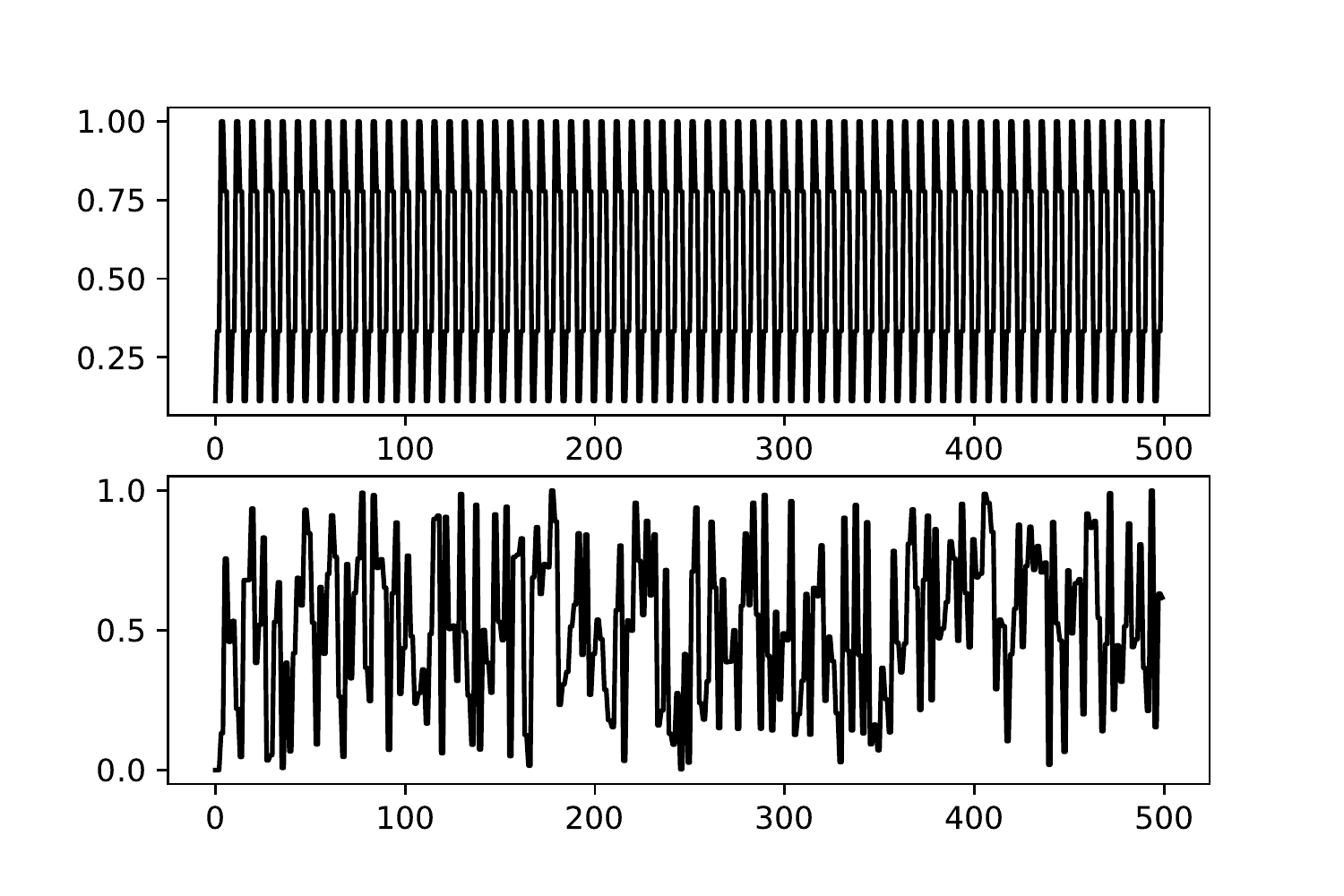}
\caption{The figure on the top shows a linear congruential generator with suboptimally chosen parameter values. The sequence, which spans 500 draws, contains a clear high-frequency period. In contrast, the figure on the bottom uses parameters that maximize the period ($a = 7^{5}$, $c = 0$, $m = 2^{31} - 1$).}
\label{fig:LCG}
\end{center}
\end{figure}

A commonly-used type of PRNG is the family of linear congruential generators introduced by \cite{Leh51}. As described in \cite{HG17}, these use the recursive formula shown in Equation (\ref{LCG}) to produce numbers that appear to be drawn from a uniform distribution. Such sequences can then be used to construct seemingly random draws from other commonly-used statistical distributions \citep{HLD04}.

\begin{equation}\label{LCG}
X_{n+1} = (a X_{n} + c) \mbox{ mod m}
\end{equation}
Note that $m>0$, $0 \leq a < m$, and $0 \leq c < m$. Selecting $a$, $m$, and $c$ will determine the period of the generator and the quality of the randomness. Furthermore, $mod$ is the modulo operator, which yields the remainder of $(a X_{n} + c) / m$. Figure \ref{fig:LCG} compares a sequence of numbers generated randomly using standard parameters for a linear congruential generator with parameters selected sub-optimally for the algorithm. In addition to containing a visible, high-frequency period, the sequence with sub-optimally chosen parameter values has a sample mean that deviates from the uniform distribution's mean by 10\%.

While such algorithms were once commonly used in statistical packages, they have recently been phased out in favor of the Mersenne Twister (MT) algorithm, introduced by \cite{MN98}. For most applications in economics and finance, the linear congruential or MT algorithms are sufficient. Because linear congruential generators have been in widespread use for decades, there is a large literature documenting their failures. \cite{HG17} provide a list of problems documented in this literature, including their tendency to cluster and become autocorrelated if a bad seed is selected.

Beyond pseudo-random number generators, there are also ``physical'' random number generators, which do not rely on algorithmic generation. These exploit random variation in the physical environment to generate a sequence of numbers or to select a seed for a PRNG. This might involve the use of random variation generated by a computer's internal processes or user's decisions, such as heat, noise, and mouse movements. Alternatively, it may instead rely on the use of random variation generated externally in the physical environment. One disadvantage of physical random number generation relative to PRNGs is that they cannot be reproduced from a seed and an algorithm. Instead, we must retain the set of numbers generated to achieve reproducibility. Due to the slow number generation rates for physical random number generation and the convenience and speed of PRNGs, physical random number generation has not gained widespread use for research applications within economics.

The approaches described above for physical RNG rely on classical physical processes, which are deterministic in nature. This is essentially the reason why ``classical'' physical RNG is so difficult: if nature is deterministic, where would the randomness come from? In contrast, quantum physics is probabilistic and, therefore, provides scope for generating randomness. We saw this in Figure \ref{fig:rng_circuit} in Section \ref{Preliminaries}, where we created a superposition and then performed measurement in the computational basis. This exact procedure can, in fact, be used to perform quantum random number generation (QRNG). Faster methods for QRNG involve the use of vacuum fluctuations, phase noise, and amplified spontaneous emission. We refer the interested reader to \cite{HG17} for further detail.

\begin{figure}[htp]
    \centering
    \includegraphics[width=0.75\textwidth,bb=0 0 2649 1492]{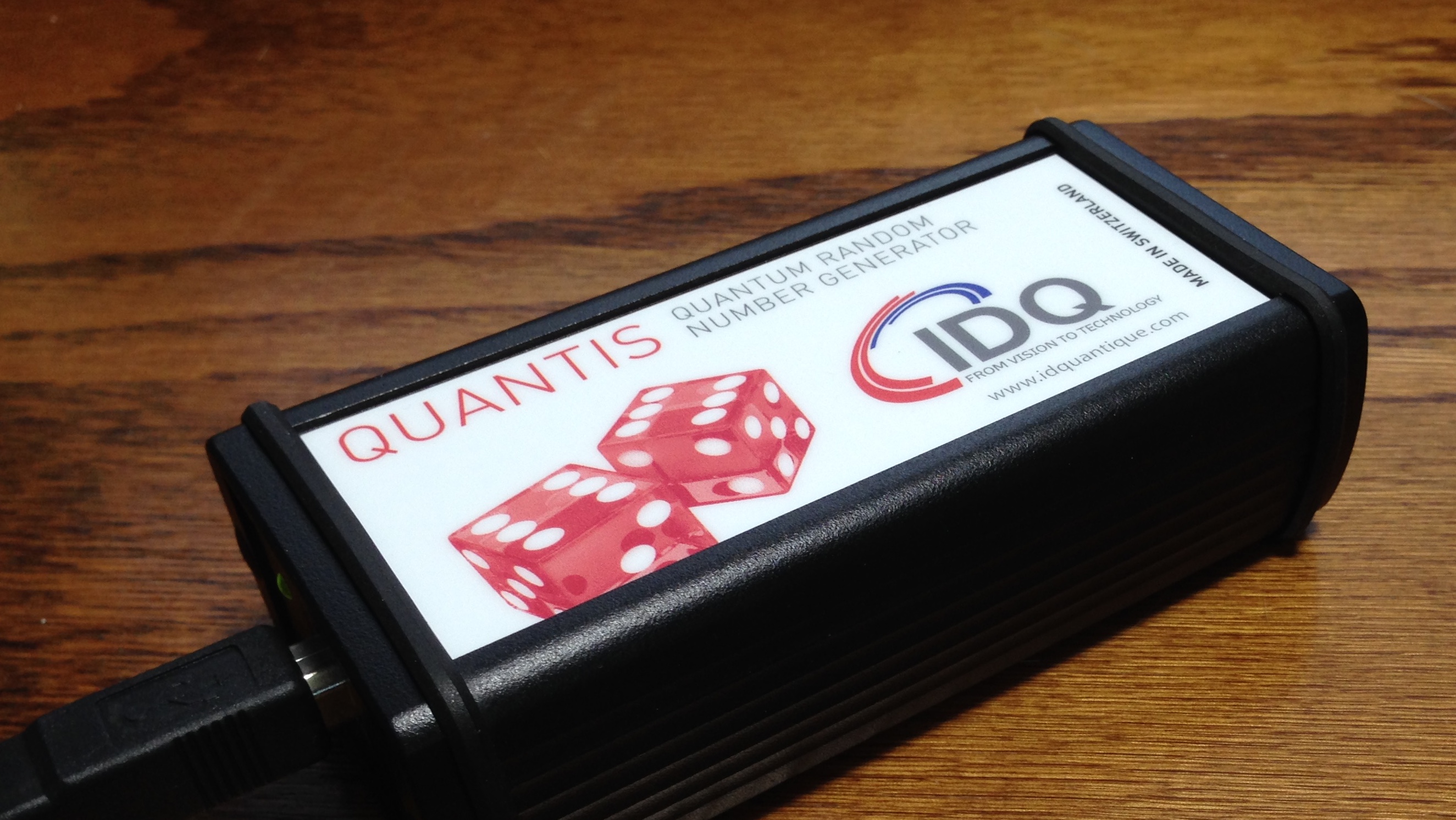}
    \caption{A commercially available QRNG device, produced by ID Quantique. Photo by John Sloan, License: CC BY-NC-SA 2.0.}
    \label{fig:qrng}
\end{figure}

QRNGs are one of the most mature quantum technologies. Beyond implementation within a quantum computing environment, they may also be generated by standalone, specialized quantum devices that are already commercially-available. Devices produced by ID Quantique, for instance, are small, easy-to-use, and affordable.\footnote{In November of 2020, ID Quantique sold QRNG devices for roughly \$1000.} They have also recently been integrated into a smartphone that makes use true random number generation to provide enhanced security features.\footnote{See \href{https://www.idquantique.com/id-quantique-integrates-its-quantum-chip-in-vsmart-aris-5g-smartphone}{https://www.idquantique.com}.} See Figure~\ref{fig:qrng} for an example of a standalone quantum random number generator.

Relative to physical random number generators, QRNGs tend to perform generation faster, making them a more viable competitor for PRNG schemes. While MT-generated random numbers may be sufficient for most research problems in economics, QRNGs are likely to provide a compelling alternative that does not suffer from problems common to PRNGs.

An even stronger notion, called certified randomness, was recently introduced: one can generate randomness, starting from a short random seed, without trusting the quantum devices which generate the randomness. For an overview of the related literature, see \citet{CR12}, \citet{VV12}, \citet{MS16}, \citet{CY14}, and \citet{HG17}.

\subsection{Experimental Implementation}
\label{Sec:Experimental Implementation}

While there has been considerable progress in the development of quantum algorithms since the 1990s, much of the experimental progress has been realized more recently. In this subsection, we discuss the historical development of quantum computers, including the current state of progress. We also discuss the limits of what we can expect to achieve with quantum computation.

\subsubsection{Experimental Progress in Quantum Computing}
\label{Sub:Quantum Supremacy}

In 1982 Richard Feynman remarked that the world is not classical \citep{Fey82},\footnote{``Nature isn't classical, dammit, and if you want to make a simulation of nature, you'd better make it quantum mechanical, and by golly it's a wonderful problem, because it doesn't look so easy."} and  that quantum devices should be used for calculating properties of quantum systems at the microscopic level  \citep{Fey82,Fey86}. This triggered the beginning of an algorithmic \citep{BBC+95,Sho94,Gro96} and hardware-related \citep{MMK+95,KMW02,MSS01,NT99} development that strongly influenced computational science and promised to revolutionize some lines of computation. 

During the 1990s, ground-breaking work toward quantum computing with ion traps \citep{MMK+95} and molecular spins \citep{VSB+01} got a head start because of the previous development of atomic clocks and magnetic resonance imaging. From 2000, superconducting circuits \citep{NT99,MSS01} emerged as potential contenders for  quantum computing through circuit quantum electrodynamics (cQED)~\citep{WSB+04} and Josephson-junction based qubits with long coherence times~\citep{KYG+07}. There followed many advanced experiments and proofs-of-concept \citep{Win13,BR12,AWB+09,DCG+09} and a 2012 Nobel Prize in Physics. Nevertheless, only ten years ago a common opinion was that quantum computing was merely a distant possibility. 

The basic reason for this pessimistic perspective was the need for quantum error correction (QEC), putting extreme demands on quantum hardware. However, in 2013 John Martinis's group published a seminal paper \citep{BKM+14} demonstrating basic operations with a five-qubit processor with error rates low enough for future successful fault-tolerant operation of the QEC surface code \citep{FMM+12}. That work was followed by advanced operations with a nine-qubit chip \citep{BSL+16} that ultimately became a fundamental building block in Google's 54-qubit Sycamore chip \citep{AAB+19}.

There is now a broad effort to scale up devices and build large systems that should be able to challenge classical computers \citep{Wen17}. These  scaled-up systems involve ion traps \citep{BKM16,WBD+19,Hon20}, superconducting circuits  \citep{AAB+19,IBM20,Rig20,Ali20,Bai20,ARL+19a,ARL+19b,OSQ20}, optical traps for cold atomic gases \citep{GB17,SGB+19}, and photonic circuits \citep{TF19}. Semiconductor qubits are still at a proof-of-concept level \citep{WPK+18}, but the compatibility with standard silicon computer technology has lead to strong efforts to scale up to large systems \citep{LPF+18,VE19}. The present efforts to scale up different kinds of quantum hardware are necessarily putting strong focus on computer architecture for optimizing performance with respect to qubit coherence, connectivity, and integration \citep{MJH19}.

So, what is the computational power and usefulness of these systems? Preskill introduced the concepts of quantum supremacy \citep{Pre12} and noisy intermediate-scale quantum (NISQ) devices \citep{Pre18}. Quantum supremacy (or quantum advantage) is a computational regime in which quantum computers can perform at least one task orders of magnitude faster than classical high-performance computers (HPC).\footnote{A ``classical'' computer is any computer that uses only classical operations, such as NAND, XOR, and OR gates in its computation. For example, our existing desktops are ``classical" computers. In general, we will use the term ``classical" to indicate that something is not quantum.} It was speculated that quantum supremacy would initially be achieved using NISQ devices -- the current vintage of quantum computers -- which have enough qubits to challenge the performance of classical computers on certain tasks, but not enough to also provide quantum error correction.

The AI-group at Google published a paper in 2016 on how to characterize quantum supremacy in near-term devices with sampling of quantum random circuits \citep{BIS+18}. Anticipating that quantum supremacy might be on the horizon, \citet{AC17} then formalised the set of conditions that would need to be met in order to technically achieve it. They also identified tasks where quantum supremacy would most likely be achievable first. The viability of the Google random-circuit approach was verified by \cite{BFN+19}, and shortly afterwards Google published a paper in which they claimed to have achieved quantum supremacy by performing a computation in 200 seconds that would have taken 10,000 years on the world's fastest classical supercomputer \citep{AAB+19}. This achievement was followed by a second demonstration of quantum supremacy by a group at the University of Science and Technology of China in Hefei \citep{ZWD+20}, which performed a boson sampling task $10^{14}$ times faster than would be possible on a state-of-the-art classical supercomputer.\footnote{In contrast to \citet{AAB+19}, \citet{ZWD+20} demonstrated quantum supremacy using a specialized quantum computing device that can only perform boson sampling.}


IBM has argued \citep{IBM_blog_Oct21_2019} that the computation Google performed on its new quantum chip, Sycamore, could actually be performed in 2.5 days, rather than 10,000 years on IBM's supercomputer, Summit, and that this classical performance could be further refined and sped up. The argument illuminates the floating division between classical performance and quantum supremacy \citep{PGN+17}, as well as the usefulness and accuracy of quantum calculations. Adding a few more functional qubits to Sycamore would ensure that it surpasses any near-future HPC on the problem of sampling random quantum circuits when it comes to memory storage. However, that would also require many more, and higher-fidelity, operations on the quantum processing unit (QPU) to demonstrate quantum supremacy. 
 
While the achievement of quantum supremacy marks an important milestone in the development of quantum computers, it does not imply that all tasks can now be more efficiently performed on quantum computers. Contrary to popular depictions, quantum computers are not simply classical computers with an expanded capacity for parallel computation. Rather, quantum computers allow us to compute with quantum physical resources, which require altogether different algorithms to perform the same tasks. Problems that can be directly mapped onto two-dimensional quantum spin systems -- quantum magnets -- can achieve a quantum speedup \citep{CMN+18}. However, there are, in fact, many computations that will almost certainly be more efficiently performed on classical computers in the foreseeable future. Furthermore, for many computational problems, there are no known quantum speedups and in some specific cases, it is possible to prove that no such speedups can be achieved. 

It is also worth emphasizing that quantum computers are unlikely to be used as standalone substitutes for classical computers. Rather, it is more likely that they will be employed as quantum processing unit (QPU) accelerators in conjunction with classical computing systems, playing a role that is similar to that of graphical processing units (GPUs) or tensor processing units (TPUs).\footnote{GPUs, which were originally developed to render graphics, have since been exploited to perform massively parallel computation of basic floating point operations. TPUs were developed to perform the computational function of a GPU, but without the capacity to render graphics. Had there not been substantial progress in the development of GPUs over the last decade, it is unlikely that machine learning would have experienced as much success as it has as a field. Similarly, it is possible that quantum computing could generate similar transformations by unlocking the solution and estimation of otherwise intractable models.}

Currently, theoretical efforts and software development are focused toward near-term ``use cases" -- useful quantum computations on current and near-term NISQ QPUs. It can, of course, be discussed what is ``useful": Are fundamental physics problems useful or does usefulness imply a ``real-life" problem? The issue with QPUs is that they are based on spin-like two-level or few-level systems (qubits, qutrits, qudits) and mapping problems onto qubit registers may be complicated. Spin systems map directly onto qubit registers, and many optimization problems are variational problems based on Ising type Hamiltonians and cost functions. A popular type of algorithm is the variational quantum approximate optimization algorithm (QAOA)~\citep{FGG14,WWJ+19,VGS+19,BVW+19}.

Variational quantum computing combines short bursts of quantum computation to execute quantum trial functions with classical pre- and post-processing of data (e.g. evaluating cost functions based on measurements for sampling of quantum trial functions). Currently, one applies the variational quantum eigensolver (VQE) algorithm \citep{PMS+14} to electron correlation problems in general, and quantum chemistry in particular  (see e.g.  \cite {SBO+19,NCP+19,TRJ+20,LRS+20}). The recipe for the quantum trial function is constructed classically by generating excitations from a molecular reference state, involving a large number of variational parameters to achieve chemical accuracy. Moreover, one must classically map fermionic operators to qubits via Jordan-Wigner-type transformations and construct the list of quantum gates describing the quantum circuit. The single, unique quantum step is the execution of the quantum gates that creates the quantum trial function in the multi-qubit register. The post-processing steps are all classical: measurements for sampling the trial state function, the calculation of expectations values, optimization, and iterative minimization. The rationale for the VQE is two-fold: (1) the quantum step is sufficiently hard in a classical setting so that the (classical) overhead will be unimportant in the end, and (2) the coherence time for the NISQ QPU is long enough for the quantum step to be executed. 

The VQE was invented  because the phase estimation algorithm \citep{CEM+98,DJS+07} requires QPU coherence times far beyond what is possible in the NISQ era. The phase estimation algorithm scales much better than the VQE because it is essentially quantum all the way: it describes the time evolution from a classically defined initial quantum reference state under the action of the quantum Hamiltonian. There are no repeated measurements, optimization, and minimization -- only final measurements and post-processing to find the energy of the molecular state. However, the quantum time evolution requires extremely long coherence times (i.e. a large number of quantum gate operations) in order to solve challenging problems and to establish quantum supremacy.

In practice, the VQE -- and perhaps most classical-quantum hybrid schemes -- also needs QPUs with extremely long coherence times for useful applications. Currently, the VQE can only be applied to fairly small molecules for proof-of-principle experimental demonstration of QC \citep{LRS+20}. On the other hand, it provides a useful platform for benchmarking NISQ quantum hardware and for developing software and user interfaces. However, for establishing quantum supremacy, quantum error corrections schemes will be needed.

\subsubsection{Limitations of Quantum Computing}
\label{Sub:Limitations}
Suppose we have access to some black-box function, $f:\{0,1\}^n\mapsto\{0,1\}$, and our objective is to find an $x$ such that $f(x)=1$. This could be viewed as finding a needle (an $x$ such that $f(x)=1$) in a haystack (of size $2^n$). Consider the following (na\"{i}ve) algorithm: we first prepare a uniform superposition over all possible $x$ values using the Hadamard gate: $H\otimes \cdot \otimes H \ket {0^n}=\frac{1}{\sqrt{2^n}} \sum{x \in \{0,1\}^n} \ket{x}$. We then add another qubit register, and evaluate $f$, so that the overall state is:
\begin{equation}
\frac{1}{\sqrt{2^n}} \sum_{x\in \{0,1\}^n} \ket{x}\otimes \ket{f(x)} = \frac{1}{\sqrt{|f^{-1}(0)|}}  \sum_{x\in f^{-1}(0)} \ket{x}\otimes \ket{0} +\frac{1}{\sqrt{|f^{-1}(1)|}}  \sum_{x\in f^{-1}(1)} \ket{x}\otimes \ket{1}    
\end{equation}

At this point, next to every $x$, we have an $f(x)$. This might give the impression that we have almost solved the problem. Unfortunately, if we measure the last qubit, the probability of it being $1$ is $|f^{-1}(1)|$. As such, we cannot expect to recover an $x$ for which $f(x)$ is $1$, even though there exists such an $x$ in the superposition.  If there is exactly one needle in the haystack, we will to run this algorithm $O(2^n)$ times to find a needle. \citet{Gro96} came up with a quantum algorithm that solves this using $O(\sqrt{2^n})$ queries to $f$, which is \textit{quadratically} faster than the best classical algorithm, and the one we presented before.

We might wonder whether there is a better alternative that could provide more than a quadratic speedup. Unfortunately, \citet{BBBV97} proved that Grover's algorithm is optimal in the black-box model. Thus, quantum computers cannot solve the search problem exponentially faster than a classical computer. But what about other types of black-box problems? Could these be solved faster on a quantum computer? Unfortunately, this is also not the case: whatever black-box problem a quantum computer solves using $T$ queries, a classical computer could also solve using $O(T^4)$ queries~\citep{ABKT20,BBCMW01}. Therefore, in order to get a super-polynomial speedup, we have to exploit some structure. This additional structure that the black-box satisfies is called a \textit{promise}, and with a promise, we know that super-polynomial speedups can be achieved (see, e.g., \citet{Sim97} and \citet{BV97}). 

The discussion above centers on the black-box model; however, analysis in the standard model, without the black-box assumption, is harder, and there are more open problems and conjectures than answers. Consider, for example, the question of whether either classical or quantum computers can efficiently solve all $\mathsf{NP}$ problems -- that is, all problems for which there is an efficient algorithm to verify a candidate proof. For example, a Sudoku puzzle has a valid solution, which can be efficiently verified given a candidate solution.\footnote{Formally, this could be done in polynomial time in the size of the puzzle.} It is  conjectured that classical computers cannot efficiently solve all the problems in the $\mathsf{NP}$ class. This is known as the $\mathsf{P} \neq \mathsf{NP}$ conjecture and is one of the Millennium Prize Problems~\citep{CJW06}. It is also conjectured, partly based on the arguments that we discussed for the black-box model, that quantum computers also cannot efficiently solve all $\mathsf{NP}$ problems. 

While substantial progress has been made in the development of quantum computers, it is worth emphasizing that quantum computing will not provide an exponential speedup for all algorithms, even after the production of quantum computers matures. Contrary to popular depictions, quantum superpositions do not allow for massive parallel computation in a trivial sense. Rather, quantum speedups typically rely on the subtle exploitation of quantum physics, rather than the brute force application of increased computational resources. This is why quantum speedups necessarily entail the development of quantum algorithms.

%% file: Conclusion.tex
\section{Conclusion}
\label{Conclusion}

With Google claiming quantum supremacy \citep{AAB+19}, we have arguably crossed the threshold from one computational regime into another. Quantum computers went from a theoretical curiosity to the best available computational tool for at least one computational task. And while this change may seem sudden and unexpected, it followed a decade of steady progress in the development of quantum computers and was anticipated by the literature \citep{AC17, Wen17}. Quantum supremacy does not mean that all tasks will be performed more efficiently on quantum computers. Rather, it suggests that quantum computers have become sufficiently powerful that we can expect that a larger variety of computational tasks will eventually be most efficiently performed on them.

Our objective in this manuscript was to explain the implications of improvements in quantum technology, including quantum computers, for economists with the intention of drawing new voices into the existing dialogue with physicists, mathematicians, and computer scientists. As part of this effort, we have included both high-level explanations of concepts and also sufficient low-level detail to enable economists to construct literatures on quantum econometrics and quantum computational economics. We also attempted to clear up common misconceptions about quantum computers: No, quantum computers are not classical computers with an expanded capacity for parallelization. And no, quantum computers will not perform all computations exponentially faster than classical computers, but in certain cases, they will.

One implication of the development of powerful quantum computers is that the financial system may eventually come under quantum attack. Such attacks are categorically different from what can be done with classical computers, since sufficiently large quantum computers will be able to perform certain computationally tasks exponentially faster. \citet{Sho94}, for instance, provides a near-exponential speedup to prime factorization, which compromises the RSA encryption algorithm. One solution to this problem is to employ post-quantum cryptography, which makes use of classical computers and classical algorithms, but employs cryptographic methods that are robust to both classical and quantum attack. Another option is to achieve security through the use quantum key distribution, which is one of the more mature quantum technologies and does not require the use of large quantum computers. In order to remain secure, forms of payment will need to be adapted to protect against such threats. Another possibility, which we examine, is to construct forms of money and payment instruments, broadly referred to as ``quantum money,'' which are robust to classical and quantum attacks.

Quantum money offers a bundle of features that is not achievable with physical cash or any digital currency scheme, including cryptocurrencies. In particular, quantum money is able to retain the beneficial features of a debit or credit transaction, including the ability to transact with minimal latency and at a distance, and the beneficial features of physical cash, such as the ability to transact offline and without the involvement of a trusted third party. It also achieves a higher level of security than is possible with any classical scheme. Furthermore, quantum money can also be issued and maintained by a central bank, which means that conventional monetary policy could be applied.

While quantum money remains technologically infeasible to implement at scale in the short-run, there has been substantial experimental progress in the past few years, including the partial implementation of quantum money and quantum credit cards. No public-key scheme has yet been implemented experimentally; and even the theoretical construction of such schemes remains a formidable analytical challenge. However, the benefits of implementing a public-key scheme would be substantial: it would allow for the possibility of combining both the anonymity and privacy of physical cash and the benefits that digital payment systems have introduced.

Finally, we conducted an exhaustive review of quantum algorithms that could potentially be used to solve or estimate economic models. We identify more than ten categories of algorithms that could provide quantum speedups to solution and estimation routines. In many cases, the quantum versions of the routines face limitations that were not present in the classical version. This is partly the result of quantum computers achieving speedups through the subtle exploitation of quantum physical phenomena, rather than brute-force parallelization. We also saw that phase kickback, phase estimation, and quantum Fourier transforms (QFTs) were common ingredients in algorithms that managed to achieve such speedups.

\section{Acknowledgements}
\label{Acknowledgements}
G.W. is supported by the European Commission through project 820363: OpenSuperQ, and by the Knut and Alice Wallenberg (KAW) foundation and Chalmers University of Technology through the WACQT project. E.D. is supported by the French National Research Agency through the project ANR-17-CE39-0005 quBIC. And O.S. is supported by the Israeli Science Foundation (ISF) grant No. 682/18 and 2137/19, and by the Cyber Security Research Center at Ben-Gurion University.

%% file: Appendix.tex
\section{Appendix}
\label{Appendix}

In order to make this introduction to quantum computing self-contained, we have included six additional short subsections in the Appendix. The first explains the concept of computational complexity and defines related notation. The following three sections explain commonly-used subroutines that appear frequently in quantum algorithms: phase kickback, phase estimation, and the quantum Fourier transform. Readers who want to understand the low-level workings of quantum algorithms will want to master these concepts. Finally, the remaining two sections are related to quantum money and provide the details of attacks on Wiesner's money and the quantum money scheme introduced in \citet{FGH+12}.

\subsection{Computational Complexity}\label{Computational Complexity}

The term ``computational complexity'' refers to the amount of resources needed to solve a problem as a function of the size of the input. We will often express speedups in terms of resource requirement reductions. The resources in question might be the number of elementary gates, the number of queries to an oracle, or the passage of time. We may denote these different notions of complexity as gate complexity, query complexity, and time complexity. 

We will typically describe the computational complexity of a problem using bounds and will use special notation to describe those bounds. Big-O notation, $O$, indicates that complexity grows no faster than some rate. For instance, if $O(N)$ is the complexity of a quantum circuit that takes $N$ bits of data as an input, then we say that the computational complexity grows no faster than linearly in $N$. If we instead want to indicate that computational complexity grows slower than linearly, then we will use little-o notation: $o(N)$.
Finally, to make a similar statements about the lower bound and about equivalence, we would use $\Omega(N)$ and $\Theta(N)$, respectively.

Note that measures of computational complexity omit constant terms and exclusively retain the highest power term. Consider the case where the number of steps needed to solve a problem is at most $10*N^{3} + N^{2} + N$. We would indicate that its computational complexity is cubic in the input size: $O(N^{3})$. We will often distinguish between problems with polynomial complexity, which are considered ``easy,'' and problems with exponential complexity, which are considered ``hard.'' In some cases, quantum algorithms will have polynomial complexity when the best classical algorithm has exponential complexity.

\subsection{Phase Kickback}\label{Phase Kickback}

Phase kickback is a common component of quantum algorithms, including the \cite{DJ92} algorithm, which was one of the first to achieve an exponential speedup over its classical counterparts. It makes use of a counterintuitive property of quantum circuits -- namely, that applying a controlled gate can actually change the state of the control qubit, rather than its target.

In the standard setup, we are given a unitary matrix, $U$, and one of its eigenvectors, $\ket{\psi}$, which is encoded as a quantum state. We want to determine the eigenphase, $\phi$, where $\ket{\phi} \in [0,1)$, which is unknown. Since $\ket{\psi}$ is an eigenvector of $U$ and $e^{2 \pi i \phi}$ is the associated eigenvalue, we know that the following relationship holds: $U \ket{\psi} = e^{2 \pi i \phi} \ket{\psi}$.

\begin{figure}
\begin{center}
\scalebox{1.5}{\mbox{
\Qcircuit
{
\ket{0}& &\gate{H} & \ctrl{1} & \qw & \qw \\
\ket{\psi} & & \qw & \gate{U} & \qw & \qw \\
}
}
}
\caption{The figure above shows the phase kickback effect in a two qubit circuit.  Note that $U$ is a unitary matrix and $\ket{\psi}$ is an eigenvector of that matrix. Counterintuitively, applying a controlled-U gate to an eigenvector of $U$ changes the state of the control qubit, but not the target qubit.}
\label{fig:phase_kickback}
\end{center}
\end{figure}
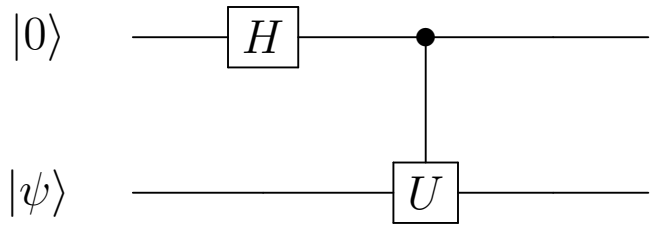

Figure \ref{fig:phase_kickback} demonstrates how phase kickback can be generated in a quantum circuit using $U$ and $\ket{\psi}$. We first initialize an ancilla qubit in state $\ket{0}$ and another qubit in state $\ket{\psi}$, yielding an initial state of $\ket{0} \ket{\psi}$. We then put the ancilla qubit in an equal superposition of states $\ket{0}$ and $\ket{1}$ by applying a Hadamard gate, $H$, yielding the state given in Equation (\ref{phase_kickback_1}). We then apply a controlled unitary, $U$, which gives us the state in Equation (\ref{phase_kickback_2}).

\begin{equation}\label{phase_kickback_1}
\frac{\ket{0}+\ket{1}}{\sqrt{2}} \ket{\psi}
\end{equation}

\begin{equation}\label{phase_kickback_2}
\frac{\ket{0}+e^{2 \pi i \phi} \ket{1}}{\sqrt{2}} \ket{\psi}
\end{equation}

Note that the state of the target qubit, $\ket{\psi}$, remains unchanged; however, the relative phase of the control qubit has picked up the eigenvalue, $e^{2 \pi i \phi}$, which is associated with the eigenvector, $\ket{\psi}$. It is applied to $\ket{1}$ because the control only executes $U$ when its state is $\ket{1}$.

Phase kickback is a valuable phenomenon that is exploited in many quantum algorithms. It is common to encode the solutions to computational problems in relative phases and then extract them by measuring the ancilla qubit. As we will discuss briefly in the following subsection, this can be done using a process called phase estimation.

\subsection{Phase Estimation}\label{Phase Estimation}

In the previous subsection, we used phase kickback to modify the relative phase of a superposition state by $e^{2 \pi i \phi}$, which is the eigenvalue associated with an eigenvector, $\ket{\psi}$, of a unitary matrix, $U$. Our intent was to extract the eigenphase, $\phi$. We did not do this, but the superposition we constructed in the ancilla qubit, $\frac{\ket{0}+e^{2 \pi i \phi} \ket{1}}{\sqrt{2}}$, can provide information about $\phi$.

More specifically, we can identify $\phi$ to arbitrarily high precision using phase estimation.\footnote{We will assume that $\phi$ is encoded as a binary fraction. A binary fraction of the form $\phi = 0.x_1 x_2 ...  x_{n-1}$ may be rewritten in base 10 as $\sum_{i=0}^{n-1} \frac{x_{i}}{2^{i+1}}$. Note that $x_{j} \in \{0,1\}$.} This can be done by expanding the number of ancilla bits from 1 to $n$, where $n$ captures the desired number of bits of precision. Each ancilla is then placed in an equal superposition by applying an $H$ gate. Finally, rather than applying a single controlled $U$, we will apply a sequence of controlled higher powers of $U$: $\{U^{2^{0}},...,U^{2^{n-1}} \}$. In each case, an ancilla qubit will act as the control and $\ket{\psi}$ will be the target.

The complete state of the system after the application of the Hadamard gates and controlled powers of $U$ is given in Equation (\ref{phase_estimation_1}).

\begin{equation}\label{phase_estimation_1}
\frac{\ket{0} + e^{2 \pi i 0.x_{n-1}} \ket{1}}{\sqrt{2}}  \otimes \frac{\ket{0} + e^{2 \pi i 0.x_{n-2}x_{n-1}} \ket{1}}{\sqrt{2}} \otimes ... \otimes \frac{\ket{0} + e^{2 \pi i 0.x_{0}...x_{n-2}x_{n-1}}\ket{1}}{\sqrt{2}} \ket{\psi}
\end{equation}

We next apply an inverse quantum Fourier transform, $F^{\dagger}$, yielding the state in Equation (\ref{phase_estimation_2}).

\begin{equation}\label{phase_estimation_2}
\ket{x_{n-1}} \ket{x_{n-2}} ... \ket{x_{0}} \ket{\phi}
\end{equation}

Recall that $x_{j} \in \{0,1\}$ and assume that $n = 5$ and that the system is in the state $\ket{1} \ket{0} \ket{1} \ket{1} \ket{0} \ket{\phi}$ after applying $F^{\dagger}$. This means that $\phi = 0.01101$, which we may rewrite as $\frac{0}{2} + \frac{1}{4} + \frac{1}{8} + \frac{1}{16} + \frac{0}{32} = 0.4375$ in base 10. For a more detailed description of phase estimation techniques, readers should consult \cite{AC12b}. 

\subsection{Quantum Fourier Transform}\label{Quantum Fourier Transform}

The quantum Fourier transform (QFT), which we will denote $F$, is the quantum equivalent of the classical discrete Fourier transform (DFT). Similar to the DFT, the QFT can be used to recover the period of a function. We will mostly use the QFT and inverse QFT, $F^{\dagger}$, as a subroutine within a quantum algorithm that serves a different purpose. There are several useful ways in which the QFT can be expressed, but we will use the one that has the most obvious direct relevance for phase estimation, which is given in Equation (\ref{qft_1}).

\begin{equation}\label{qft_1}
F(\ket{x_{n-1}} ... \ket{x_{0}}) = \frac{\ket{0} + e^{2 \pi i 0.x_{n-1}} \ket{1}}{\sqrt{2}}  \otimes  ... \otimes \frac{\ket{0} + e^{2 \pi i 0.x_{0}...x_{n-1}}\ket{1}}{\sqrt{2}}
\end{equation}

\noindent You may notice that Equation (\ref{qft_1}) differs from Equation (\ref{phase_estimation_2}) only by the presence of the eigenvector, $\ket{\phi}$. Consequently, we may invert $F^{\dagger}$ and apply it to the state given in Equation (\ref{qft_1}), allowing us to recover the eigenphase, as we did in the previous subsection.

\begin{figure}
\begin{center}
\scalebox{1.25}{\mbox{
\Qcircuit
{
\ket{x_2} & & \gate{H} &\gate{R_2} & \qw & \gate{R_3} & \qw & \qw &\qw \\
\ket{x_1} & & \qw & \ctrl{-1} & \gate{H} & \qw & \gate{R_2} & \qw & \qw\\
\ket{x_0} & & \qw & \qw & \qw & \ctrl{-2} & \ctrl{-1} & \gate{H} & \qw \\
}
}
}
\caption{The figure above shows a quantum Fourier transform applied to three qubits. It involves the application of Hadamard gates and controlled rotation gates. It is the quantum counterpart of the discrete Fourier transform and can be used to identify the period of a function.}
\label{fig:quantum_fourier_transform}
\end{center}
\end{figure}
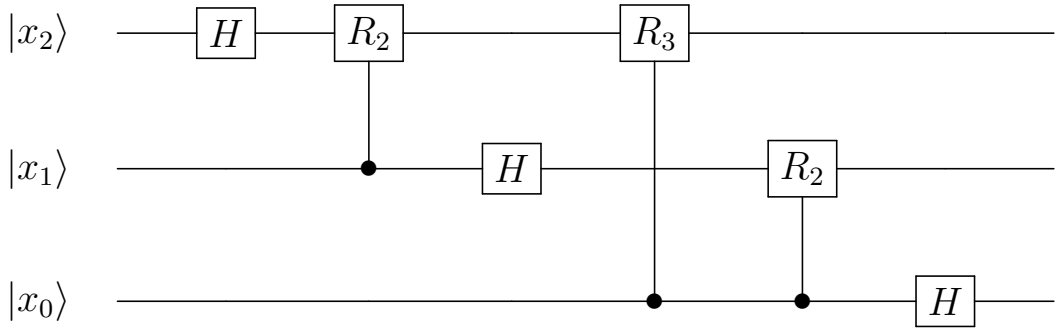

The quantum Fourier transformation can be implemented in a quantum circuit using a sequence of Hadamards and controlled rotation gates of the form given by Equation (\ref{rotation_gate}). An example QFT on three qubits is given in Figure \ref{fig:quantum_fourier_transform}. Note that the $R_{k}$ gates are defined as in Equation (\ref{rotation_gate}).

\begin{equation}\label{rotation_gate}
R_{k} = 
\begin{bmatrix}
1 & 0 \\
0 & e^{2 \pi i / 2^{k}}
\end{bmatrix}
\end{equation}

\subsection{Attacks on Wiesner's Scheme}
\label{Attacks on Wiesner}

\cite{Aar09}, \cite{Lut10}, \cite{NSBU16}, and \cite{MVW12} show that \cite{Wie83} and its early extensions are subject to adaptive attacks. \cite{Lut10} and \cite{Aar09} independently provided a simple adaptive attack scheme for \cite{Wie83} that works in linear time. They assume that the bank returns the post-measured state, both when the quantum money is valid and invalid. The attack is outlined below.
\begin{enumerate}
\item Assume a counterfeiter has a quantum bill in the following product state:
\begin{equation}
\ket{\$_{s}} = \ket{\psi_{1}} \ket{\psi_{2}} ...  \ket{\psi_{n}}
\end{equation}
\item If sent for verification and the state is correct, the central bank will return VALID, the serial number, $s$, and the quantum state, $\ket{\$_{s}}$. Otherwise, it will return INVALID, the serial number, $s$, and the post-measurement quantum state.
\item A counterfeiter can identify the underlying quantum state in linear time by guessing one qubit at a time. To do this for the $i$th qubit, she would send the state $(s,X_{i}\ket{\$_{s}})$ to the central bank. If she received INVALID as a response, then she would know that the state was either $\ket{0}$ or $\ket{1}$, since the other bases are eigenstates of $X_{i}$. Thus, the state returned would be as follows:
\begin{equation}
\ket{\psi_{1}}...\ket{\psi_{i}^{\perp}}...\ket{\psi_{n}}
\end{equation}
\item The counterfeiter now knows that $\ket{\psi_{i}^{\perp}}$ is an eigenstate of Z. She can then apply $X_{i}$ to recover $\ket{\$_{s}}$ and then measure $\ket{\psi_{i}}$ in the Z basis to determine whether the state is $\ket{0}$ or $\ket{1}$.
\item To the contrary, if the central bank answers VALID, then the counterfeiter has now learned that $\ket{\psi_{i}}$ must be either $\ket{-}$ or $\ket{+}$. The counterfeiter would receive $\ket{\$_{s}}$ back from the central bank, which she could then measure to determine whether the state was $\ket{-}$ or $\ket{+}$. 
\item The counterfeiter then repeats the process for all remaining qubits. This yields a complete description of $\ket{\$_{s}}$, which she can use to counterfeit an unlimited number of bills. 
\end{enumerate}

\noindent To prevent successful attacks, Lutomirski and Aaronson suggest that the bank return the quantum money state to the user only when the verification succeeds. \cite{NSBU16} showed that their proposal is also insecure: there is a way to reconstruct the quantum money state even when valid states are returned (and invalid states are not returned). \cite{NSBU16} proposed replacing the old quantum money state with a new quantum money state after each valid verification. \cite{MVW12} analyzed the optimal forging strategy for Wiesner's scheme in the non-adaptive setting, and proved that the probability of successfully counterfeiting a note decreases exponentially fast in the number of qubits.

\subsection{Knot-Based Quantum Money}
\label{Knot-Based Quantum Money}

\cite{FGH+12} propose a quantum lightning scheme that is difficult to copy and can be verified locally and without the use of a third party. They achieve this by coupling knot theory with exponentially-large superpositions. Unfortunately, they were not able to prove the security of their scheme. Rather, \cite{Lut11} showed that a problem \textit{related} to counterfeiting quantum money from knots is as hard as solving a knot theory computational problem. We still lack a full security proof for this scheme, and therefore have limited confidence in its security.

The scheme has the following properties:

\begin{enumerate}
\item The central bank produces pairs of serial numbers and quantum states, $(p,\ket{\$_{p}})$.
\item If a merchant receives a bill, $(p, \ket{\$_{p}})$, she can run a verification algorithm on $\ket{\$_{p}}$ that outputs either VALID or INVALID and leaves $\ket{\$_{p}}$ (almost) unchanged.
\item Given $(p, \ket{\$_{p}})$, it is difficult to make two states, $\ket{\psi}$ and $\ket{\psi'}$, each of which passes the verification algorithm.
\end{enumerate}

\noindent We first present a blueprint for the construction and then show how to instantiate it. Let $G$ be a large set and $P$ be a smaller set, where $f: G \rightarrow P$ is an efficiently computable function. A quantum bill state is generated using the following procedure:

\begin{enumerate}
\item Construct the initial state:
\begin{equation}
\ket{\mbox{initial}} = \frac{1}{\sqrt{|G|}} \sum_{g\in G} \ket{g} \ket{0}
\label{eq:knot_init}
\end{equation}
\item Compute the function, $f$, into the second register:
\begin{equation}
\frac{1}{\sqrt{|G|}} \sum_{g\in G} \ket{g} \ket{f(g)}
\end{equation}
\item Measure the second register. If the observed value is p and $N = |f^{-1}(p)|$, then the state is now:
\begin{equation}
\frac{1}{\sqrt{N}} \sum_{g \in G: f(g) = p} \ket{g} \ket{p}
\end{equation}
\item The first register contains the bill's state:
\begin{equation}
\ket{\$_{p}} = \frac{1}{\sqrt{N}} \sum_{g \in G: f(g) = p} \ket{g}
\end{equation}
\end{enumerate}

\noindent \cite{FGH+12} propose using tools from knot theory to instantiate the blueprint above. For the sake of simplicity, we will mostly avoid giving formal definitions for the notions that are used and instead provide specific examples for the relevant objects. 

\begin{figure}[!htb]
\minipage{0.25\textwidth}
  \includegraphics[width=\linewidth,bb=0 0 600 500]{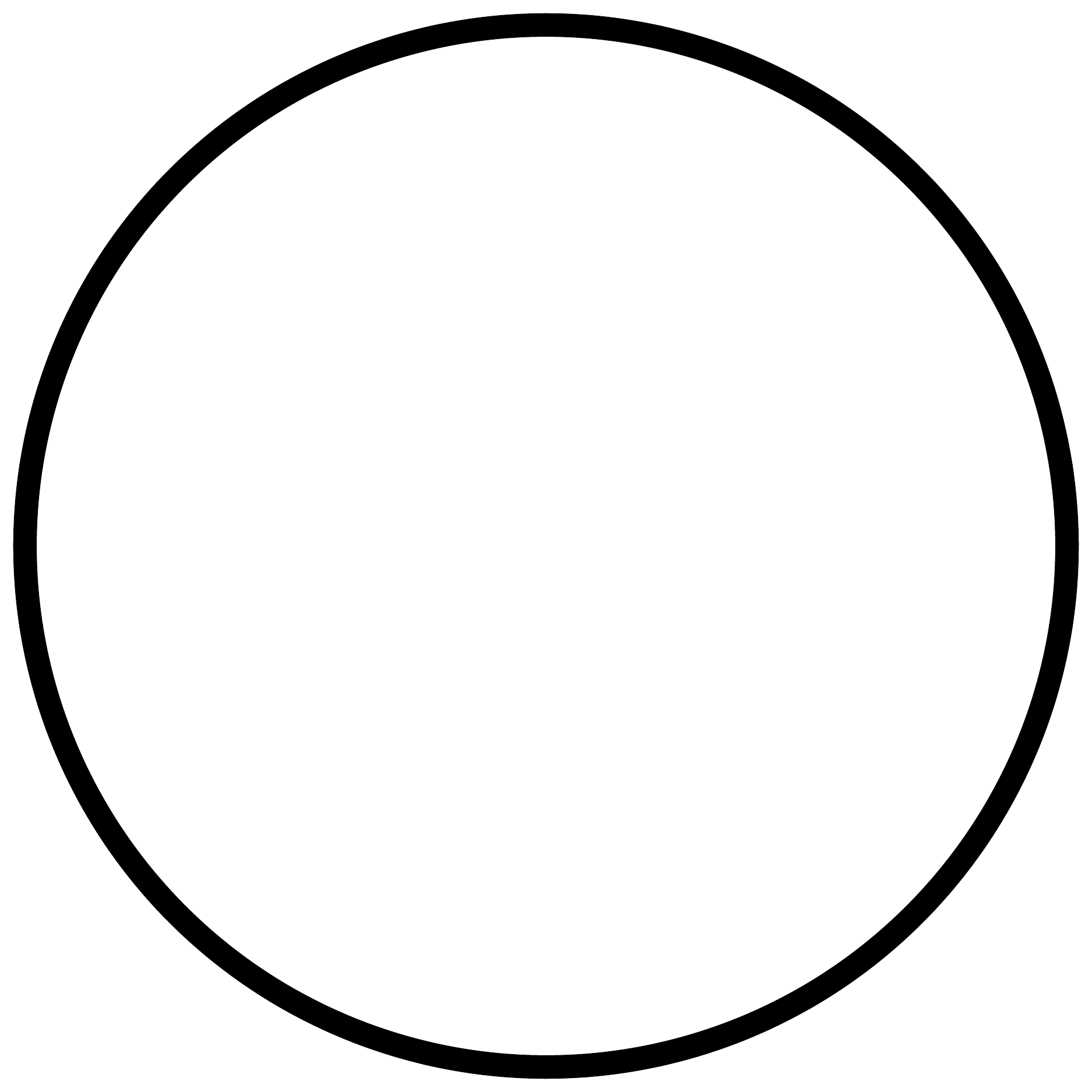}
  \caption*{Unknot}
\endminipage\hfill
\minipage{0.25\textwidth}
  \includegraphics[width=\linewidth,bb=0 0 600 500]{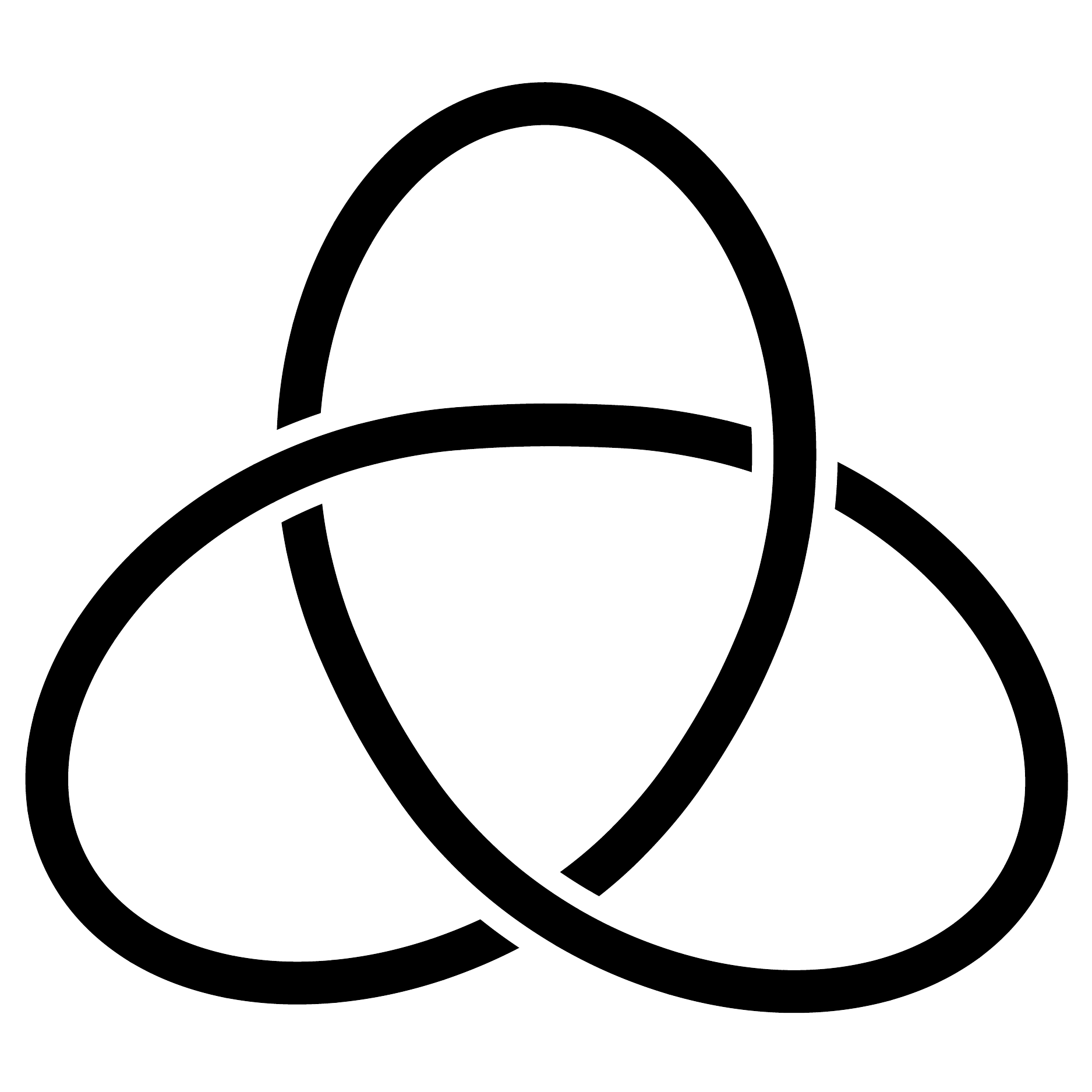}
  \caption*{Trefoil}
\endminipage\hfill
\minipage{0.25\textwidth}
  \includegraphics[width=\linewidth,bb=0 0 600 500]{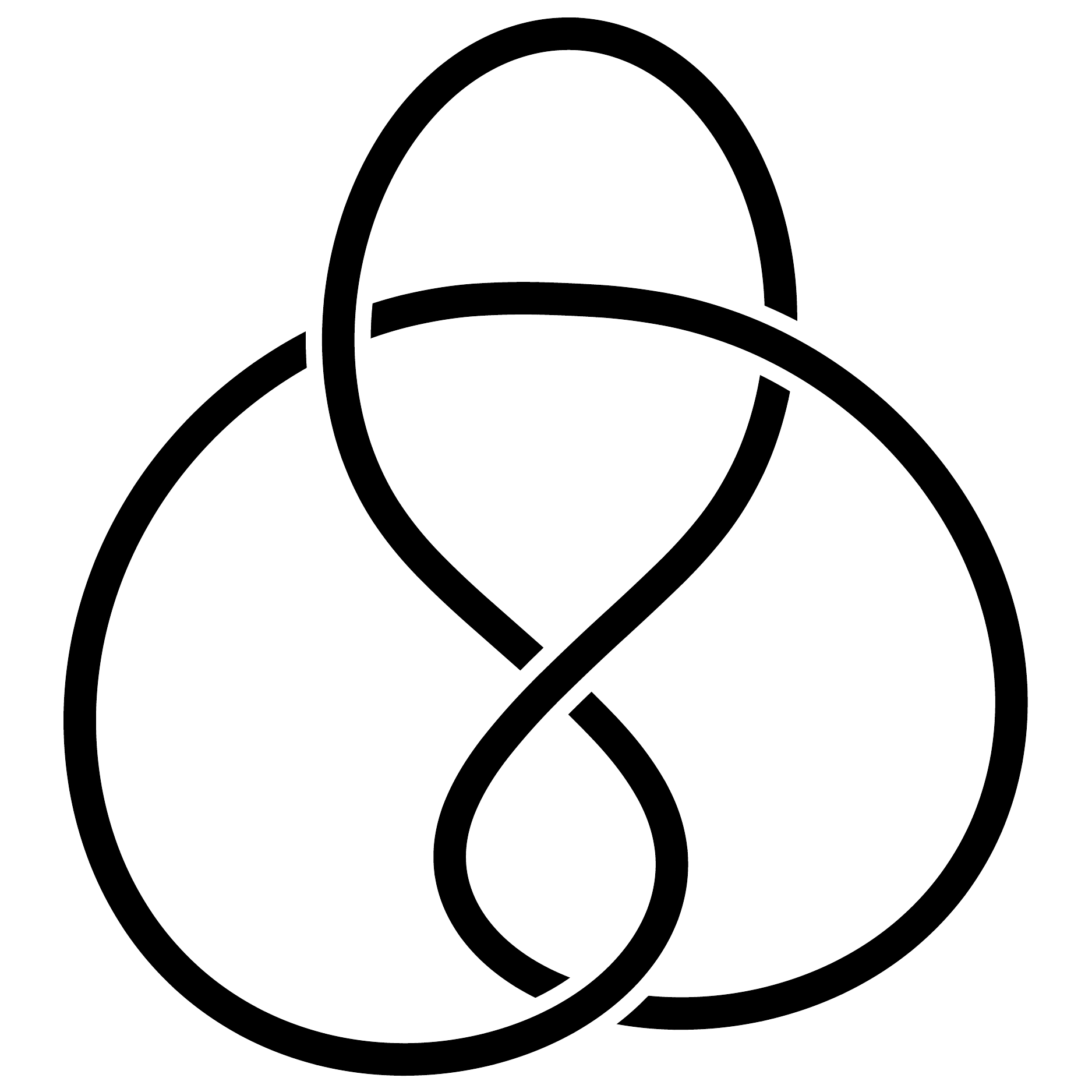}
  \caption*{Figure-Eight}
\endminipage
\caption{This figure shows the three simplest knots by number of crossings: the unknot, which has zero crossings; the trefoil knot, which has three crossings; and the figure-eight knot, which has four crossings. All other knots have at least five crossings.}
\label{fig:example_knots}
\end{figure}

\begin{figure}
\begin{center}
\includegraphics[width=0.25\textwidth,bb=0 0 500 500]{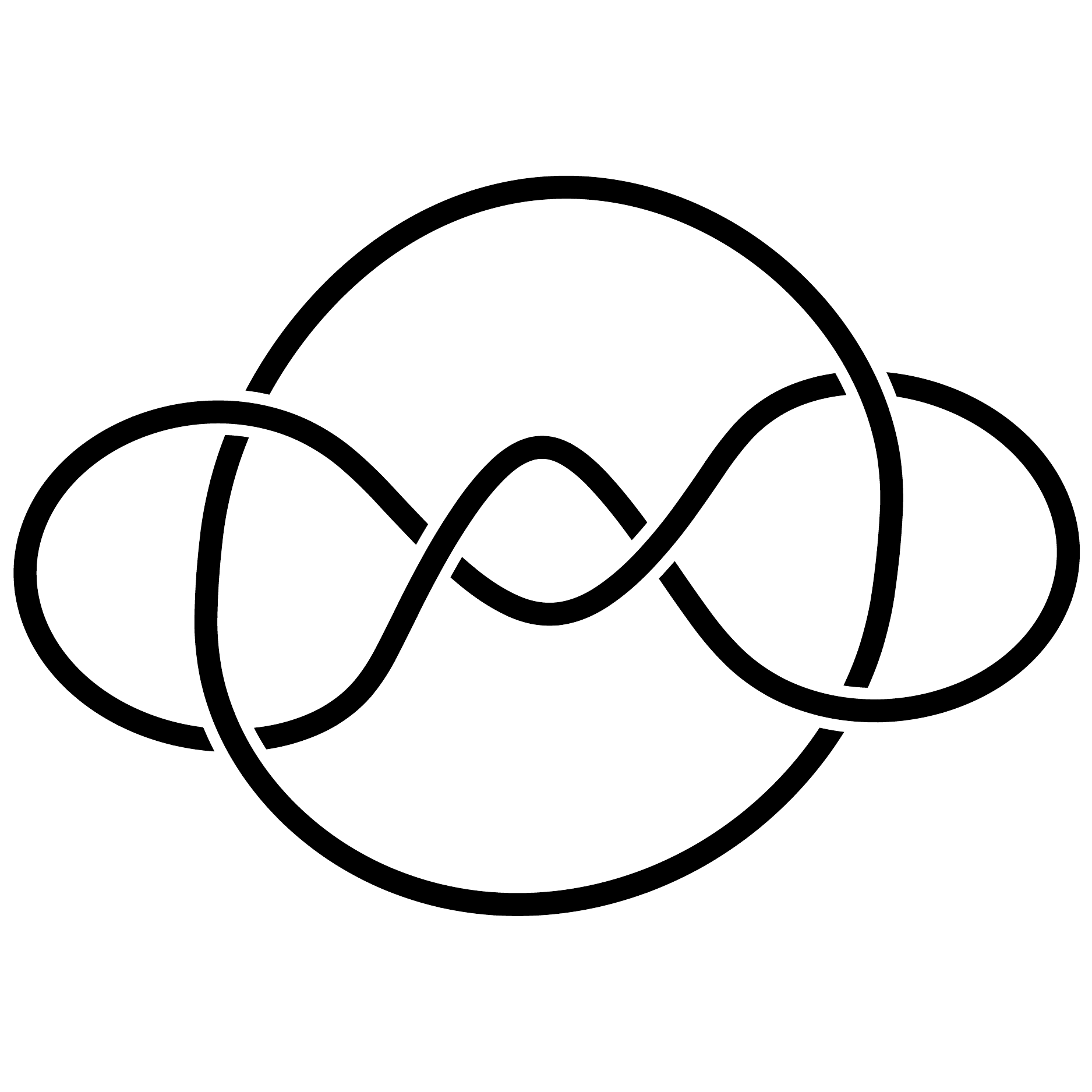}
\caption{The figure above shows an example of a link. A link is a group of knots that may be connected by knots, but do not intersect with each other.}
\label{fig:link_example}
\end{center}
\end{figure}

\begin{figure}[!htb]
\minipage{0.27\textwidth}
  \includegraphics[width=\linewidth,bb=0 0 500 175]{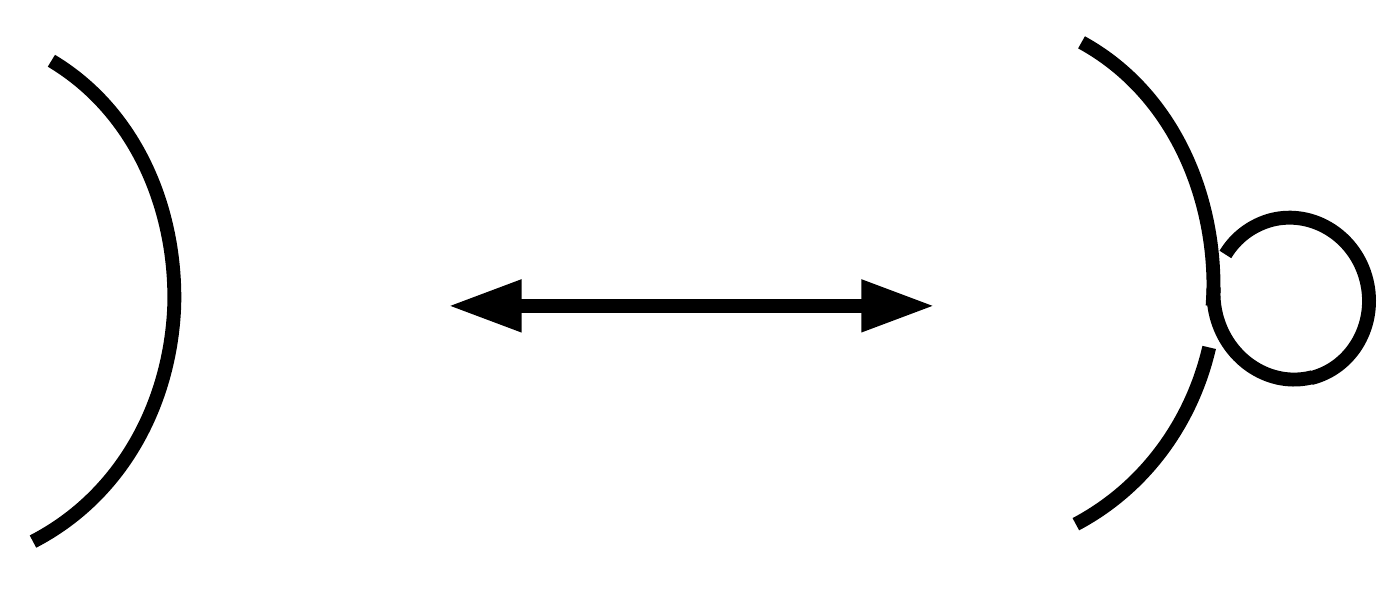}
  \caption*{I: Twist}
\endminipage\hfill
\minipage{0.27\textwidth}
  \includegraphics[width=\linewidth,bb=0 0 500 175]{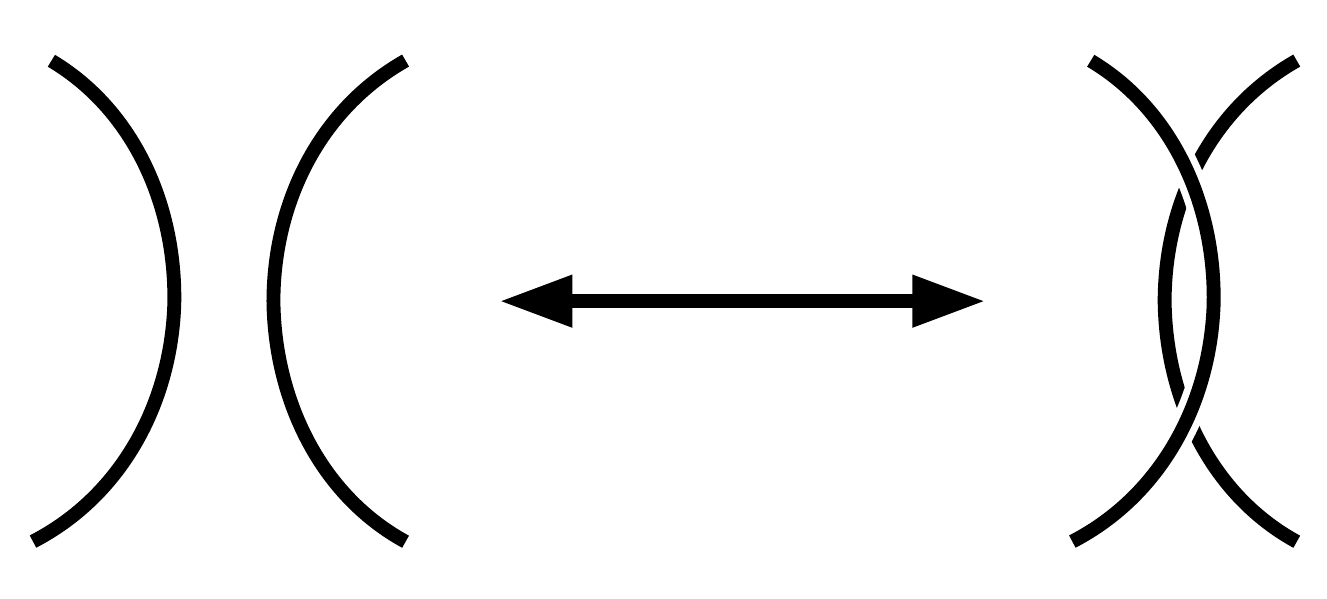}
  \caption*{II: Poke}
\endminipage\hfill
\minipage{0.27\textwidth}
  \includegraphics[width=\linewidth,bb=0 0 500 175]{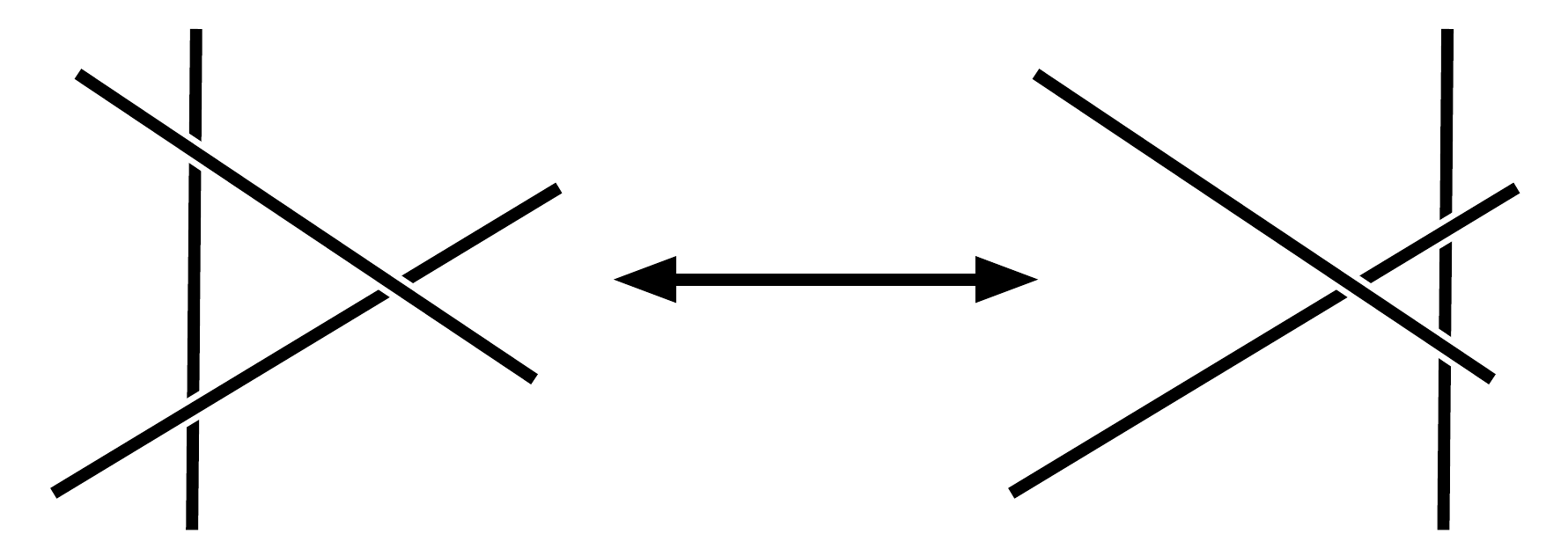}
  \caption*{III: Slide}
\endminipage
\caption{This figure shows the three Reidemeister moves. If two knots are equivalent, then one may be deformed into the other using a sequence of Reidemeister moves.}
\label{fig:reidemeister_moves}
\end{figure}

\begin{figure}[!htb]
\includegraphics[width=\linewidth,bb=0 0 400 220]{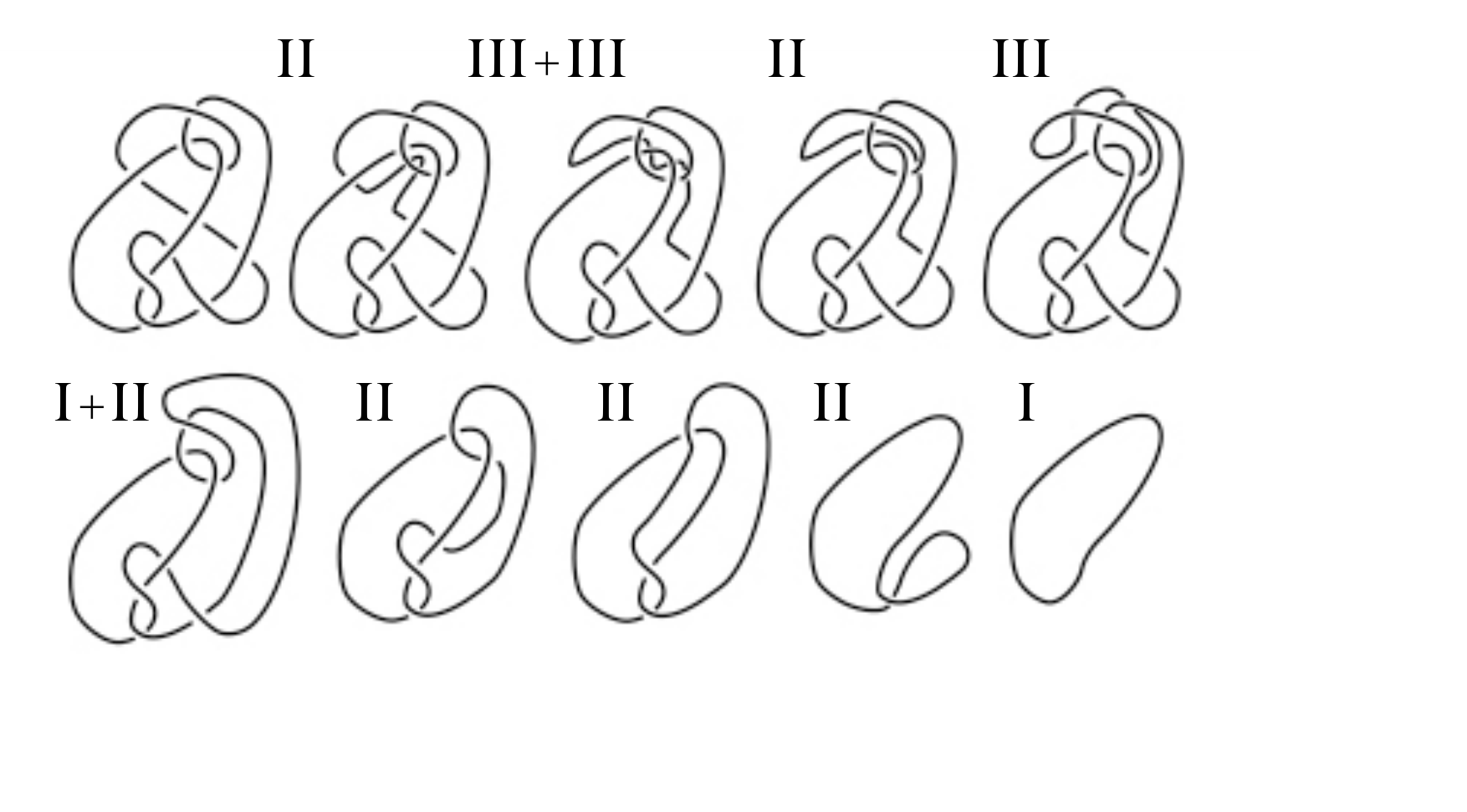}
\caption{This figure is reproduced from \cite{HK11} and modified to include labels for the Reidemeister moves. It shows the transformation of a knot called ``the culprit'' into the unknot. The transformation is counterintuitive because it requires the use of a first move that increases the number of crossings.}
\label{fig:culprit_unknot}
\end{figure}

In mathematics, a knot is a circle in $\mathbb{R}^3$. The three simplest knots are shown in Figure~\ref{fig:example_knots}. If we think of a knot as a connected piece of string, two knots are equivalent if they can be transformed into each other without cutting the string. A link, which is depicted in Figure~\ref{fig:link_example}, is a set of intertwined knots. There are three Reideimeister moves, shown in Figure~\ref{fig:reidemeister_moves}. Two links are equivalent if and only if there is a sequence of Reidemeister moves that transforms one link into the other. Figure~\ref{fig:culprit_unknot}, for instance, demonstrates the transformation of ``the culprit'' into the unknot using Reidemeister moves. The Alexander (Laurent) polynomial is an efficiently computable link invariant: it maps two equivalent links to the \textit{same} Laurent polynomial.

Farhi et al. instantiate their blueprint by picking $G$ to be the set of all links and $P$ the set of all  polynomials.\footnote{Here, we oversimplify the construction. In practice, they choose all links which are not too ``large'' according to their representation in a \textit{grid diagram}.} The function $f$ is the Alexander polynomial. 
This yields bill states of the following form:
\begin{equation}
\ket{\$_{p}} \propto \sum_{g\in G:A(G) = p} \ket{g}
\end{equation}
The summation is taken over all links such that $A(G)=p$. Finally, note that $\ket{\$_{s}}$ is an exponentially large superposition over all links that have the same Alexander polynomial.

We may now use the knot invariance property of Alexander polynomials to perform verification. To see how this works, suppose $\hat{P_s}$ is a unitary transformation that applies the Reidemeister move $s\in\{I,II,III\}$ and note that $\hat{P_s}\ket{\$_p}=\ket{\$_p}$ for all $s$ and $p$. The verification procedure effectively checks this invariance. This is done by adding an ancillary qubit in the $\ket{+}$ state, applying a controlled-$\hat{P_s}$ unitary, measuring the first qubit in the $\{\ket{+},\ket{-}\}$ basis, and accepting if the outcome is $+$. Note that $\textrm{controlled-}\hat{P_s}\ket{+}\otimes \ket{\$_p} = \ket{+}\otimes \ket{\$_p}$, and therefore the verification will accept valid money. This is repeated many times with all 3 Reidemeister moves.\footnote{Our presentation is intentionally simplistic and differs from the original presentation in \cite{FGH+12}. For example, the set of all links, $G$, is infinite, and therefore one cannot generate the state in Equation~\eqref{eq:knot_init}. \cite{FGH+12} also uses a specific representation of links called grid diagrams, along with grid moves, which are analogous to Reidemeister moves that can be efficiently applied to grid diagrams. This allows them to work with grid diagrams up to certain size, so that the set $G$ is finite.}

\pagebreak